\pdfoutput=1 
\documentclass[reprint,superscriptaddress,
amsmath,amssymb,aps,prx,
]{revtex4-2}
\usepackage{braket}
\usepackage{amsmath}
\usepackage{amssymb} 
\usepackage{graphicx} 
\usepackage{dcolumn}
\usepackage{bm} 
\usepackage{bbm}
\usepackage[sort&compress]{natbib}
\usepackage{hyperref}
\hypersetup{colorlinks=true,linkcolor=blue,citecolor=blue,urlcolor=black}
\usepackage{etoolbox}

\newcommand{\nn}[0]{\nonumber}

\allowdisplaybreaks
\begin{document}

\title{Directional superradiance in a driven ultracold atomic gas in free-space}

\author{Sanaa Agarwal}
\email{sanaa.agarwal@colorado.edu}
\affiliation{JILA, NIST, Department of Physics, University of Colorado, Boulder, CO 80309, USA
}
\affiliation{Center for Theory of Quantum Matter, University of Colorado, Boulder, CO 80309, USA
}
\author{Edwin Chaparro}
\affiliation{JILA, NIST, Department of Physics, University of Colorado, Boulder, CO 80309, USA
}
\affiliation{Center for Theory of Quantum Matter, University of Colorado, Boulder, CO 80309, USA
}
\author{Diego Barberena}
\affiliation{JILA, NIST, Department of Physics, University of Colorado, Boulder, CO 80309, USA
}
\affiliation{Center for Theory of Quantum Matter, University of Colorado, Boulder, CO 80309, USA
}
\author{A. Pi\~neiro Orioli}
\affiliation{QPerfect, 23 Rue du Loess, 67000 Strasbourg, France
}
\affiliation{University of Strasbourg and CNRS, CESQ and ISIS (UMR 7006), 67000 Strasbourg, France
}
\author{G. Ferioli}
\affiliation{Universite Paris-Saclay, Institut d’Optique Graduate School,
CNRS, Laboratoire Charles Fabry, 91127, Palaiseau, France}
\author{S. Pancaldi}
\affiliation{Universite Paris-Saclay, Institut d’Optique Graduate School,
CNRS, Laboratoire Charles Fabry, 91127, Palaiseau, France}
\author{I. Ferrier-Barbut}
\affiliation{Universite Paris-Saclay, Institut d’Optique Graduate School,
CNRS, Laboratoire Charles Fabry, 91127, Palaiseau, France}
\author{A. Browaeys}
\affiliation{Universite Paris-Saclay, Institut d’Optique Graduate School,
CNRS, Laboratoire Charles Fabry, 91127, Palaiseau, France}
\author{A. M. Rey}
\email{arey@jilau1.colorado.edu}
\affiliation{JILA, NIST, Department of Physics, University of Colorado, Boulder, CO 80309, USA
}
\affiliation{Center for Theory of Quantum Matter, University of Colorado, Boulder, CO 80309, USA
}

\date{\today}
\begin{abstract}
Ultra-cold atomic systems are among the most promising platforms that have the potential to shed light on the complex behavior of  many-body quantum systems. One prominent example is the case of a dense ensemble illuminated by a strong coherent drive while interacting via dipole-dipole interactions. Despite being subjected to intense investigations, this system retains many open questions. A recent experiment carried out in a pencil-shaped geometry \cite{ferioli_non-equilibrium_2023} reported measurements that seemed consistent with the  emergence of strong collective effects in the form of   a ``superradiant'' phase transition  in free space,  when looking at the light emission properties in the forward direction.
Motivated by the experimental observations, we carry out a systematic theoretical analysis of the system's steady-state properties  as a function of the driving strength and atom number, $N$. We observe signatures of  collective  effects in the  weak drive regime, which disappear with increasing drive strength as the system evolves into a single-particle-like  mixed state comprised of randomly aligned dipoles.  Although the steady-state features some similarities to the reported superradiant to normal non-equilibrium  transition, also known as cooperative resonance fluorescence, we observe significant qualitative and quantitative differences, including  a different scaling of the critical drive parameter (from $N$ to $\sqrt{N}$).  We validate the applicability of a mean-field treatment to capture the steady-state dynamics under currently accessible conditions. Furthermore, we develop a simple theoretical model that explains the scaling properties by accounting for interaction-induced inhomogeneous effects and spontaneous emission, which are intrinsic features of interacting disordered arrays in free space.

\end{abstract}
\maketitle


\section{\label{sec:intro}Introduction}

Atoms and photons are among the fundamental building blocks  of our universe. Their interactions rule the behavior of our physical world and understanding them is an essential need. However, it is  a challenging task as atom-light interactions can be extremely complex, especially in the context of  many-body quantum systems \cite{Gross1982,HammererRevModPhys2010}.

In most relevant situations, a large number of   electromagnetic  modes remain in the vacuum and the  photons just act  as passive mediators of excitations between atoms. The net effect is a  virtual exchange of  excitations which gives rise to   dipole–dipole interactions between atoms with   both dispersive and dissipative contributions \cite{stephen_1964,LehmbergPRA1970,FRIEDBERG1973101,JamesPRA1993}. Despite eliminating the vast  number of  electromagnetic vacuum modes from the picture and simplifying the complexity to only the  atomic degrees of freedom, understanding  the consequences of dipolar interactions in dense atomic samples is extremely complicated \cite{JenkinsPRA2012,javanainen_light_2016} and  apart from special limiting cases \cite{Gross1982} it remains a   long-standing problem in physical sciences.  

\begin{figure}[ht!]
\includegraphics[width=\linewidth]{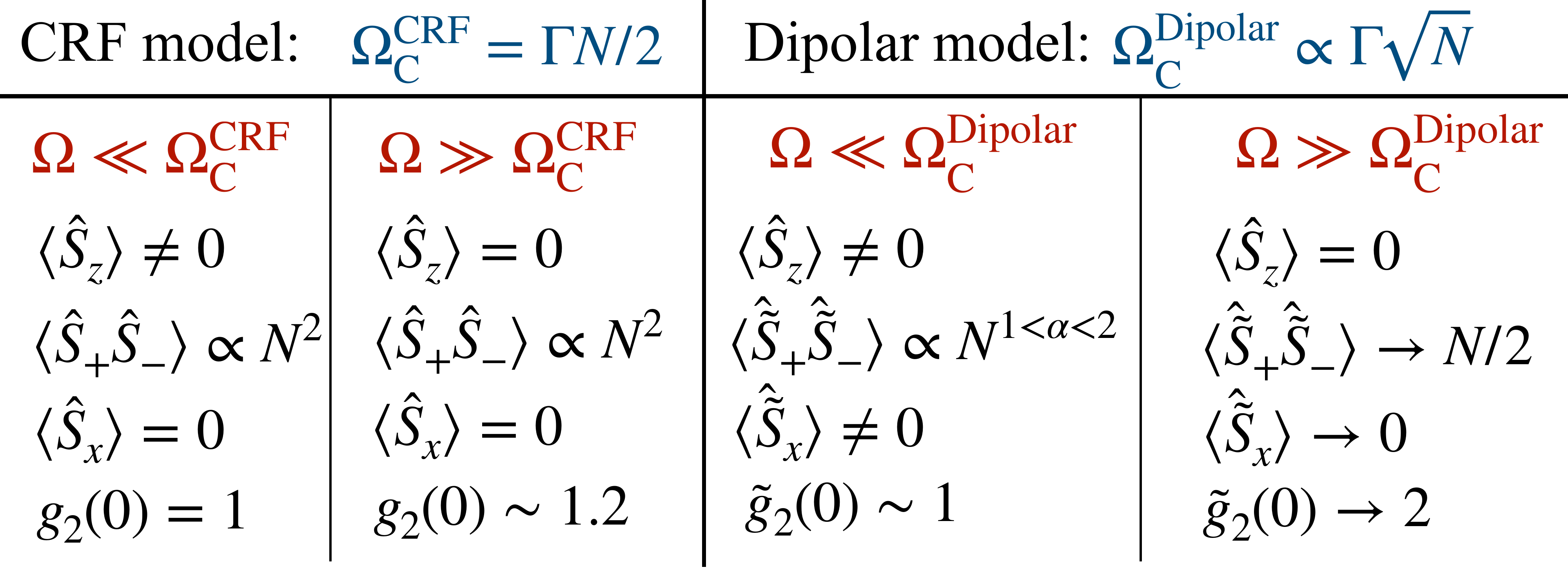}
\caption{Summary of the important steady-state observables in the weak ($\Omega\ll \Omega_{\rm C}$) and strong ($\Omega\gg \Omega_{\rm C}$) driving regimes in the CRF (left) and dipolar (right) models, with respect to a ``critical'' drive $\Omega_{\rm C}$ for each model. For the CRF model, we have the atomic inversion, $\langle \hat{S}_z\rangle$, as the order parameter of the regimes, the intensity, $\langle \hat{S}_+ \hat{S}_- \rangle$, the non-driven part of the atomic coherence, $\langle \hat{S}_x \rangle$, and the total equal-time two-photon correlation function, $g_2(0)$. For the dipolar model, the corresponding observables are in the spiral basis (defined in Eq.~(\ref{eq:tilted_basis_def})).}
\label{fig:Bloch_sph_pic}
\end{figure}

For the   case of effective two-level atoms, the weak excitation limit admits a simple semi-classical description and a great deal of theoretical \cite{PorrasPRA2008,SvidzinskyPRL2008} and experimental progress has been made in recent years, including the observation of collective level shifts \cite{IdoPRL2005,Scully2009,RöhlsbergerScience2010,KeaveneyPRL2012,MeirPRL2014,JavanainenPRL2014,bromley_collective_2016,RoofPRL2016,rui_subradiant_2020,ma2023collective,ma2023collective,gjonbalaj2023modifying,JenneweinPRA2018,GlicensteinPRL2020}, line broadening \cite{IdoPRL2005,ScullyDir2006,scully_correlated_2007,RöhlsbergerScience2010,berman_2010,OliveiraPRA2014,AraujoPRL2016,SutherlandPRA2016cloud,bromley_collective_2016,ZhuPRA2016,rui_subradiant_2020,gjonbalaj2023modifying,ma2023collective,GUERIN2023253,PellegrinoPRL2014,JenneweinPRL2016},  and  cooperative subradiant responses \cite{Pavolini_1985,BienaimePRL2012,Bienaime2013a,ScullySubrPRL2015,GuerinPRL2016,Rubies-Bigorda_2023,GUERIN2023253,FerioliPRX2021}  in optically thick \cite{Rouabah2014} and spatially ordered arrays \cite{plankensteiner_selective_2015,BettlesPRA2016,FacchinettiPRL2016,SutherlandPRA2016coll,JenPRA2016,rui_subradiant_2020}.

Away from the weak excitation limit, the problem  becomes  theoretically intractable, at least under current numerical capabilities \cite{mink2023collective}  and many open questions remain. One particular exception is the limit when only permutationally symmetric states, also known as Dicke states \cite{Dicke_1954} are populated. This situation arises naturally  in optical cavities, where a single   cavity mode  talks to all atoms independent of their location in the array. In this limit, the theoretical  treatment is significantly simpler and has been a focus of theoretical investigations for decades. One particular  case is the so called Cooperative Resonance Fluorescence (CRF) \cite{H_J_Carmichael_1977,Walls1978,drummond_volterra_1978,DFWalls_1980} or collective atomic emission \cite{Narducci_1978} that treats  the behavior of a group of two-level atoms coupled identically to a single radiation mode.  As first proposed by Dicke \cite{Dicke_1954}, it leads to 
a collective decay mechanism, superradiance,  which, as was later theoretically  shown, can be stabilized by driving the system below  a critical drive strength.  In the thermodynamic limit  of a large number of atoms (or a large 
cooperation number),  the competition between a coherent drive and decay gives rise to a non-equilibrium second-order phase transition.  Below  a threshold drive strength, the collective dipole reaches a highly pure steady-state  characterized by a collective Bloch vector below the equator, pointing at a polar angle at  which  the superradiant decay  and the external drive compensate each other.  On the other hand, above a critical drive, collective dissipation is not enough to stabilize the strongly driven system and at the mean-field  (MF) level, the system remains oscillating. Beyond-MF  effects dampen the oscillations via phase diffusion and the system becomes a highly mixed state with a distribution centered around the equator. 

A  recent experiment \cite{ferioli_non-equilibrium_2023} reported  signatures of the above collective Dicke transition  while interrogating a pencil-shaped cloud of $N$ Rb atoms optically excited by a laser propagating along its main axis. The experiment observed clear manifestations  of the two non-equilibrium phases depending on the ratio between the drive's single atom Rabi frequency, $\Omega$,  and the collective dissipation rate, $\Gamma N_{\rm eff}$, characterized by    the single particle decay rate, $\Gamma$, and an effective atom number, $N_{\rm eff}$, which accounted  for the finite extent of the cloud's
diffraction mode, $N_{\rm eff}\propto N$.  Above a critical drive $\Omega >\Omega_{\rm C}$, a scaling of the photon emission rate consistent with  $N_{\rm eff}^2$  was observed. This  scaling  was modified below $\Omega_{\rm C} $. As the system crossed the critical point,  the characteristics of the superradiant light changed as well. 

Regardless of these clear signatures, as explained in that work \cite{ferioli_non-equilibrium_2023}, the  applicability of a  fully  collective  model scaled by an  effective atom number to describe the light scattering of   an elongated sample in free-space is highly  unexpected.  A  justification of its  validity starting from a microscopic model  remains an open question.
In this paper we solve this issue by  performing  a detailed study of the light scattering properties of  pencil-shaped disordered arrays of two-level atoms  by directly solving a master equation. Our analysis   accounts for  the spatial extent of the cloud and the dipole-dipole interactions across the array, including  spatial fluctuations, elastic  dipole-dipole interactions, and single particle decay. Starting from a mean-field description of the master equation,  expected to be valid in the weak and strong excitation limits, and complementing it with a Moving Average Cluster Expansion method (MACE-MF) and the cumulant method, we  reproduce the experimental observations in all reported parameter regimes. Extended  calculations over a larger atom number window  reveal  a modified  scaling,  which is not proportional to an effective atom number $(N_{\rm eff})$ but to the square root of the total number of atoms $(\sqrt{N})$ in the array. 

Our study  demonstrates that the observed collective behavior is  not  valid over a broad parameter regime (see Fig.~\ref{fig:Bloch_sph_pic}) since it  misses important key features of free-space emission
such as  single  atom spontaneous emission and the frequency shifts arising from inhomogeneous elastic dipolar interactions. By combining  the latter mechanisms with   collective decay plus drive, we are able to qualitatively reproduce all the features observed in the mean-field calculations that account for microscopic details. 
In the future, it will be interesting to push experiments to more dense regimes where the mean-field model becomes invalid and  a genuine quantum treatment would be required in order to properly capture the role of  photon-mediated dipolar interactions in driven ultra-cold atoms. 

In the spirit of quantum simulation, our analysis takes advantage of state-of-the-art  experimental capabilities in regimes challenging for theory, uses them  to develop new theoretical insights  that   shed light on long standing problems,  and makes predictions to further inspire experimental work. 

In Sec.~\ref{sec:models}, we introduce the dipolar model, which describes our system at the microscopic level. We further discuss the extremely dilute and dense limiting cases, which are analytically solvable and useful for understanding the role of interactions in different driving regimes. In Sec.~\ref{sec:MF_dipolar_model}, we study the steady-state of the dipolar model using mean-field numerics in the weak and strong driving regimes, and we compare its emergent properties with the CRF and non-interacting models. Using these insights, we propose a phenomenological model in Sec.~\ref{sec:Mod_CRF_model}, which accounts for the inhomogeneity of dipolar interactions, is analytically tractable at the mean-field level, and qualitatively reproduces the dipolar model. In Sec.~\ref{sec:compare_with_exp}, we compare our steady-state and dynamics results with the experimental data \cite{ferioli_non-equilibrium_2023}, and find fair agreement for most observables. Lastly, in Sec.~\ref{sec:conclusions}, we discuss some concluding remarks and future directions.

\section{\label{sec:models}Quasi-1D gas of dipolar-interacting atoms}

\begin{figure}[ht!]
\includegraphics[width=\linewidth]{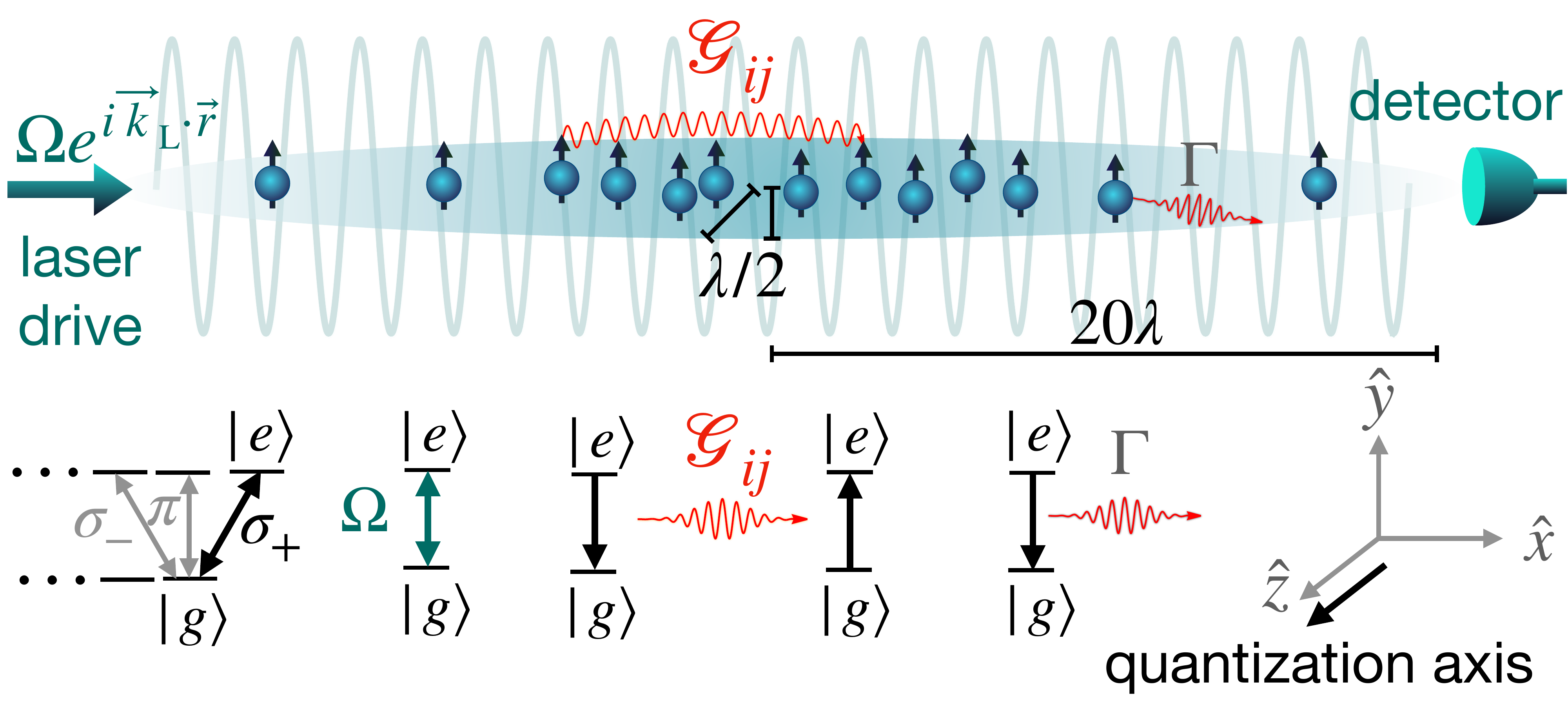}
\caption{\textbf{Schematic}: A pencil-shaped ultracold gas of frozen two-level atoms interacting via photon-mediated interactions, with elastic (${\rm Re} \mathcal{G}_{ij}$) and inelastic (${\rm Im} \mathcal{G}_{ij}$) components. A continuous laser drive excites the atoms on-resonance with Rabi frequency $\Omega$, wavevector $\vec{k}_{\rm L}=2\pi/\lambda \hat{x}$, and polarization $\hat{e}_{\rm L}=\hat{y}$ (perpendicular to the quantization axis, $\hat{z}$). Atoms spontaneously emit photons into free-space at rate $\Gamma$.}
\label{fig:schematic}
\end{figure}

\subsection{\label{subsec:dipolar_model} Dipolar model}

We consider a 3D elongated (pencil-shaped) gas of $N$  point-like atoms fixed at their Gaussian-sampled positions $\{\Vec{r}_k\}$, as shown in Fig.~\ref{fig:schematic}. The atoms have linewidth (single atom decay rate) $\Gamma$ and are driven by a coherent laser drive with Rabi strength $\Omega=|\vec{E}\cdot \vec{d}|/\hbar$ ($\vec{E}$ is the electric field amplitude of the laser and $\vec{d}$ is the transition dipole moment of the atom), detuning $\Delta=\omega_{\rm L}-\omega_0$ from the atomic transition (frequency $\omega_0$), wavevector $\Vec{k}_{\rm L}$, and polarisation $\hat{e}_{\rm L}$. The quantization axis $\hat{e}_0$ is along the $\hat{z}$-axis and the axial direction of the gas is along the $\hat{x}$-axis. As in Ref.~\cite{ferioli_non-equilibrium_2023}, we consider the case where $\hat{e}_{\rm L}=\hat{e}_y=\hat{y}$, i.e., the driving laser is linearly polarized and excites the $\sigma_+$-transition between the hyperfine levels $|F=2,m_F=2\rangle \leftrightarrow |F=3,m_F=3\rangle$. The atoms interact via induced dipole-dipole interactions mediated by the vacuum electromagnetic modes \cite{Gross1982}. 

We   assume that the atomic cloud is cold enough that  we can neglect motion of the atoms, treat  them as frozen  during the dynamics, and only focus on the internal state dynamics   spanned by the two relevant levels in each atom ($\ket{g}\equiv |F=2,m_F=2\rangle$ ground and $\ket{e} \equiv |F=3,m_F=3\rangle$ excited), which define a spin-$1/2$ system.   Under these conditions, which are similar to  the system realized in Ref.~\cite{ferioli_non-equilibrium_2023}, the state of an  atom can be described by  spin-$1/2$ Pauli operators: $\hat \sigma^z_k = \ket{e_k}\bra{e_k} - \ket{g_k}\bra{g_k},\, \hat \sigma^+_k = \ket{e_k}\bra{g_k},\, \hat \sigma^-_k = \ket{g_k}\bra{e_k}$, for an atom $k$. The collective spin operators are denoted as $\hat{S}_\alpha = \sum_{k=1}^N \hat\sigma_k^\alpha / 2$ for $\alpha\in\{x,y,z\}$ and $\hat{S}_\pm = \sum_{k=1}^N \hat\sigma_k^\pm$.

The dynamics of the system is governed by the master equation obtained upon adiabatically eliminating the photonic degrees of freedom, $\dot{\hat{\rho}} = -i[\hat{H}, \hat{\rho}]+\mathcal{L}(\hat{\rho})$, where $\hat{H}$ is the Hamiltonian, which leads to unitary evolution, and $\mathcal{L}(\hat{\rho})$ is the Lindbladian super-operator, which accounts for  all the dissipative processes. The Hamiltonian for the system is given as $\hat{H} = \hat{H}_0 + \hat{H}_{\rm int}$, where (setting $\hbar=1$)
\begin{align}
    \hat{H}_{0} &= - \frac{\Omega}{2} \sum_k (e^{i(\Vec{k}_{\rm L}\cdot \Vec{r}_k -\pi/2)} \hat\sigma^+_k + h.c.) -\Delta \hat{S}_z,
    \label{eq:H_0}
\end{align}
is the single-particle laser drive and
\begin{align}
    \hat{H}_{\rm int} &= -\sum_{j\neq k} \mathcal{R}_{kj} \hat\sigma^+_k \hat\sigma^-_j,
    \label{eq:H_1}
\end{align}
accounts for the dipole-dipole interaction. It sums over pairwise  exchange processes among two different atoms in the array with the interaction strength given by the free-space electromagnetic Green's tensor ${G}(\mathbf{r}) = ({3\Gamma}/{4})\left[(1-\hat{r} \otimes \hat{r})\frac{e^{i k_0 r}}{k_0 r} + (1-3\hat{r} \otimes \hat{r})\left(\frac{i e^{ik_0 r}}{(k_0 r)^2} - \frac{e^{i k_0 r}}{(k_0 r)^3}\right) \right]$, where  $\mathbf{r}$  is the vector connecting the two interacting atoms and  $k_0 = 2\pi/\lambda$ ($\lambda$ is the  atomic transition wavelength).
The elastic part   is determined by the real part  of $G(\mathbf{r})$,  $\mathcal{R}_{kj} = \hat{e}_+^{*^{\rm T}} \cdot {\rm Re \, [G} (\vec{r}_{kj})] \cdot \hat{e}_+$. Self-interactions are set to zero: $\mathcal{R}_{kk}=0$. The Lindbladian for the system is expressed as
\begin{align}
    \mathcal{L}(\hat{\rho}) = \sum_{j, k} \mathcal{I}_{kj} \left( 2 \hat\sigma^-_j  \hat{\rho} \hat\sigma^+_k - \{\hat\sigma^+_k \hat\sigma^-_j, \hat{\rho}\} \right),
    \label{eq:Lindbland}
\end{align}
where $\mathcal{I}_{kj} = \hat{e}_+^{*^{\rm T}} \cdot {\rm Im \, [G} (\vec{r}_{kj})]  \cdot \hat{e}_+$ is the inelastic dipolar interaction coefficient and $\mathcal{I}_{kk}=\Gamma/2$ is the spontaneous emission decay rate. We define the total dipolar interaction coefficient as $\mathcal{G}_{kj} = \mathcal{R}_{kj}+i\mathcal{I}_{kj}$. 

For simplicity, we define a ``spiral'' basis, in which we absorb the drive phase in the coherences, as 
\begin{align}\label{eq:tilted_basis_def}
    \hat{\tilde{\sigma}}_j^\pm  =  {\hat\sigma}_j^\pm  e^{\pm i(\vec{k}_{\rm L}\cdot\vec{r}_j -\pi/2)}.
\end{align} Then, the spiral collective spin operators are $\hat{\tilde{S}}_\pm = \sum_{j=1}^N \hat{\sigma}_j^\pm e^{\pm i(\vec{k}_{\rm L}\cdot\vec{r}_j -\pi/2)}$, $\hat{\tilde{S}}_x = (\hat{\tilde{S}}_++\hat{\tilde{S}}_-)/2$, $\hat{\tilde{S}}_y = -i(\hat{\tilde{S}}_+-\hat{\tilde{S}}_-)/2$, and $\hat{\tilde{S}}_z = \hat{{S}}_z$. Accordingly, we also define new interaction coefficients $\tilde{\mathcal{G}}_{kj} = \mathcal{G}_{kj} e^{-i\vec{k}_{\rm L}\cdot\vec{r}_{kj}}$, $\tilde{\mathcal{R}}_{kj}=\mathcal{R}_{kj} e^{-i\vec{k}_{\rm L}\cdot\vec{r}_{kj}}$, and $\tilde{\mathcal{I}}_{kj}=\mathcal{I}_{kj} e^{-i\vec{k}_{\rm L}\cdot\vec{r}_{kj}}$. In the spiral basis, the Hamiltonian and the Lindbladian are given as
\begin{align}
    \hat{H}_{0} &= - \Omega \hat{\tilde{S}}_x - \Delta \hat{S}_z, \label{eq:H_0_tilted}\\
    \hat{H}_{\rm int} &= -\sum_{j\neq k} \tilde{\mathcal{R}}_{kj} \hat{\tilde{\sigma}}^+_k \hat{\tilde{\sigma}}^-_j,\label{eq:H_int_tilted}\\
    \mathcal{L}(\hat{\rho}) &= \sum_{j, k} \tilde{\mathcal{I}}_{kj} \left( 2 \hat{\tilde{\sigma}}^-_j  \hat{\rho} \hat{\tilde{\sigma}}^+_k - \{\hat{\tilde{\sigma}}^+_k \hat{\tilde{\sigma}}^-_j, \hat{\rho}\} \right). \label{eq:Lindblad_tilted}
\end{align}
Thus, the drive acts collectively along the spiral-$\hat{x}$ direction in the Bloch sphere. Hereafter, we will work in the spiral basis and also set $\Delta=0$, unless otherwise mentioned.

Similar to the experimental protocol in Ref.~\cite{ferioli_non-equilibrium_2023}, we initialize all the atoms in the ground-state $\ket{g}^{\otimes N}$ and continuously drive them on resonance with the atomic transition, i.e., $\Delta=0$. The excitation of atoms by the drive is counteracted by the free-space  single particle and collective emission, which generates damping, allowing the system to eventually, reach  its steady-state. To characterize the system across a wide-range of drive strengths, $\Omega$, we look at the collective spin observables in the steady-state, namely, the atomic inversion $\langle \hat{S}_z\rangle$, the absolute value of the spiral atomic coherence $|\langle \hat{\tilde{S}}_+\rangle|$, and its real part $|\langle \hat{\tilde{S}}_x\rangle|$. 

The intensity operator describing the  photon emission from the atomic sample along a direction $\vec {k}$ can be written in terms of the spin operators  as \cite{MollowPR1969,LehmbergPRA1970,loudon2000quantum},
\begin{align}\label{eq:int_k_def}
\hat{\mathcal I}(\vec {k}) 
\equiv I_0(\vec {k}) \sum_{i, j} \hat{{\sigma}}^+_i\hat{{\sigma}}^-_j e^{i\vec{k} \cdot(\vec{r}_i - \vec{r}_j)},
\end{align} where $I_0(\vec{k}) $ is  a proportionality factor that accounts for the geometry of the dipolar emission pattern. In the spiral basis, the expectation value of the intensity  can be expressed  as  $ I(\vec {k})\equiv \langle \hat {\mathcal I}(\vec {k})\rangle 
\equiv I_0(\vec {k}) \sum_{i, j} \langle{\hat{\tilde{\sigma}}^+_i\hat{\tilde{\sigma}}^-_j}\rangle e^{i(\vec{k}-\vec{k}_{\rm L}) \cdot(\vec{r}_i - \vec{r}_j)}$. 
The intensity along the forward direction, $\vec{k}=\vec{k}_{\rm L}$,
features very interesting properties and will be the focus of this study. This direction is special because it is the  direction along which the driving laser imposes coherence and therefore, the intensity can be enhanced due to constructive interference. 

We also look into the equal-time two photon correlation function, defined as \cite{loudon2000quantum}
\begin{align}\label{eq:g2_k_kp_def}
    \tilde{g}_2(\vec{k},\vec{k'})=\frac{\langle : {\hat {\mathcal I} }(\vec {k}) \hat { {\mathcal I} }(\vec {k'}) :\rangle}{I(\vec {k}) I(\vec {k'}) },
\end{align}
which gives the likelihood of simultaneously emitting a photon along $\vec{k}'$ and another photon along the $\vec{k}$-direction. The  $::$  in the above expression  implies it needs to be evaluated using normal-ordered operators. 
In the experiment \cite{ferioli_non-equilibrium_2023} in question, the detector was placed in the forward direction to measure the above mentioned observables. Even though the experiment did  not measure  the atomic coherence,  we will  include it in our study  as it is useful for gaining an intuitive understanding of the system. In addition to the steady-state, we also compute the dynamics of  the forward intensity, $I(\vec{k}_{\rm L},t)$, and the excitation fraction, 
\begin{align}
    n_e (t) = \frac{1}{2N}\sum_{j=1}^N \left(\langle \hat \sigma_j^z (t) \rangle + 1 \right).
\end{align}

In the following, we use these observables to investigate distinct  key features of the system as a function of the strength of the drive, from the  weak  to the  strong driving regimes. To understand the role of interactions, we consider the cases of the extremely dense and dilute limits, which are analytically tractable and well-studied. Then, we study  the moderately dense ensemble of the experiment (Fig.~\ref{fig:schematic}) and compare it  with the extremely dilute and dense limits as a means to investigate the behaviors emergent from  dipolar interactions. Similar to the experiment, we keep the size of the atomic cloud fixed at an rms axial length $l_{\rm ax} = 20\lambda$ and radial length $l_{\rm rad} = \lambda/2$, as shown in Fig.~\ref{fig:schematic}. By varying $N$ we can tune the density of the cloud and thereby, the strength of the dipole-dipole interactions between the atoms. 
\subsection{\label{subsec:DDM} Cooperative Resonance Fluorescence (CRF)}

In the limit in which  all the atoms are confined within a single wavelength, ${|{r}_{0,N}|}<\lambda$, we have $\mathcal{R}_{kj}\sim 1/(|r_{kj}|/\lambda)^3, \, \mathcal{I}_{kj} \to \Gamma/2$. 
If we choose to neglect the elastic part of the interactions ($\mathcal{R}_{kj}=0$), and freeze the motion of the atoms, we can emulate the situation found in an optical cavity, where all the atoms interact via a single electromagnetic mode. When the cavity mode is set on resonance with the atomic transition, the elastic interactions are suppressed and the dominant interactions are the ones responsible for collective decay or superradiance, via which atoms emit collectively at an enhanced emission rate $\Gamma N$. In the presence of an additional drive, $\Omega$, the system reduces to the  well-studied Cooperative-Resonance-Fluorescence (CRF) model, which describes a system of $N$ atoms  driven with a resonant laser drive at a Rabi frequency $\Omega$, and subject to collective  decay (superradiance) described by the jump operator  $\sqrt{\Gamma} \hat{S}_{-}$  with  $\hat{S}_\pm = \sum_i \hat{\sigma}^\pm_i$ \cite{Bonifacio1971}. 
Here we use the notation CRF to distinguish from the so-called Dicke model \cite{HEPP1973360,WangPRA1973,Roses_2020} which contains both the rotating and counter-rotating terms. In the CRF model, the counter-rotating terms are irrelevant and therefore neglected.  
The master equation governing the dynamics of the CRF model is given by:
\begin{align}
\partial_{t}\hat{\rho}=-i\Omega\left[\hat{S}_x,\hat{\rho}\right]+\frac{\Gamma}{2}\left(2\hat{S}_{-}\hat{\rho}\hat{S}_{+}-\left\{ \hat{S}_{+}\hat{S}_{-},\hat{\rho}\right\} \right)\label{meq}
\end{align} 
where the drive, $\Omega$, is applied along the $x$-direction of the Bloch-sphere.

\begin{figure}[ht!]
\includegraphics[width=\linewidth]{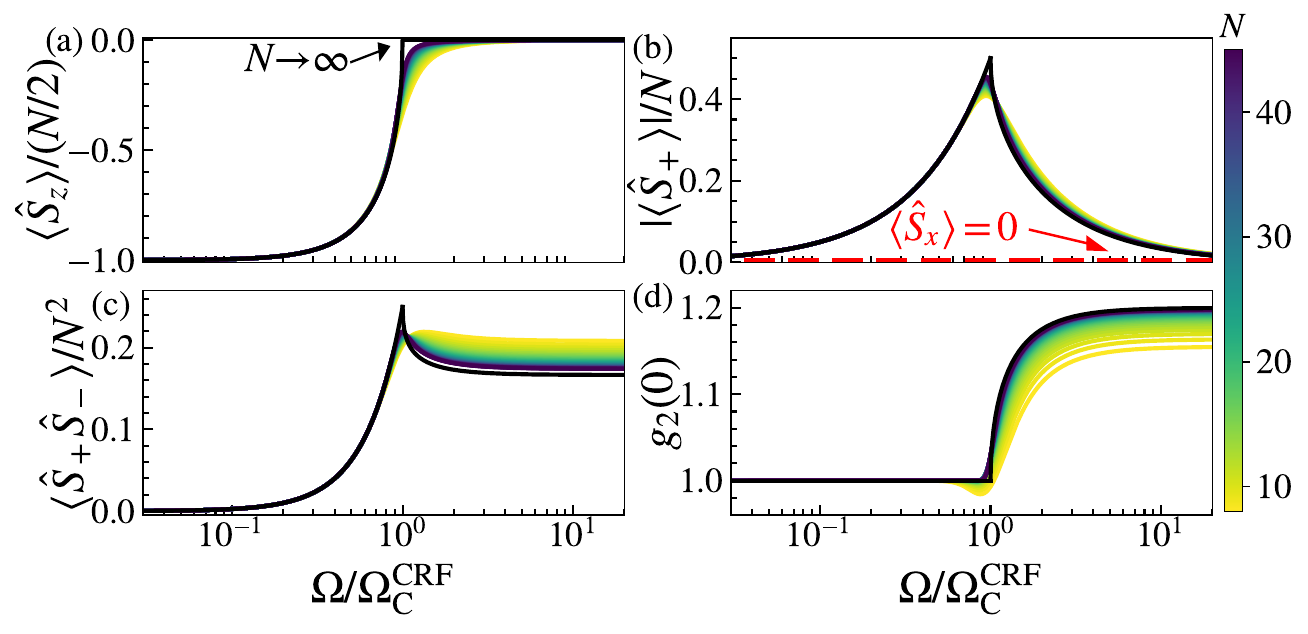}
\caption{Steady-state of the CRF model using exact diagonalization (extremely dense limit, $r_{ij}\to 0$, and artificially setting $\mathcal{R}_{ij}=0$): (a) atomic inversion $\langle \hat{S}_z\rangle/(N/2)$ shows the superradiant phase transition in the thermodynamic limit, (b) total coherence $|\langle \hat{S}_+\rangle|/N = |\langle \hat{S}_y\rangle|/N$ driven by the laser drive, which is along the $\hat{S}_x$-direction, such that $\langle \hat{S}_x\rangle=0$, (c) intensity scales as $N^2$ and remains constant as $\Omega$ is varied in the strong-drive regime (due to preservation of $\langle \mathbf{\hat{S}}^2\rangle = \langle \hat{S}_+ \hat{S}_- \rangle + \langle \hat{S}_z^2 \rangle = N(N+1)/4$), (d) two-photon correlation function ${g}_2(0)$ does not reach the maximally mixed state (${g}_2(0)=2$) due to preserved quantum correlations. $N\to \infty$ curves correspond to Eqs.~(\ref{eq:Sz_Dicke})-(\ref{eq:g2_Dicke}).}
\label{fig:SS_CRF_4plots}
\end{figure}

The CRF exhibits a steady-state phase transition at a critical frequency  $\Omega_{\rm C}^{\rm CRF}=N\Gamma/2$  \cite{carmichael_analytical_1980}, which delineates two distinct steady-state behaviors depending on the dimensionless parameter $\beta=\Omega/\Omega_{\rm C}^{\rm CRF}$. For $\beta<1$, the system is in the ``superradiant phase'', where the drive is balanced by the collective emission, canceling the total electric field experienced by the collective dipole. We use quotation marks to remark that this regime was initially referred to instead as ``superfluorescent'' by the authors \cite{carmichael_analytical_1980,drummond_volterra_1978, H_J_Carmichael_1977}.
In this phase, the collective dipole lies in the $y-z$ plane of the Bloch sphere at an angle $\theta$ from the south pole, given by $\sin\theta=\beta$. For $\beta>1$, the system transitions into a highly mixed steady-state, known as the ``normal'' phase. The statistical mixture arises from the collective emission not being strong enough to compensate the excitation from the drive. As the atomic self-radiated field is not canceled by the drive, the dipoles undergo collective Rabi flopping. Although Rabi oscillations persist at long times at the mean-field level, quantum fluctuations lead to phase diffusion, which dampens the oscillations and destroys the coherences, although the system always remains in the collective manifold by construction. These two distinct steady-state behaviors are separated by a second-order phase transition at $\beta=1$, as shown in Fig.~\ref{fig:SS_CRF_4plots}a.

Here it is important to note that  Ref.~\cite{ferioli_non-equilibrium_2023} used the opposite convention and denoted the $\beta > 1$ regime as the superradiant phase. This is inspired by Refs.~\cite{Hannukainen2018,Link2019} and  the fact that Eq.~(\ref{meq})
has a discrete symmetry characterized by its invariance under a mirror reflection $\hat{S}_x\to -\hat{S}_x$ followed by complex conjugation. This  mirror symmetry of the Lindblad generator is
spontaneously broken in  the $\beta > 1$ phase and therefore under this view, this is the phase featuring spontaneous symmetry  breaking.  For the purpose of this work, this is just a different convention and does not affect our conclusions. We will stick to the more standard CRF convention that  uses  $ \hat{S}_z$ as the physically motivated  order parameter and denotes  the $\beta<1$ regime as the superradiant phase. 

By doing a Holstein-Primakoff expansion around the mean-field steady-state in the superradiant regime as well as 
introducing a semi-classical method to average in the normal phase \cite{Barberena_averages}, we can express the steady-state analytically across all driving regimes by varying $\beta$ (with $\eta=\sqrt{\beta^2-1}$) as (see Appendix~\ref{APP_sec:CRF_calc} for derivation):
\begin{align}\label{eq:Sz_Dicke}
\langle \hat{S}_{z}\rangle  & =\begin{cases}
-\frac{N}{2}\sqrt{1-\beta^{2}} \text{,} & \beta<1\\
0, & \beta\geq1
\end{cases}\\
\label{eq:Sx_Dicke}
\langle \hat{S}_{x}\rangle  & =0,\\
\label{eq:Sy_CRF}
\langle \hat{S}_{y}\rangle  & =\begin{cases}
\frac{N}{2}\beta \text{,} & \beta<1\\
\frac{N}{2}\left(\beta-\frac{\eta}{\beta\arctan\left(\frac{1}{\eta}\right)}\right) , & \beta\geq1
\end{cases}\\
\label{eq:SpSm_Dicke}
\langle \hat{S}_{+}\hat{S}_{-}\rangle  & =\begin{cases}
\frac{N^{2}}{4}\beta^{2} \text{,} & \beta<1\\
\frac{N^{2}}{4}\left(\beta^{2}-\frac{\eta}{\arctan\left(\frac{1}{\eta}\right)}\right), & \beta\geq1
\end{cases}\\\label{eq:g2_Dicke}
g_{2}\left(0\right) & =\begin{cases}
1 \text{,} & \beta<1\\
\eta\left(\frac{\beta^{4}}{\eta}\arctan\left(\frac{1}{\eta}\right)-\beta^{2}-\frac{2}{3}\right)\\
\times\arctan\left(\frac{1}{\eta}\right)\left(\beta^{2}\arctan\left(\frac{1}{\eta}\right)-\eta\right)^{2} , & \beta\geq1
\end{cases}
\end{align}

For this phase transition, the order parameter is the population inversion, $\langle \hat{S}_z\rangle$, which is non-zero in the superradiant phase and zero in the normal phase (Eq.~(\ref{eq:Sz_Dicke})). $\langle \hat{S}_z\rangle$ is continuous but features an abrupt change in its derivative at the critical point in the thermodynamic limit ($N\to\infty$), as shown in Fig.~\ref{fig:SS_CRF_4plots}a. The finite-$N$ steady-state in Fig.~\ref{fig:SS_CRF_4plots} is obtained using exact diagonalization.

The observable $\langle\hat{S}_x\rangle$ is conserved and zero in all regimes (Eq.~(\ref{eq:Sx_Dicke})). This is a distinguishing feature of this model, which will become important when comparing with the other models in this paper. Thus, the atomic coherence is purely imaginary, $\langle \hat{S}_+ \rangle = i \langle\hat{S}_y\rangle$. In the superradiant phase, $\langle\hat{S}_y\rangle$ grows linearly with $\beta$. On the other hand, as $\beta$ is increased in the normal phase, the steady-state becomes more and more mixed, and $\langle\hat{S}_y\rangle$ goes to zero, as shown in Fig.~\ref{fig:SS_CRF_4plots}b. 
 
The photon emission rate (intensity), $I=\langle\hat{S}_+\hat{S}_-\rangle$,  at a fixed $\beta$, increases as $N^2\beta^2$ in the superradiant phase. 
In the normal phase, the intensity still scales as $N^2$, reflecting the collective nature of the system, as $\langle \hat{\bf S}\cdot \hat{\bf S}\rangle=N/2(N/2+1)$, where $\hat{\bf S}=\{\hat S_x,\hat S_y,\hat S_z\}$. However, the normal phase intensity is independent of $\beta$ due to the mixed nature of the state in  the normal phase. In fact, for $\beta\gg 1$, the intensity reaches the asymptotic value of $I=N^2/6$, as shown in Fig.~\ref{fig:SS_CRF_4plots}c. The coherent nature of the steady-state in the superradiant phase is also evident in the value of the two-photon correlation function (Fig.~\ref{fig:SS_CRF_4plots}d), as  $g_2\left(0\right)=1$ for $\beta<1$, in the large-$N$ limit. In the normal phase, on the contrary, the system enters a regime where $g_2\left(0\right)>1$, suggesting the mixed nature of the steady-state and the build-up of classical correlations. 
\subsection{Non-interacting model}

In the extremely dilute limit, the mean inter-atomic distance is large, ${\bar|{r}_{ij}|}\gg \lambda$, and thus to a good approximation we can neglect the interactions in Eq.~(\ref{eq:H_int_tilted}) and Eq.~(\ref{eq:Lindblad_tilted}) by setting $\mathcal{R}_{kj}=0$ and $\mathcal{I}_{k\neq j}=0$. This gives a simplified master equation for the non-interacting model, $\dot{\hat{\rho}} = -i[\hat{H}_0, \hat{\rho}]+\mathcal{L}_0(\hat{\rho})$, that describes an array of independent atoms coherently interacting with a classical laser drive while  spontaneously emitting photons at a rate $\Gamma$.  The Lindbladian, $\mathcal{L}_0(\hat{\rho})= (\Gamma/2)\sum_{j} \left( 2 \hat{\tilde{\sigma}}^-_j  \hat{\rho} \hat{\tilde{\sigma}}^+_j - \{\hat\sigma^z_j, \hat{\rho}\}/2 - \hat{\rho}  \right)$, captures the single-particle spontaneous emission.  The laser drive acts as an effective global magnetic field, $\vec{B}_{\rm L}=( {\Omega}/{2},0,{\Delta}/{2})$.  The corresponding  dynamics  of the Bloch vector,
$\langle\Vec{\hat{\tilde{\sigma}}}\rangle = (\langle\hat{\tilde{\sigma}}^x\rangle,\langle\hat{\tilde{\sigma}}^y\rangle,\langle\hat{\tilde{\sigma}}^z\rangle)$, can 
be described by the well-known Bloch equations as
\begin{align}\label{eq:Bloch_eq_non-int}
    \frac{d\langle \Vec{\hat{\tilde{\sigma}}}\rangle}{dt} &= 2  \langle \vec{\hat{\tilde{\sigma}}}\rangle \times   \vec{B}_{\rm L}  - \vec{f}(\langle \vec{\hat{\tilde{\sigma}}}\rangle),
\end{align}
where $\vec{f}(\langle \vec{\hat{\tilde{\sigma}}}\rangle)=- ({\Gamma}/{2}) [ \langle \vec{\hat{\tilde{\sigma}}}\rangle + (\langle\hat{\tilde{\sigma}}^z \rangle+2)\hat{z}  ]$ accounts for the damping.

For the on-resonant case ($\Delta=0$), the steady-state Bloch vector is obtained as
\begin{align}\label{eq:non_int_ss_sp_sz}
\langle\hat  {\tilde \sigma}^+ \rangle = i\frac{\Omega}{\Gamma}  \langle \hat \sigma^z \rangle ,\quad \langle \hat \sigma^z \rangle = -R \equiv -\frac{1}{1+2(\Omega/\Gamma)^2} ,
\end{align} 
and the steady-state intensity is
\begin{align}
I(\vec{k}) = I_0(\vec{k})\frac{(\Omega/\Gamma)^2}{1+2(\Omega/\Gamma)^2}\bigg[N +  \sum_{j,i\neq j} \frac{e^{i(\vec{k} - \vec{k}_{\rm L})\cdot(\vec{r}_i - \vec{r}_j)}}{1+2(\Omega/\Gamma)^2}\bigg],\label{spi}
\end{align} 
which is dominated by the coherences in the weak-driving regime ($\Omega \ll \Gamma$) and by the incoherent single-particle-like term in the strong-driving regime ($\Omega \gg \Gamma$). In the large-$N$ limit, the steady-state intensity can be obtained analytically by converting the sum above to an integral (with $\hat{k} = (\cos \theta, \sin \theta \cos \phi, \sin \theta \sin \phi)$), as 
\begin{align}
    I(\hat{k}) &\propto \frac{2\left(\frac{\Omega}{\Gamma}\right)^4 N}{\big[1+2\left(\frac{\Omega}{\Gamma}\right)^2\big]^2}\bigg[ 1  + \frac{N e^{-(2\pi)^2 \left(\sigma_{\rm ax}^2 (1-\cos\theta)^2 + \sigma_{\rm rad}^2 \sin^2\theta \right)}}{2(\Omega/\Gamma)^2}  \bigg],
\end{align}
where $\sigma_{\rm ax}$ and $\sigma_{\rm rad}$ are the axial and radial extents of the cloud, respectively.
As discussed earlier, when the laser-wavevector aligns with the direction of observation ($\vec{k} = \vec{k}_{\rm L} \Rightarrow \theta = 0$), the imprinted phases interfere constructively. In the weak-drive limit, the forward emission is dominated by its coherent part, $I(\vec{k}_{{\rm L}}) \approx I_0(\vec{k}_{{\rm L}})  |\langle \hat{\tilde{\sigma}}^+ \rangle|^2 N(N-1)$. This $N^2$-enhancement of the emission rate is only seen in the forward direction and vanishes exponentially fast with increasing $\theta$ when $\theta\ll 1$ (Eq.~(\ref{eq:int_non_int_k_dir_SS_close_to_forw})). Along other $\vec{k}$, the phases in Eq.~(\ref{spi}) do not cancel and the intensity, dominated by its incoherent part, scales as $N$ far away from $\vec{k}_{\rm L}$.   
It is worth noting that this $N^2$-enhancement is not due to quantum correlations and arises purely from the coherences of the atoms.

For comparison with the other models in this work, we define the notion of a ``critical'' drive strength, $\rm\Omega_C^{Non-int.}$, that maximizes the ``forward'' intensity. 
There is, however, no phase transition in the non-interacting model. This definition of the critical drive is inspired by the CRF model, in which the total coherence and the total intensity peak at the critical drive strength and a phase transition is observed (see Sec.~\ref{subsec:DDM} for details). For the non-interacting model, we set $d\langle\hat{\tilde \sigma}^+\rangle/d\Omega=0$ to obtain the critical drive strength as $\rm\Omega_C^{Non-int.}=\Gamma/\sqrt{2}$.

\begin{figure}[ht!]
\includegraphics[width=\linewidth]{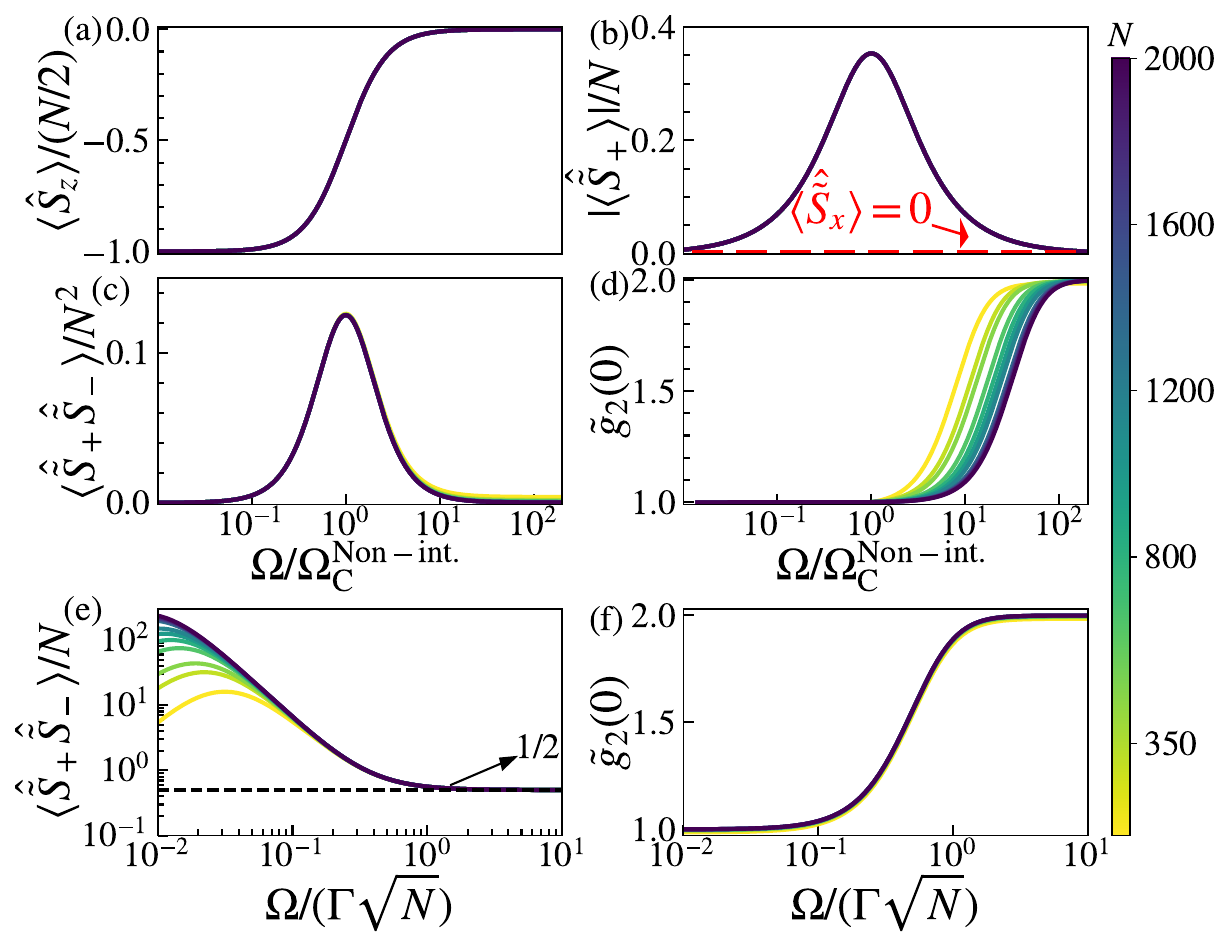}
\caption{Steady-state in the non-interacting (dilute) limit across driving regimes: (a) atomic inversion $\langle \hat{S}_z\rangle/(N/2)$, (b) total coherence $|\langle \hat{\tilde{S}}_+\rangle|/N = |\langle \hat{\tilde{S}}_y\rangle|/N$ driven by the laser in the $\hat{\tilde{S}}_x$-direction such that $\langle \hat{\tilde{S}}_x\rangle=0$, (c) ``forward'' intensity (along the axial direction) scales as $N^2$ in the weak-drive regime and as (e) $N$ in the strong-drive regime ($\langle \hat{S}_z\rangle/(N/2)\to 0$), (d) two-photon correlation function $\tilde{g}_2(0)$ in the ``forward'' direction. (e), (f): Due to the constructive interference of coherences in the ``forward'' direction, an $N$-dependent stronger drive ($\Omega \gtrsim \Gamma \sqrt{N}$) is needed to reach the thermal state ($\langle \hat{\tilde{S}}_+ \hat{\tilde{S}}_-\rangle \to N/2$, $\tilde{g}_2(0)\to 2$).}
\label{fig:SS_Non_int_4plots}
\end{figure}

As shown in Fig.~\ref{fig:SS_Non_int_4plots}a, in the weak-driving limit ($\Omega\ll\rm\Omega_C^{Non-int.}$), the forward intensity scales as  $I(\vec{k}_{\rm L}) \propto ({ N\Omega}/{{\Gamma}})^2$, as shown in Fig.~\ref{fig:SS_Non_int_4plots}c. 
Moreover, we find that $\tilde{g}_2(0)=1$ (Eq.~(\ref{eq:g2_MF_lowOm})) in this limit, as shown in Fig.~\ref{fig:SS_Non_int_4plots}d, which reflects that the system is in a coherent state with the majority of the atoms in the ground state. 

In the strong-driving limit ($\Omega\gg\rm\Omega_C^{Non-int.}$), radiative decay disrupts the coherences and leads to a completely mixed steady-state ($R\to 0$ in Eq.~(\ref{eq:non_int_ss_sp_sz})). In this mixed state, the Bloch vector of each spin-1/2 (atomic dipole) is reduced to a point at the center of the Bloch sphere such that $\langle\hat{\sigma}^z\rangle\to 0$ (Fig.~\ref{fig:SS_Non_int_4plots}a) and $|\langle {\hat { \tilde \sigma}}^+\rangle|\to 0$ (Fig.~\ref{fig:SS_Non_int_4plots}b). In the absence of a finite coherence, there is no constructive interference in the forward direction, i.e., $I(\vec{k}_{\rm L}) \propto N/2$ (Fig.~\ref{fig:SS_Non_int_4plots}e). We obtain $\tilde{g}_2(0)=2$ in this limit, as shown in Fig.~\ref{fig:SS_Non_int_4plots}d, since the system is in a thermal state. As shown in Fig.~\ref{fig:SS_Non_int_4plots}f, a very strong drive, $\Omega\gtrsim \Gamma \sqrt{N}$, is needed to fully reach this condition. This $N$-dependent drive scaling emerges due to the $N^2$ enhancement from the constructive interference of coherences, and is further explained in Appendix \ref{APP_sec:non_int_model}.

Lastly, the on-resonant drive Hamiltonian commutes with $\hat{\tilde{\sigma}}^x$, so starting from $\langle\hat{\tilde{\sigma }}^x\rangle =0$ in the initial state, it remains zero at all times, irrespective of $\Omega$, as shown in Fig.~\ref{fig:SS_Non_int_4plots}b. 
\section{\label{sec:MF_dipolar_model} Mean-field Dipolar model}

\subsection{\label{subsec:MF_dipolar_short_time} General description}

For a general distribution of atomic positions, the interacting system  is free to explore an exponentially large Hilbert space ($2^N$) and its exact dynamics is not tractable. For simplicity, we take the mean-field (MF) approximation here to characterize the steady-state of the system.  In Appendix~\ref{APP_sec:validity_MF}, we compare MF with beyond-MF approximation methods to include the effects of correlations, namely the MACE-MF (see Appendix~\ref{APP_sec:MACE-MF}) and cumulant approximation methods. We show that at the atomic densities considered in this paper, the MF approximation is valid when describing the observables under consideration.

We obtain the MF equations of motion (see Appendix \ref{APP_sec:MF_dipolar_eqns}) and express them as the Bloch equation, same as Eq.~(\ref{eq:Bloch_eq_non-int}), but here the components of the effective magnetic field, $ \vec{B}^{\rm MF}_i $ at atom $i$, are site-dependent due to the inclusion of the dipolar interactions as
\begin{align}
     {B}^{\rm MF}_{i,x}  &= \frac{\Omega}{2} + \frac{1}{4} \sum_{\substack{j\neq i}} \left({\rm Re} \tilde{\mathcal{G}}_{ij} \langle\hat{\tilde{\sigma}}_j^x\rangle + {\rm Im} \tilde{\mathcal{G}}_{ij} \langle\hat{\tilde{\sigma}}_j^y\rangle \right), \\
     B^{\rm MF}_{i,y}  &= \frac{1}{4} \sum_{\substack{j\neq i}} \left({\rm Re} \tilde{\mathcal{G}}_{ij} \langle\hat{\tilde{\sigma}}_j^y\rangle - {\rm Im} \tilde{\mathcal{G}}_{ij} \langle\hat{\tilde{\sigma}}_j^x\rangle \right),\\
     B^{\rm MF}_{i,z}  &= \frac{\Delta}{2} .
\end{align}

{\it Short-time physics}: Although the MF treatment above significantly reduces the complexity of the problem, the non-linearity of these equations makes them difficult to solve analytically  and apart from simple limiting cases, we need to solve them numerically. The early time dynamics ($\Gamma t\ll 1$) is one such regime where it is possible to gain analytical insights.

To understand the role of interactions at short-times, we include the inelastic interaction term in the Hamiltonian $\hat{H}$ (Eqs.~(\ref{eq:H_0_tilted}), (\ref{eq:H_int_tilted})) and obtain the non-Hermitian Hamiltonian in the spiral basis as $\hat{H}_{\rm NH} = \hat{H} - i \sum_{k,j\neq k} \tilde{\mathcal{I}}_{kj} \hat{\tilde{\sigma}}_k^+ \hat{\tilde{\sigma}}_j^- = \hat{H}_0 - \sum_{k,j\neq k} \tilde{\mathcal{G}}_{kj} \hat{\tilde{\sigma}}_k^+ \hat{\tilde{\sigma}}_j^- $, 
where $\hat{H}_0$ describes the laser drive. Given that the  system is initialized in a state where all the Bloch vectors are identical in the tilde basis, i.e., $\langle \hat{\vec{\tilde{\sigma}}}_k \rangle \equiv \langle \hat{\vec{\tilde{\sigma}}} \rangle$,  at short times, each Bloch vector commutes with  terms of the form  $(\hat{\vec{\tilde{\sigma}}}_k \cdot \hat{\vec{\tilde{\sigma}}}_j)$, since   $\sum_{k,j\neq k}\langle[(\hat{\vec{\tilde{\sigma}}}_k \cdot \hat{\vec{\tilde{\sigma}}}_j),\hat{\vec{\tilde{\sigma}}}_i ]\rangle \sim \sum_{j\neq i} \langle \hat{\vec{\tilde{\sigma}}}_j  \rangle \times \langle \hat{\vec{\tilde{\sigma}}}_i \rangle \sim 0$.

The emergence of a density shift  can be elucidated by adding such terms, $(\hat{\vec{\tilde{\sigma}}}_k \cdot \hat{\vec{\tilde{\sigma}}}_j)$, to $\hat{H}_{\rm NH}$ without altering the physics of the original system at  short times as 
\begin{align}
    \hat{\mathcal H} = \hat{H}_{\rm NH} + \frac{1}{4}\sum_{k,j\neq k} {\rm Re} \tilde{\mathcal{G}}_{kj} (\hat{\vec{\tilde{\sigma}}}_k \cdot \hat{\vec{\tilde{\sigma}}}_j),
\end{align}
which allows us to rewrite the  real part of the new Hamiltonian as   $ {\rm Re} \hat{\mathcal H} = \hat{H}_0 + \frac{1}{4}\sum_{k,j\neq k} {\rm Re} \tilde{\mathcal{G}}_{kj} \hat{{{\sigma}}}^z_k \hat{{{\sigma}}}^z_j $, and at the mean field as,    $ {\rm Re} {\hat{\mathcal H}}^{\rm MF}=  -\sum_i{\vec {\mathcal B}_i}^{\rm MF} \cdot {\hat{\vec{\tilde{\sigma}}} }_i$, where ${\vec {\mathcal B}}^{\rm MF}_i = \vec{B}_{\rm L} - (\delta_i/2) \hat{z} =\left( \Omega/2,0, \Delta/2 - \delta_i/2 \right)$, and $\delta_i=\sum_{j\neq i} {\rm Re}  \tilde{\mathcal{G}}_{ij} \langle{{\sigma}}_j^z\rangle/2  $ is the interacting part of ${\vec {\mathcal B}}^{\rm MF}_i$. This acts as a self-adjusting magnetic field along $\hat z$ that depends on the atomic inversion, $\langle{{\sigma}}_j^z\rangle $. This effective field generated by other atoms in the array,  leads  to a precession of each atomic dipole $i$ about the $\hat{z}$-axis of the Bloch sphere, generating  what is known as a density shift  or a collective Lamb-shift \cite{Chang2004}. 
As the number of atoms in the cloud is increased, with the spatial extent fixed, the cloud gets denser and the induced frequency shift at each atom gets larger. The frequency shift has two components -- a homogeneous component, $\Delta_{\rm MF}$, which is the average shift across atoms, and an inhomogeneous component, associated with the random distribution of the atoms in the array. The homogeneous component is non-zero when the cloud is not spherically symmetric, and it pushes the atoms out of resonance, suppressing the growth of  coherences and the excitation fraction. The inhomogeneous component leads to  dephasing. These two key features will be an important consideration when we derive a simplified model of the dipolar Hamiltonian in Sec. \ref{sec:Mod_CRF_model}.

The average frequency shift can be measured  in Rabi spectroscopy  by scanning the detuning for the new resonance condition, i.e., $\Delta_{\rm MF}=\sum_i \delta_i/N=\sum_{i, j\neq i} {\rm Re}  \tilde{\mathcal{G}}_{ij} \langle{{\sigma}}_j^z\rangle/(2N)$, where the RHS has been averaged over all atoms $i$ \cite{Chang2004}.  The density shift can also be measured via Ramsey spectroscopy \cite{Hutson2024}. In this case, even at  $\Delta=0$, there is a residual  precession   which leads to a non-zero $\langle \hat { {\tilde S}}_x \rangle $ during the dynamics of the system even when  the system is initialized with $\langle \hat { {\tilde S}}_x \rangle =0$. This feature distinguishes the dipolar model from the CRF and non-interacting models, where $\langle \hat { {\tilde S}}_x \rangle =0 $ at all times.

{\it Weak-drive limit: \label{subsec:MF_dipolar_uniform_gas_weak_drive}} Another simplified case is when the laser drive is very weak, $\Omega \ll \Gamma$, such that individual dipoles remain close to the south-pole of the Bloch sphere, i.e., $\langle \hat{\sigma}^z_j \rangle \approx -1$, even in the steady-state. In this regime, correlations are suppressed by factors of $\Omega/\Gamma$ and the system can be described almost exactly using mean-field theory  by setting $\langle \hat{\sigma}^z_j\rangle=-1$ at all times and only considering the dynamics of the coherences $\langle \hat{\tilde{\sigma}}^+_j\rangle$. This regime has been intensively studied for dilute samples of dipolar-interacting atomic gases \cite{ScullyDir2006,Scully2009,Bienaime2013a,bromley_collective_2016,ZhuPRA2016,RoofPRL2016,AraujoPRL2016,SutherlandPRA2016cloud,GUERIN2023253,ma2023collective,gjonbalaj2023modifying}. 

In the weak-drive limit ($\langle \hat{\sigma}^z_j\rangle=-1$) of  our pencil-shaped Gaussian cloud, the average frequency shift is non-zero and as discussed above, can be obtained at short times as $\Delta_{\rm MF}=-\sum_{j\neq i}^N {\rm Re}  \tilde{\mathcal{G}}_{ij}/(2N)$. In the far-field limit ($|r_{ij}| \gg \lambda \Rightarrow\tilde{\mathcal{G}}_{ij}\propto 1/|r_{ij}|$), we find that $\sum_{j\neq i}^N {\rm Re}  \tilde{\mathcal{G}}_{ij}/(2N) \propto N$, with a small ($\ll \Gamma$) proportionality constant that depends on the cloud extent and can be estimated numerically (see Appendix~\ref{APP_sec:analytical_model} for details). 

In the weak-drive limit, the  mean-field steady-state coherence, can also be obtained  for a dilute gas   by treating the interactions perturbatively  as \cite{bromley_collective_2016,ZhuPRA2016} 
\begin{align}
    \langle \hat{\tilde{\sigma}}_j^-\rangle & = \frac{i\Omega/2}{i\Delta+\Gamma/2} \left[1 - \frac{i}{i\Delta+\Gamma/2} \sum_{k\neq j} \tilde{\mathcal{G}}_{jk}\right],
\end{align}
and the forward intensity is obtained as
\begin{align}
    \frac{I(\vec{k}_{\rm L})}{ I_0 (\vec{k}_{\rm L})} & = \frac{N(N-1) \Omega^2/4}{\Delta^2+\Gamma^2/4} \left[1 - \frac{\sum_{j,k\neq j} \left(\Gamma {\rm Im}\tilde{\mathcal{G}}_{jk}-2\Delta {\rm Re}\tilde{\mathcal{G}}_{jk}\right)}{(\Delta^2+\Gamma^2/4)N}   \right].
\end{align}
The steady-state density shift and the linewidth broadening can be obtained from the lineshape as $\bar{\Delta}=-\sum_{j,k\neq j}{\rm Re}\tilde{\mathcal{G}}_{jk}/N$ and $\bar{\Gamma}=2\sum_{j,k\neq j}{\rm Im}\tilde{\mathcal{G}}_{jk}/N$, respectively \cite{bromley_collective_2016,ZhuPRA2016}. On resonance ($\Delta=0$), the $x$-component of the coherence is proportional to the density shift, $\langle \hat{\tilde S}_x\rangle/N=2\Omega \bar{\Delta} /\Gamma^2 $ and the forward intensity is reduced by the inhomogeneous broadening as 
\begin{align}\label{eq:int_forw_dil_gas_zhu_on_reso}
    {I(\vec{k}_{\rm L})} \propto N(N-1) (\Omega/{\Gamma})^2 (1 - 2\Bar{\Gamma}/\Gamma ).
\end{align}
However, in a highly dense gas, multi-photon scattering processes become significant, making the steady-state more complicated and beyond the perturbative limit \cite{ZhuPRA2016}.

\subsection{\label{sec:critical_drive_} Dynamical Regimes in the Steady State}

Outside the special limits discussed above, we resort to numerical solutions for obtaining the steady-state of the system. We find that our system exhibits different dynamical behaviors characterized by distinct light emission properties, as we vary the laser driving strength, $\Omega$. Here, we draw parallels and distinctions among the steady-state properties of the dipolar, the non-interacting, and the CRF models, across different driving regimes.

\begin{figure}[ht!]
\includegraphics[width=\linewidth]{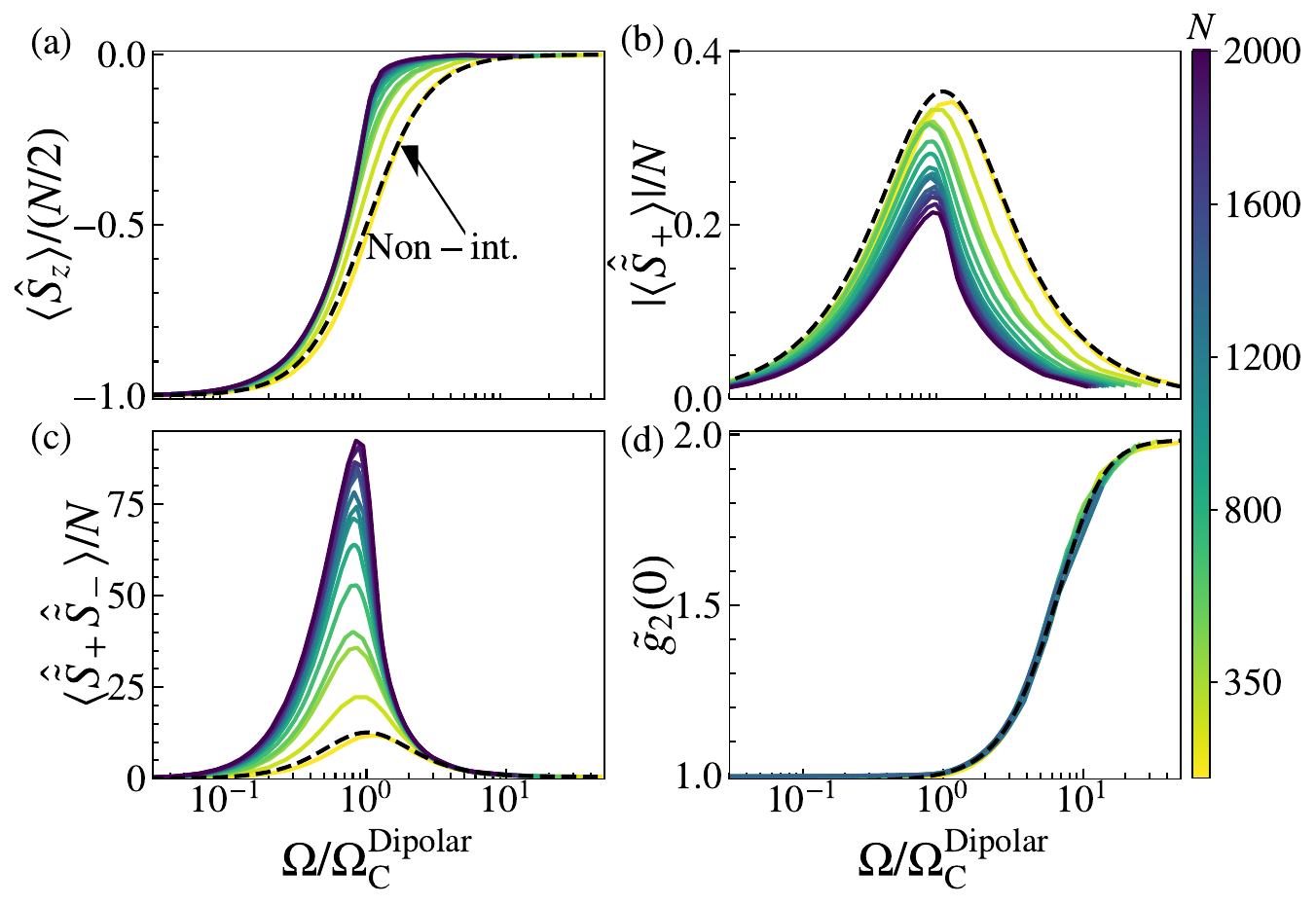}
\caption{Steady-state of the MF Dipolar model: (a) atomic inversion $\langle \hat{S}_z\rangle/(N/2)$, (b) contrast $|\langle \hat{S}_+\rangle|/N$, (c) ``forward'' intensity (along the axial direction), (d) two-photon correlation function $\tilde{g}_2(0)$ in the ``forward'' direction. (a)-(d) The dipolar model (solid) converges to the non-interacting model (black dashed) in the dilute limit (small $N$). The $x$-axis of the non-interacting curve is scaled as $\Omega/\Omega_{\rm C}^{\rm Non-int.}\sim \mathcal{O}(1)$ in (a)-(c) and as $\Omega/\Omega_{\rm C}^{\rm Dipolar} \propto 1/\sqrt{N}$ in (d). In (a), (b), and (d), the non-interacting curve is independent of $N$; in (c), the non-interacting curve is for $N=100$.}
\label{fig:SS_Dip_4plots}
\end{figure}

\subsubsection{ Inversion,  forward intensity, and two-photon correlation function  } In Fig.~\ref{fig:SS_Dip_4plots}a, we show the atomic inversion of the dipolar model across a range of $\Omega$. Unlike the CRF model, the dipolar model does not exhibit a sharp phase transition and the system crosses over smoothly from the weak to the strong driving regime. Nevertheless, we find it useful to define a ``critical" driving strength, $\rm\Omega_C^{Dipolar}$, for the dipolar model, as the drive strength at which the forward intensity, $\langle \hat{\tilde{S}}_+\hat{\tilde{S}}_- \rangle$, peaks for a given $N$. The intensity peak separates the dynamical behaviors in the different driving regimes based on their emission properties, similar to the CRF model. We obtain ${\rm\Omega_C^{Dipolar}}=c\Gamma \sqrt{N}$ ($c \approx 0.08$) from the numerical steady-state of the system. By plotting the steady-state atomic inversion as a function of a normalized  drive strength, $\rm \Omega/\rm \Omega_C^{Dipolar}$, we observe that the curves for different $N$ appear to collapse, as shown in Fig.~\ref{fig:SS_Dip_4plots}a. 

 As $N$ decreases, the cloud becomes dilute and the dipolar interaction coefficients become less important. At very small $N$, the steady-state of the non-interacting model is smoothly recovered for all the observables, as shown in Fig.~\ref{fig:SS_Dip_4plots}.
 
In Fig.~\ref{fig:SS_Dip_4plots}c of the dipolar model, we show that the variation of the forward intensity $\propto \langle \hat{\tilde{S}}_+ \hat{\tilde{S}}_- \rangle$ curve with the normalized drive strength is very similar to that of the non-interacting model (Fig.~\ref{fig:SS_Non_int_4plots}c). The main difference between these two models is evident in the $N$-scaling of $\langle \hat{\tilde{S}}_+ \hat{\tilde{S}}_-\rangle$. In the weak-driving regime of the CRF and non-interacting models, where phase matching leads to a perfect $N^2$ enhancement, we show that the forward intensity curves for different $N$ collapse when divided by $N^2$, in Fig.~\ref{fig:SS_CRF_4plots}c and Fig.~\ref{fig:SS_Non_int_4plots}c, respectively. But for the dipolar model, we find that the curves do not collapse when divided by $N^2$. At weak drive intensities, the site-dependent shifts due to the dipolar interactions imprint random phases on the coherences and thereby suppress the $N^2$ enhancement of $\langle \hat{\tilde{S}}_+ \hat{\tilde{S}}_- \rangle$ \cite{Bienaime2013a,ZhuPRA2016}. Hence, the $N$-scaling of $\langle \hat{\tilde{S}}_+ \hat{\tilde{S}}_- \rangle$ decreases as $N$ increases, as later shown in Fig.~\ref{fig:alpha_2_plots}b. 

To further understand the emission properties of the steady-state, we look at the forward two-photon correlation function $\tilde{g}_2(0) = \frac{\langle \hat{\tilde{S}}_+\hat{\tilde{S}}_+\hat{\tilde{S}}_-\hat{\tilde{S}}_- \rangle}{\langle \hat{\tilde{S}}_+\hat{\tilde{S}}_- \rangle^2}$. As shown in Fig.~\ref{fig:SS_Dip_4plots}d, in the weak driving regime $(\rm \Omega\ll\rm \Omega_C^{Dipolar})$, we have $\tilde{g}_2(0)\approx 1$, which corresponds to a coherence state. This is expected because the steady-state of the system resembles a coherent state on the Bloch sphere with a finite coherence $|\langle \tilde{S}_+ \rangle|/N$, as seen in Fig.~\ref{fig:SS_Dip_4plots}b  and later validated analytically using a simpler model (Sec.~ \ref{sec:Mod_CRF_model}). Given that the CRF, non-interacting, and dipolar models all feature almost pure Gaussian-like steady-states in this regime, $\tilde{g}_2(0)$ is not able to distinguish their subtle differences and all the three models have $\tilde{g}_2(0)\sim 1$ in the weak excitation limit (see Figs.~\ref{fig:SS_CRF_4plots}d, \ref{fig:SS_Non_int_4plots}d, and \ref{fig:SS_Dip_4plots}d).

When the drive strength is above the  ``critical" drive strength,  i.e., $\Omega \gg \rm \Omega_C^{Dipolar}$, the drive  quickly dephases the array. As a consequence,   dipolar interactions   become subdominant  compared to the rapid rotation induced by  the drive.  
Consistently, the steady-state starts to become fully mixed,  with  suppressed coherences, signaled by the fact that $\tilde{g}_2(0)\sim 2$ approaches its thermal value (see Fig.~\ref{fig:SS_Dip_4plots}d).

This is in strike contrast to  the  CRF case, where the collective nature of the master equation enforces the preservation of the collective nature of the state even in the large driving limit ($\Omega \gg \rm\Omega_C^{CRF}$). Therefore,   the intensity keeps scaling  as $N^2$ (Eq.~(\ref{eq:SpSm_Dicke})), as shown in Fig.~\ref{fig:SS_CRF_4plots}c and $\tilde{g}_2(0)$   remains always below the thermal value and saturates as $\tilde{g}_2(0)\to 1.2$ (see Fig.~\ref{fig:SS_CRF_4plots}d). 
These key distinctions imply  that a free-space atomic cloud does not behave as a   CRF model in the strong drive regime.

\subsubsection{\label{sec:weak_drive_freq_shift} Steady-state frequency shift and contrast}

\begin{figure}[ht!]
\includegraphics[width=\linewidth]{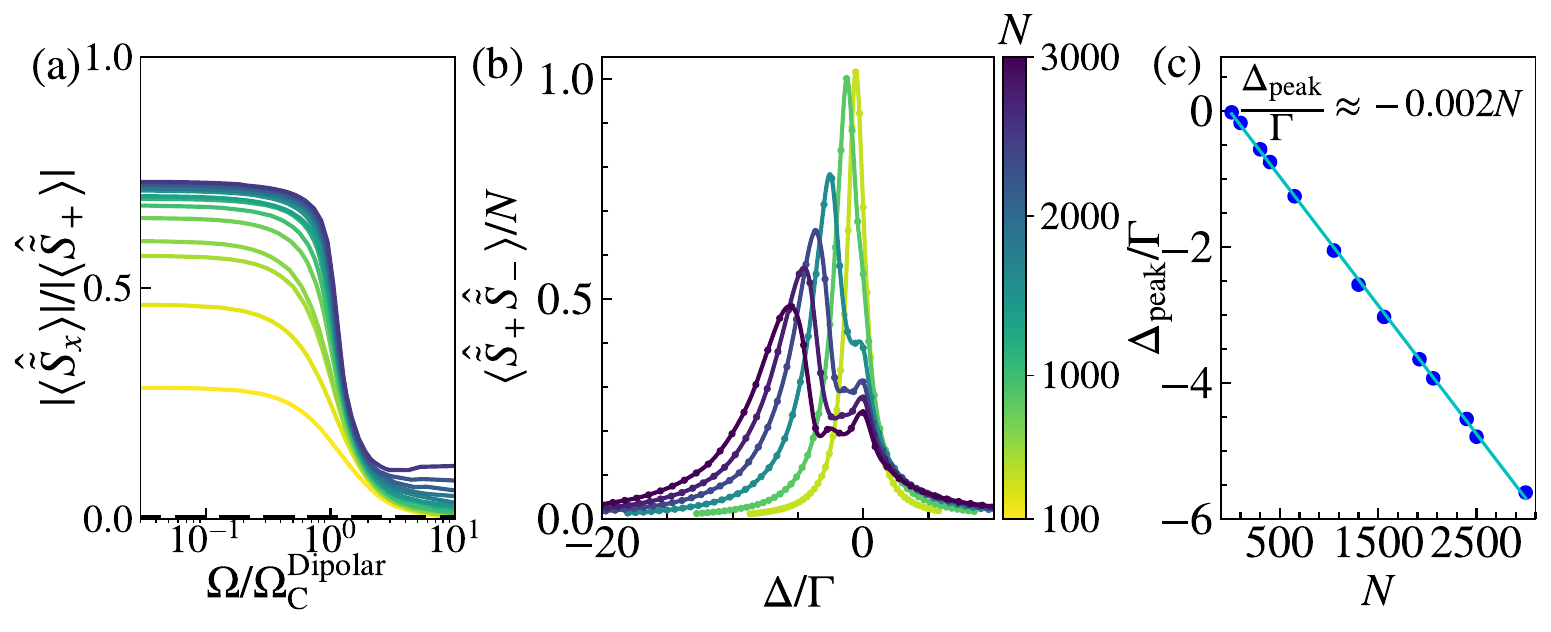}
\caption{(a) The $x$-component of the coherence, $|\langle \hat{S}_x\rangle|/|\langle \hat{S}_+\rangle|\neq 0$, arising from frequency shifts due to the elastic dipolar interactions. For the non-interacting model, $\langle \hat{\tilde{S}}_x\rangle = 0$ (black dashed). (b) Steady-state ``forward'' intensity for a range of laser detunings, $\Delta$: MF dipolar model numerics (dots), interpolated data (solid lines). (c) Steady-state frequency shift computed as the detuning at which the maximum intensity is reached (circles) which follows a linear dependence with $N$ as can see from the  linear best fit (solid line). Data from numerics of the MF dipolar model in the weak-drive regime ($\Omega/\Gamma=0.1$), averaged over $~10$ realizations.}
\label{fig:freq_shift}
\end{figure}

In Fig.~\ref{fig:freq_shift}a, we see that the steady-state of the dipolar model has $\langle \hat{\tilde{S}}_x\rangle \neq 0$, unlike the non-interacting (Fig.~\ref{fig:SS_Non_int_4plots}b) and CRF (Fig.~\ref{fig:SS_CRF_4plots}b) models where the $x$-component of the collective coherence is zero. The resonant laser drive commutes with $\langle \hat{\tilde{S}}_x\rangle$ and the finite $\langle \hat{\tilde{S}}_x\rangle$ arises from the dispersive (elastic) part of the dipolar interactions, which leads to frequency shifts, as discussed in Sec.~\ref{subsec:MF_dipolar_short_time}. The latter is  most prominent  in the weak-drive limit, where the inversion remains close to its initial minimum value. In this limit,  the fractional $x$-component of the coherence, $|\langle \hat{\tilde{S}}_x\rangle|/|\langle \hat{\tilde{S}}_+\rangle|$, increases with $N$, which suggests that denser clouds have larger collective shifts. 

Beyond the critical drive, as shown in Fig.~\ref{fig:SS_Dip_4plots}b, the contrast, $|\langle \hat{\tilde{S}}_+\rangle| = \sqrt{\langle \hat{\tilde{S}}_x\rangle^2 + \langle \hat{\tilde{S}}_y\rangle^2}$,  starts decaying  as  the state begins  to get mixed and the Bloch vector length gets smaller. This behavior continues with increasing $\Omega/\Gamma$ until eventually the system's  Bloch vector reduces  to the one expected for the non-interacting system. The approach to  the non-interacting regime as the system reaches  the strong drive regime is also signaled by   the   dominant $\langle \hat{\tilde{S}}_{y,z}\rangle$  spin projections, and the reduced $\langle \hat{\tilde{S}}_{x}\rangle$ one.
This in turn leads to a reduced value of  $|\langle \hat{\tilde{S}}_x\rangle|/|\langle \hat{\tilde{S}}_+\rangle|$, as shown in Fig.~\ref{fig:freq_shift}a.

In Fig.~\ref{fig:freq_shift}b, we look at the lineshape of the steady-state intensity in the weak-driving regime $(\Omega/\Gamma=0.1)$ by varying the laser detuning, $\Delta$, of the MF dipolar model. We find that increasing $N$ leads to a bimodal distribution with a peak at $\Delta=0$ and a second resonance at $\Delta_{\rm peak}$. 
Note the second peak arises from dipolar interactions,  and scales linearly with $N$, as shown in Fig.~\ref{fig:freq_shift}c.

The  linear $N$-scaling is consistent with  previous work done in the weak drive limit \cite{Morice1995a,Chomaz2012a,Bienaime2013a,ZhuPRA2016,bromley_collective_2016} that observed a frequency shift  proportional to the atomic density. In our  quasi-1D configuration along the $\hat{x}$-axis, the atomic density scales as, $\rho \sim N/(a_{ho}^2 L)$, where $L=2l_{\rm ax}$ is the axial length of the pencil-shaped cloud and $a_{ho}\ll x_{ij} $ sets the radial confinement. The existence of a  prominent second peak   can be naively understood from the fact that  the inter-atomic distances are determined  by the axial spacing, i.e.,  $k_0 |\vec{r}_{ij}| = k_0 \sqrt{x_{ij}^2+y_{ij}^2+z_{ij}^2} \approx k_0 |x_{ij}|$. As such,
the pairwise interactions between atoms along the $\hat{x}$-axis contribute constructively to the shift. Mathematically, this is because   the phases imprinted by the laser cancel  the ones of the  Green's function   $\tilde{\mathcal{G}}_{i>j} \propto \exp(ik_0 |\vec{r}_{ij}| - i \vec{k}_{\rm L} \cdot \vec{r}_{ij}) \sim 1$ (see Appendix~\ref{APP_sec:MF_dipolar_eqns} for more details). Hence, by aligning the laser wavevector along the elongated geometry of the cloud,  one can induce a non-zero global frequency shift even in a disordered configuration \cite{GlicensteinPRL2020}. 

As discussed earlier in Sec.~\ref{subsec:MF_dipolar_uniform_gas_weak_drive}, the steady-state density shift of the lineshape in the weak-drive limit for a dilute gas is $\bar{\Delta}=-\sum_{k\neq j}{\rm Re}\tilde{\mathcal{G}}_{jk}/N$. For simplicity, we consider the far-field regime $(\propto 1/r)$ of the interactions, which is the dominant term in the dilute case, and obtain $\bar{\Delta}$ in the large-$N$ limit by integrating over the pencil-shaped distribution of the cloud, as $\bar{\Delta} = -c_{\rm R} N \Gamma$, where $c_{\rm R}\approx 0.003$ is a constant depending on the spatial extent of the cloud. Similarly, we obtain the linewidth broadening as $\bar{\Gamma} = c_{\rm I} N \Gamma$ with $c_{\rm I} \approx 0.004$. In the regime where $2\bar{\Gamma} > \Gamma$, i.e., $N > 250$, the dilute gas description is no longer valid on resonance as, under this assumption, the predicted forward intensity becomes unphysical (Eq.~(\ref{eq:int_forw_dil_gas_zhu_on_reso})), and the perturbative treatment fails. Thus, even though the prediction of $\bar{\Delta}$ is very close to $\Delta_{\rm peak}$, the weak-drive physics in our system is beyond the first-order expansion in interactions.

Of course, our analysis neglects   motional effects in thermal samples,  laser forces, and dipolar forces.  As shown in Ref.~\cite{ZhuPRA2016}, motional decoherence can induce a reduction of the density shift and wash out the double peak structure compared to what is expected for frozen atoms.

\subsection{\label{sec:phases}\textit{N}-Scaling  for Different Driving Strengths}

\begin{figure}[ht!]
\includegraphics[width=\linewidth]{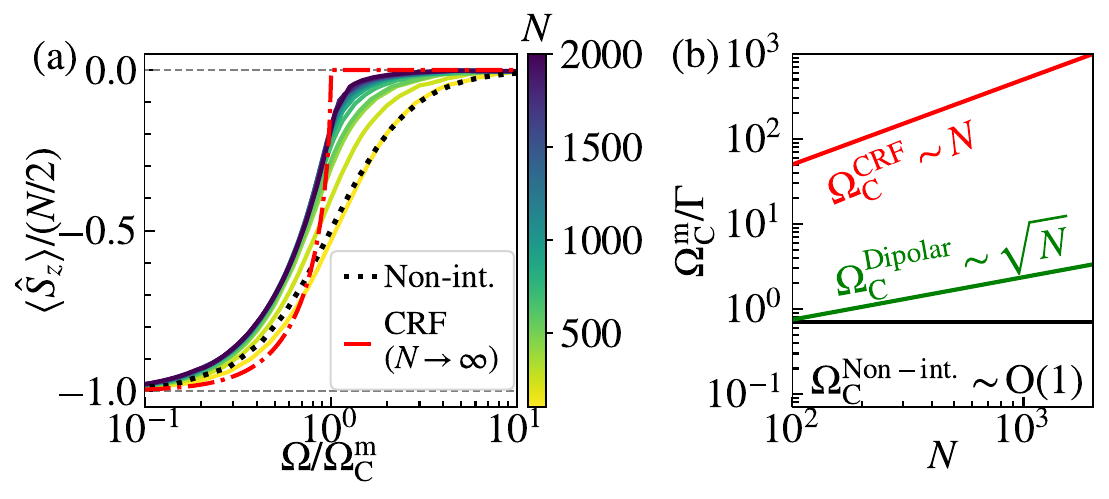}
\caption{(a) Steady-state atomic inversion $\langle \hat{S}_z\rangle/(N/2)$ for the dipolar (solid), non-interacting (dotted), and CRF (dashed dotted) models. The $x$-axis is scaled with the critical drive $\rm\Omega_{C}^m$ for each model, m. 
(b) $N$-scaling of $\rm\Omega_{C}^m$ for the dipolar ($\sim \sqrt{N}$), non-interacting ($\sim \mathcal{O}(1)$), and CRF ($\sim N$) models.}
\label{fig:Jz_Om_3_models}
\end{figure}

To emphasize the collapse of different $N$ curves and the different critical properties across models, we compare the steady-state atomic inversion across a wide range of driving strengths for the dipolar model (solid), the CRF model (dashed-dotted), and the non-interacting model (dotted), in  Fig.~\ref{fig:Jz_Om_3_models}a. The $x$-axis of Fig.~\ref{fig:Jz_Om_3_models}a has been scaled differently for each model, corresponding to the scaling of the ``critical" driving strength $\rm \Omega_C^m$, where $\rm m \in $ \{CRF model, Dipolar model, Non-int. (Non-interacting) model\}. The $N$-scaling of $\rm \Omega_C^m$ for these models has been shown in Fig.~\ref{fig:Jz_Om_3_models}b. The $N$-scaling of the critical drive for the dipolar model $(\propto \sqrt{N})$ clearly differs from that of the non-interacting $(\sim \mathcal{O}(1))$ and CRF $(\propto {N})$ models. 

\begin{figure}[ht!]
\includegraphics[width=\linewidth]{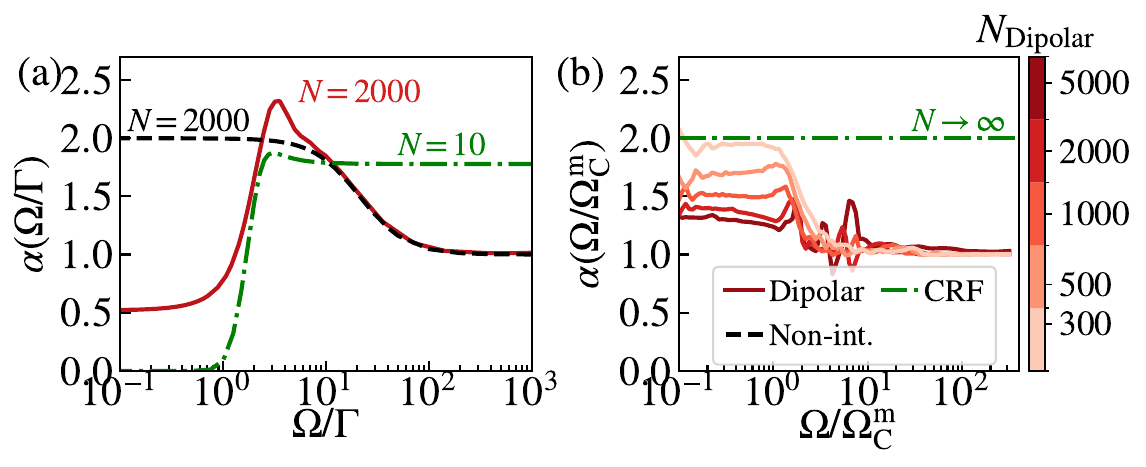}
\caption{$N$-scaling exponent of steady-state ``forward'' intensity, $\langle \hat{\tilde{S}}_+\hat{\tilde{S}}_-\rangle\propto N^\alpha$: (a) $\alpha$ computed at a fixed Rabi frequency, $\rm \Omega/\Gamma$, similar to Ref.~\cite{ferioli_non-equilibrium_2023}. $\alpha$ is similar for the CRF ($N=10$) and dipolar model ($N=2000$) in the intermediate driving regime for a small range of $\Omega/\Gamma$, which was probed in Ref.~\cite{ferioli_non-equilibrium_2023}. At large $\Omega/\Gamma$, $\alpha=2$ for the CRF due to the preservation of $\langle\hat{\mathbf{S}}^2\rangle$, but the dipolar model becomes single-particle-like ($\alpha=1$) due to spontaneous emission. Due to finite-size effects, $\alpha < 2$ for $N=10$ (CRF). (b) $\alpha$ computed at a fixed $\rm \Omega/\Omega_C^m$, the ratio of the Rabi frequency and the critical drive, for each model. 
MF dipolar (solid), non-interacting (black dashed), and CRF (green dashed-dotted).}
\label{fig:alpha_2_plots}
\end{figure}

In Fig.~\ref{fig:alpha_2_plots}, we show the $N$-scaling of the steady-state intensity, i.e., we define $\alpha$ such that $I(\vec{k}_{\rm L}) \propto \langle \hat{\tilde{S}}_+ \hat{\tilde{S}}_- \rangle \propto N^\alpha$ at fixed $\Omega/\Gamma$, across a wide-range of driving strengths. In Fig.~\ref{fig:alpha_2_plots}a, we choose the $x$-axis to be the bare driving strength $\Omega/\Gamma$, to compare our predictions with the measurements of a recent experiment \cite{ferioli_non-equilibrium_2023}. In the weak-driving regime, we find $\alpha<1$ for the dipolar and the CRF models. Usually, $\alpha<1$ is understood as a signature of subradiance but in our case this is simply an artifact due to the $N$-scaling being calculated at a fixed bare driving strength, $\Omega/\Gamma$. At a fixed value of $\Omega/\Gamma$, the system crosses over from the strong $(\Omega \gg \rm \Omega_C^{Dipolar},\, \Omega \gg \rm \Omega_C^{CRF})$ to the weak $(\Omega \ll \rm \Omega_C^{Dipolar},\,\Omega \ll \rm \Omega_C^{CRF})$ driving regime as $N$ is increased, leading to an inaccurate $N$-scaling. Similarly, we see a diverging value of $\alpha$, i.e., $\alpha>2$, for the mean-field dipolar model at intermediate driving strengths in Fig.~\ref{fig:alpha_2_plots}a, which is again an artifact of fixing $\Omega/\Gamma$. These artifacts are not seen for the non-interacting model because ${\rm \Omega_C^{Non-int.}}\sim\mathcal{O}(1)$ (in the large-$N$ limit), so fixing $\Omega/\Gamma$ also fixes the regime for all $N$.

In Fig.~\ref{fig:alpha_2_plots}a, we see that $\alpha$ of the dipolar model coincides with that of the non-interacting model in the regime of strong driving for all $N\leq 2000$, i.e., $\Omega \gg {\rm \Omega_C^{Dipolar}} = c\Gamma \sqrt{N} \approx 3.6\Gamma$. This confirms our expectation that in the strong driving regime, the dipolar model behaves like the non-interacting model. 
In Fig.~\ref{fig:alpha_2_plots}a, we also see the clear difference between the CRF and the dipolar models in the strong driving regime. We show the CRF model for $N=10$ to compare with the experimental prediction \cite{ferioli_non-equilibrium_2023}, where it is reported that a cloud of $N=2000$ atoms can equivalently be described by the CRF model of a reduced effective atom number, $N_{\rm eff} =10$. Close to the peak, the $\alpha$ of the dipolar model agrees with that of the CRF model in a small range of $\Omega/\Gamma$. Even though,  the two models have very different $\alpha$ in general, this was the regime probed by the experiment, partly justifying  the conclusions drawn in Ref.~\cite{ferioli_non-equilibrium_2023}.  Reaching the genuine strong drive limit requires large enough Rabi frequencies  which are not so accessible in current experiments.

In Fig.~\ref{fig:alpha_2_plots}b, we calculate the $N$-scaling by properly fixing the scaled driving strength, $\rm\Omega/\Omega_C^{Dipolar}$, and as expected, we find no signatures of subradiance. Moreover, the divergence of $\alpha$ seen in Fig.~\ref{fig:alpha_2_plots}a vanishes. Instead at  weak driving, $\alpha$ decreases from 2 (perfect phase matching) to close to 1 (randomized phases) for the dipolar model as $N$ is increased and the collective shifts get stronger, suppressing the $N^2$-enhancement. In the strong driving regime, $\alpha\to 1$ for all $N$, as the system becomes single-particle-like. For the CRF model, in the large-$N$ limit, $\alpha$ scales as $N^2$ in all driving regimes.

\section{\label{sec:Mod_CRF_model}Modified-CRF Model }

In this section, we propose a simplified theoretical model at the mean-field level to describe the emergent properties of our system. In the microscopic picture of the dipolar model, the dispersive part of the dipole-dipole interactions between atoms leads to a frequency shift in the transition frequency of each atom, as discussed in Sec.~\ref{subsec:MF_dipolar_short_time}. This shift can equivalently be captured  by  an additional term $\sum_{i=1}^N \delta_i(t)\sigma^z_i$ in the CRF Hamiltonian as
\begin{align}
    \hat{H}_{\rm Mod-CRF} &= - \Omega \hat{S}_x + \sum_{i=1}^N \delta_i(t)\sigma^z_i ,
    \label{eq:H_mod_DDM}
\end{align}
where the frequency shift of an atom $i$ is described as $\delta_i(t)= \sum_{j\neq i} {\rm Re}\tilde{\mathcal{G}}_{ij} \langle\hat{\sigma}^z_j(t)\rangle/4$, inspired from the short-time dynamics of the MF dipolar model. 
To distinguish the collective slow varying part  of the interaction-induced shift from the inhomogeneous, fast time varying  effects, we express the frequency shift as $\delta_i(t)= \bar{\delta}(t) + h_i(t)$ where, $ \bar{\delta}(t) = \sum_{i,j\neq i} {\rm Re}\tilde{\mathcal{G}}_{ij}\langle\hat{\sigma}^z_j(t)\rangle/(4 N)$ is the average shift across all the atoms in the cloud at any time $t$ and $h_i(t)$ is the fast time varying inhomogeneous component, which describes beyond short-time dynamics. The homogeneous component, $\bar{\delta}(t)$, is dominated by the shifts that add up along the axial direction due to the constructive interference of the laser-induced phases with the interaction-induced phases. To capture the constructive shifts in a simplified way, we assume that $\bar{\delta}(t)$ for our cloud can be described by a collective system, such as a translationally invariant atomic array, giving $\bar{\delta}(t) \approx \sum_{i} {\rm Re}\tilde{\mathcal{G}}_{i0}\langle\hat{S}_z(t)\rangle / (2N)$. The interaction-induced time varying inhomogeneous component, $h_i(t)$, accounts for the random and dynamic    evolution of the atomic dipolar phases. $h_i(t)$ has a zero mean, is time-dependent, and is therefore not removable by a simple echo pulse.  
For simplicity, these conditions are roughly incorporated by  considering the $h_i$ functions as stochastic white noise variables with  spectral function: $ \overline{h_i (t) h_j (t')}= (\gamma_d/2)\delta_{ij} \delta(t-t') $ \cite{ReyPRA2007}.

The net  dephasing arising from the $h_i(t)\hat{\sigma}^z_i$ term can  be accounted for by local jump operators $\sqrt{\gamma_d} \hat{\sigma}^z_i$ in the master equation formulation, where ${\gamma_d} $ is a common dephasing rate for all the atoms, which we set to be  proportional to the variance of the frequency shifts, which scales linearly with the density of the atomic ensemble \cite{FRIEDBERG1973101,ZhuPRA2016,bromley_collective_2016}. Thus, we define the dephasing rate as $\gamma_d=c_d N\gamma_s$, where $c_d$  is a phenomenological constant that depends on the overall volume and geometry of the atomic cloud and $\gamma_s$ is the spontaneous emission decay rate. 

Accounting for the overall elastic interactions, the collective superradiant emission from nearby atoms, and the single-particle spontaneous emission (since the purely collective behavior is only possible in the high density limit, not achieved in the experiment), we obtain the modified CRF model described by a master equation of the form $\dot{\hat{\rho}} = -i[\hat{H}_{\rm Mod-CRF}, \hat{\rho}]+\mathcal{L}_{\rm Mod-CRF}(\hat{\rho})$, where the Hamiltonian is  
\begin{align}
    \hat{H}_{\rm Mod-CRF} = - \Omega \hat{S}_x + \chi  \langle\hat{S}_z\rangle \hat{S}_z,
\end{align}
where $\chi=2\bar{\delta}(t)/\langle\hat{S}_z(t)\rangle=\sum_{i\neq 0} {\rm Re}\tilde{\mathcal{G}}_{i0}/N$ is a constant. $\chi \langle\hat{S}_z (t)\rangle$ acts as a time-dependent global magnetic field, which describes the shearing of the collective Bloch vector via one-axis twisting (OAT) at the mean-field level. The Lindbladian is expressed as
\begin{align}
    \mathcal{L}_{\rm Mod-CRF}(\hat{\rho}) & = \frac{\Gamma_{\rm D}}{2} \left( 2 \hat{S}_-  \hat{\rho} \hat{S}_+ -\{\hat{S}_+ \hat{S}_-, \hat{\rho}\} \right) \nonumber \\
     &+\frac{\gamma_s}{2} \sum_{j} \left( 2 \hat\sigma^-_j  \hat{\rho} \hat\sigma^+_j - \{\hat\sigma^z_j, \hat{\rho}\}/2 - \hat{\rho} \right) \nonumber \\
     &+\gamma_d\sum_{j} \left( \hat\sigma^z_j  \hat{\rho} \hat\sigma^z_j  -\hat{\rho} \right),
    \label{eq:Lindbland_mod_DDM}
\end{align}
where the first line is the collective dissipation in the CRF model with rate $\Gamma_{\rm D}=\sum_{j\neq 0} {\rm Im}\tilde{\mathcal{G}}_{0j}/N$, the second line is the single-particle spontaneous emission with decay rate $\gamma_s$, and the third line is the single-particle dephasing with rate $\gamma_d$. Unlike the CRF model, this modified model does not preserve $\langle\hat{\mathbf{S}}^2\rangle$.  The case without dephasing ($\gamma_d=0$) and OAT ($\chi=0$) has been predicted to show a first-order phase transition and bistability \cite{Walls1978,carmichael_analytical_1980,roberts2023exact} at a critical driving strength given by  ${\rm\Omega_C} \approx \Gamma_{\rm D} N /(2\sqrt{2})$. Hence, spontaneous emission only changes the critical drive by a factor of $1/\sqrt{2}$ compared to the original CRF model (${\rm\Omega_C}^{\rm CRF} = \Gamma N/2$). When we include the effect of dephasing ($\gamma_d\neq 0$) and dispersive interactions ($\chi\neq 0$), we find that the modified model still undergoes a first-order phase transition with bistability, but the critical point depends on the $N$-scaling of $\gamma_d$. We obtain the mean-field steady-state in the superradiant phase for $N\gg 1$ from the following self-consistent equations (see Appendix \ref{APP_sec:analytical_model} for derivation) --
\begin{align}
    &\frac{\langle \hat{S}_z \rangle}{N/2} = -\frac{1}{2} \pm \frac{1}{2} \sqrt{1 - 8 \left( 1+\frac{4\gamma_d}{\gamma_s} \right) \left( \frac{\langle \hat{S}_x \rangle^2 + \langle \hat{S}_y \rangle^2}{N^2} \right)} \nonumber, \\ 
    &\frac{\langle \hat{S}_x \rangle^2+\langle \hat{S}_y \rangle^2}{N^2} = \frac{\Omega^2 - \Gamma_{\rm D}\gamma_s N(1+\langle \hat{S}_z \rangle/(N/2))}{N^2(\Gamma_{\rm D}^2 + \chi^2)} \nonumber.
\end{align}
The critical point is defined as the driving strength at which the solution above ceases to be valid and is obtained as (with $\gamma_{d}=c_dN\gamma_s/4$)
\begin{align}\label{eq:OmCrit_toy_model}
    {\rm \Omega_C^{Mod-CRF}} = \Gamma_{\rm D} \sqrt{N} f
\end{align}
where $f \approx \frac{1}{2 \sqrt{2}}\sqrt{\frac{1 + (\chi/\Gamma_{\rm D})^2}{c_d} + \frac{4\gamma_s}{\Gamma_{\rm D}}}$ is a constant.

\begin{figure}[ht!]
\includegraphics[width=\linewidth]{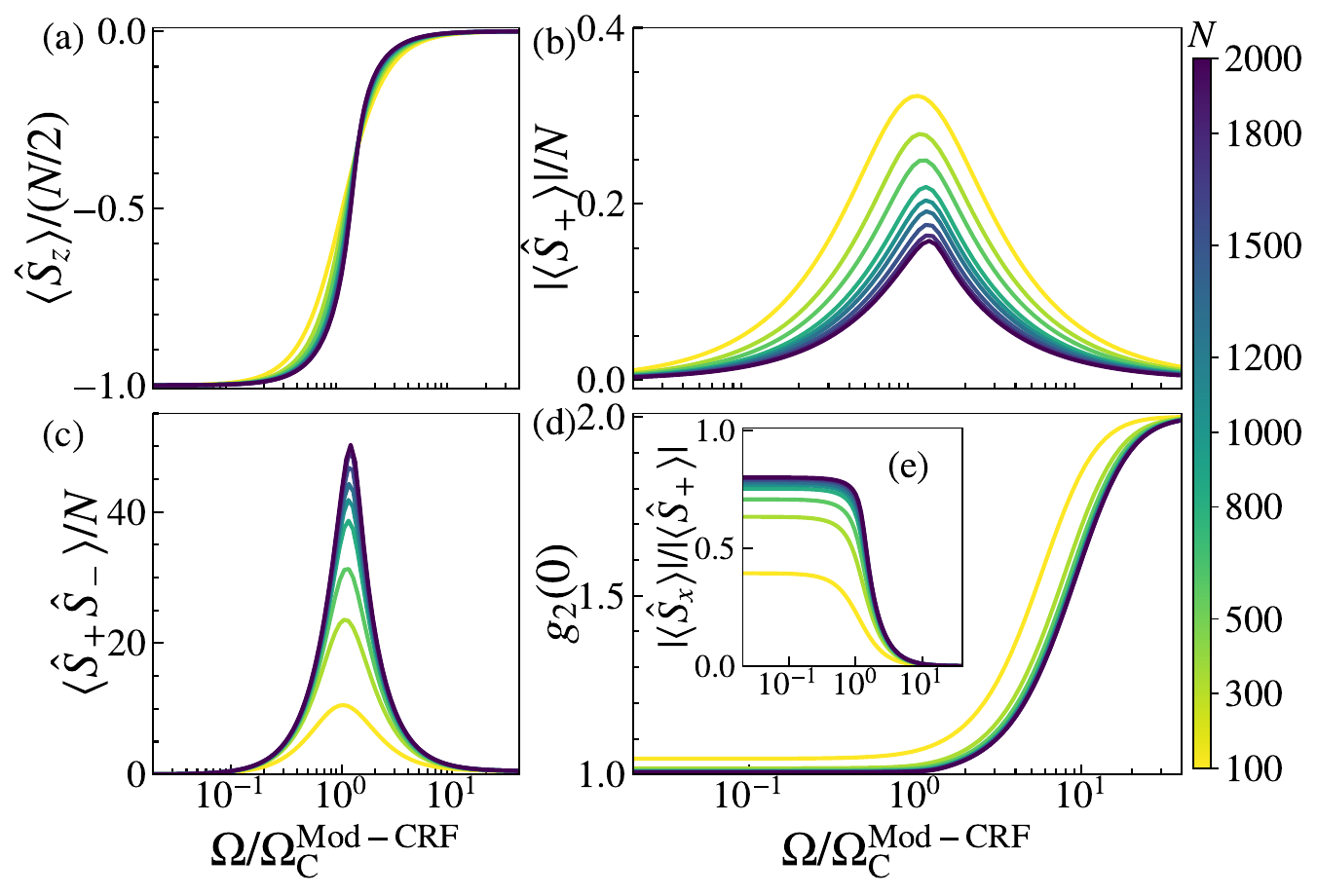}
\caption{Steady-state of the modified CRF model from MF numerics: (a) atomic inversion $\langle \hat{S}_z\rangle/(N/2)$, (b) total coherence $|\langle \hat{S}_+\rangle|/N$, (c) intensity $\langle \hat{S}_+\hat{S}_-\rangle$, which largely scales as $\sim N$ due to dephasing, (d) two-photon correlation function ${g}_2(0)$. (e) The $x$-component of the coherence, $|\langle \hat{S}_x\rangle|/|\langle \hat{S}_+\rangle|\neq 0$, arising from the mean-field frequency shift. In the large $N$ limit, ${\rm\Omega_C^{Mod-CRF}}\sim \Gamma\sqrt{N}$ is the critical drive for this model (Eq.~(\ref{eq:OmCrit_toy_model})).}
\label{fig:SS_4_plots_DDM_SpEm_Deph}
\end{figure}

In Fig.~\ref{fig:SS_4_plots_DDM_SpEm_Deph}, we show the steady-state values of the Mod-CRF model obtained from mean-field numerics, where we have set $c_d=0.002$. This specific value of $c_d$ is not special and we find qualitatively similar results for other values as long as $c_d\ll 1$. We have $\chi= \sum_{j\neq 0} {\rm Re}\tilde{\mathcal{G}}_{0j}/N \approx 0.003$ and $\Gamma_{\rm D} = \sum_{j\neq 0} {\rm Im}\tilde{\mathcal{G}}_{0j}/N \approx 0.002$ for our pencil-shaped cloud (see Appendix~\ref{APP_sec:analytical_model} for details). The steady-state of the Mod-CRF model shows qualitatively similar behaviors to the MF dipolar model (Fig.~\ref{fig:SS_Dip_4plots}). The critical drive strength for the Mod-CRF model, $\rm\Omega_C^{Mod-CRF}$, scales as $\sim \sqrt{N}$, similar to $\rm\Omega_C^{Dipolar}$. This $\sqrt{N}$-scaling is also recovered in the numerics as the scaled numerical curves ($x$-axis scaling: $\rm\Omega/\Omega_C^{Mod-CRF}$) of the atomic inversion become indistinguishable at large $N$ (Fig.~\ref{fig:SS_4_plots_DDM_SpEm_Deph}a). Similarly, the scaled numerical curves of the intensity (Fig.~\ref{fig:SS_4_plots_DDM_SpEm_Deph}c), $g_2(0)$ (Fig.~\ref{fig:SS_4_plots_DDM_SpEm_Deph}d), and the atomic coherences (Figs.~\ref{fig:SS_4_plots_DDM_SpEm_Deph}b and \ref{fig:SS_4_plots_DDM_SpEm_Deph}e) also look indistinguishable. In the weak-drive regime, the real-part of the atomic coherence, $\langle \hat{S}_x \rangle$, is non-zero and increases with $N$ due to collective shifts (Fig.~\ref{fig:SS_4_plots_DDM_SpEm_Deph}e). In the strong-driving regime ($\Omega\gg\rm\Omega_C^{Mod-CRF}$), the intensity of the Mod-CRF becomes single-particle-like, i.e., $\langle \hat{S}_+\hat{S}_-\rangle=N/2$, as the coherences go to zero (Fig.~\ref{fig:SS_4_plots_DDM_SpEm_Deph}b). In this regime, the atomic inversion goes to zero as well (Fig.~\ref{fig:SS_4_plots_DDM_SpEm_Deph}a), and $g_2(0)$ reaches its thermal value of 2 (Fig.~\ref{fig:SS_4_plots_DDM_SpEm_Deph}d). These characteristics describe a departure from the CRF model and qualitatively resemble the properties of the MF dipolar model (Fig.~\ref{fig:SS_Dip_4plots}, Fig.~\ref{fig:freq_shift}a).

\section{\label{sec:compare_with_exp}Comparisons with Experiment}

\begin{figure}[ht!]
\includegraphics[width=\linewidth]
{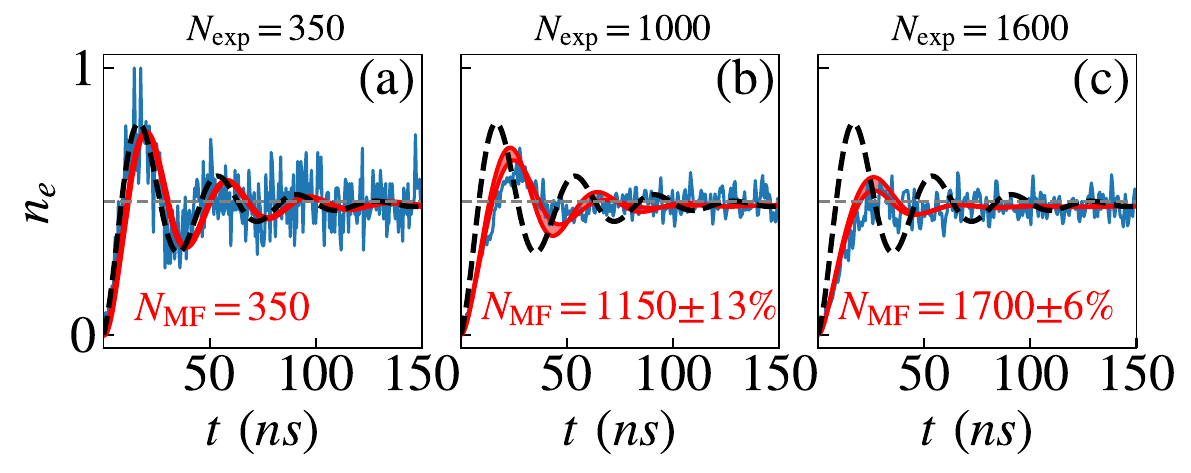}
\caption{Excitation fraction $n_e$ vs time  for fixed $\Omega/\Gamma = 4.5$: MF dipolar numerics for $N_{\rm MF}$ atoms (red thick) on top of experimental data (blue traces) for $N_{\rm exp}$ taken from Ref.~\cite{ferioli_non-equilibrium_2023}. The shaded red regions show the numerics data in the specified $N_{\rm MF}$ ranges. The dashed black line is the solution of the non-interacting model.} 
\label{fig:Exp_n_vs_t}
\end{figure}

In this section  we compare our mean-field numerics  for the dipolar model with experimental  results \cite{ferioli_non-equilibrium_2023}. The experimental system and measurement protocols are all described in Ref.~\cite{ferioli_non-equilibrium_2023}. The  participating states correspond to the $\sigma_+$-polarised atomic transition between the two levels $|5 S_{1/2},F=2,m_F=2 \rangle \to |5P_{3/2},F=3,m_F=3\rangle$ in $\rm ^{87}Rb$ with linewidth $\rm\Gamma=2\pi \times 6 MHz$. The Clebsch-Gordan coefficient for this transition is 1. This system is the same as the one described in Fig.~\ref{fig:schematic}.

In Fig.~\ref{fig:Exp_n_vs_t}, we show the dynamics of the total excitation fraction $n_e(t) = \sum_i (\langle \sigma_z^i (t)\rangle + 1)/(2N)$. To do this, we evolve the system under a continuous drive with Rabi frequency $\Omega = 4.5\Gamma$ from the ground state to the steady-state. We find that the mean-field numerics (red lines) agree more or less with the experiment (blue traces), for different values of $N$. The shaded red region shows $n_e$ within a range of $N$ values to account for experimental uncertainties. Both the numerics and the experimental data feature a dephasing of the Rabi oscillations with increasing particle number, which is not seen in the non-interacting model (black dashed).

\begin{figure}[ht!]
\includegraphics[width=\linewidth]{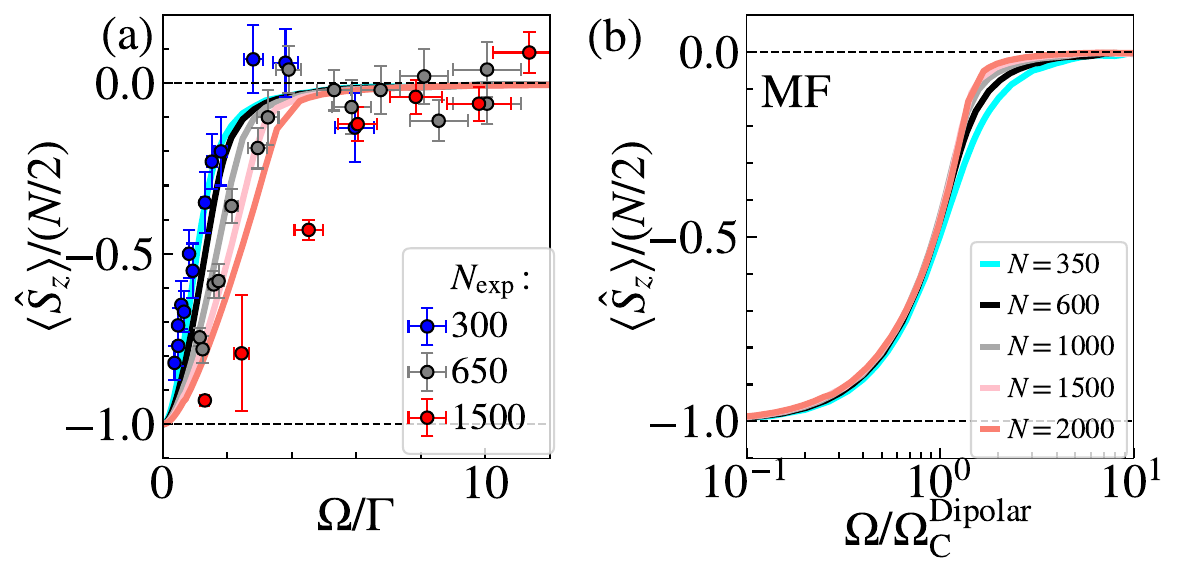}
\caption{$\langle \hat{S}_z \rangle /(N/2)$: Steady-state atomic inversion. (a) MF Dipolar numerics (solid lines) and experimental data (symbols) from Ref.~\cite{ferioli_non-equilibrium_2023}. (b) MF data with $x$-axis scaling $\rm \sim\Omega/\Omega_C^{Dipolar}$, showing collapse of curves for different $N$.}
\label{fig:Exp_Sz_Om}
\end{figure}

\begin{figure}[ht!]
\includegraphics[width=\linewidth]{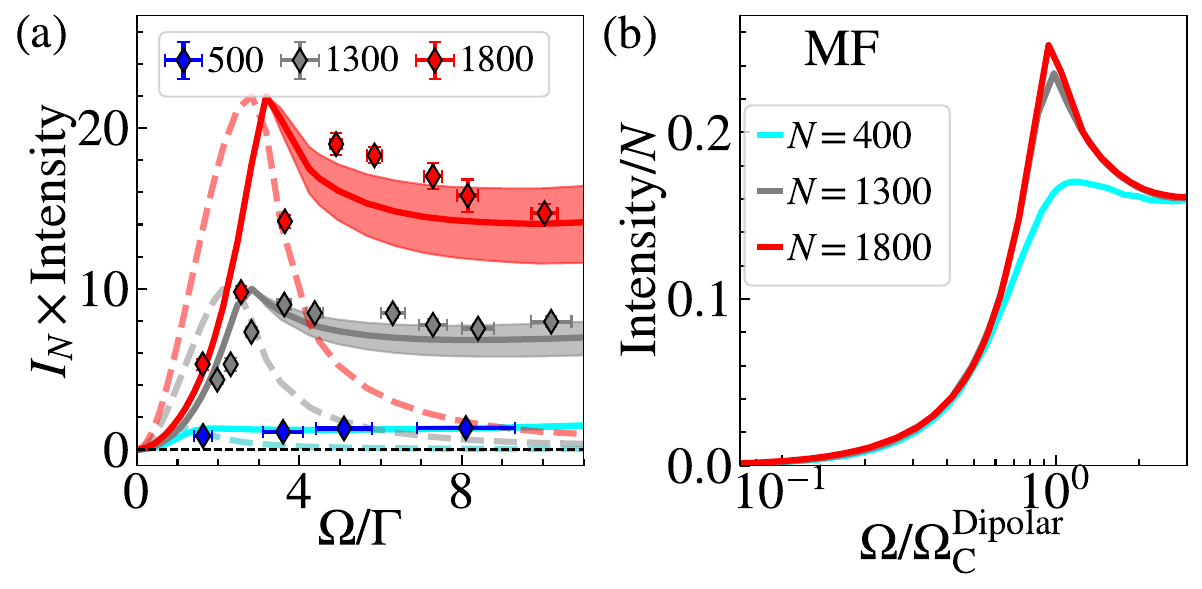}
\caption{(a) Steady-state intensity of the MF dipolar model in the forward direction ($\theta=0$, dashed), averaged over a solid angle ($\Delta\theta=\theta_{\rm out}-\theta_{\rm in}$) around the forward direction with $\theta_{\rm in}=\pi/22$ and $\theta_{\rm out}=\pi/7\pm\pi/50$ (solid lines, shaded region), and experimental data (dots) from Ref.~\cite{ferioli_non-equilibrium_2023}. $I_N$ is a fitting factor for the MF dipolar model chosen to match the theory peak with the experimental values. 
(b) MF data ($\theta_{\rm in}=\pi/22$ and $\theta_{\rm out}=\pi/7$) with $x$-axis scaling $\rm \sim\Omega/\Omega_C^{Dipolar}$ shows collapse of curves, as expected.} 
\label{fig:Exp_Intensity_Om}
\end{figure}

In Fig~\ref{fig:Exp_Sz_Om}a, we look at the steady-state atomic inversion and find very good agreement of our numerics (continuous lines)  with the experimental data (symbols) for $N=300$. An experimental data point in Fig~\ref{fig:Exp_Sz_Om}a for $N=1500$ at $\Omega/\Gamma\approx 4.5$ ($\langle\hat{S}_z\rangle/(N/2) \approx -0.45$) is not consistent with the steady-state in Fig.~\ref{fig:Exp_n_vs_t}c ($n_e (t\to \infty) \approx 0.5 \Rightarrow \langle\hat{S}_z\rangle/(N/2) \approx 0$). This could be due to calibration errors or other experimental systematics. Other than these quantitative differences, it is clear that there is a qualitative agreement with the theory in the trends observed in the experiment. 
In Fig~\ref{fig:Exp_Sz_Om}b, we plot the numerical data with our scaling of the Rabi frequency, ${\rm \Omega/\Omega_C^{Dipolar}} \sim \Omega/(0.08\Gamma \sqrt{N})$, and the data collapses to a single curve, as previously discussed. 

In Fig~\ref{fig:Exp_Intensity_Om}a, we compare the steady-state intensity from MF numerics with the experimental data \cite{ferioli_non-equilibrium_2023} and find fair agreement. Here, we look at the steady-state intensity, $I_0(\vec{k}) \sum_{ij} \langle \hat{\sigma}_i^+ \hat{\sigma}_j^- \rangle e^{i\vec{k}\cdot\vec{r}_{ij}}$, strictly in the forward direction ($\vec{k}=\vec{k}_{\rm L}$, dashed) and averaged over a solid angle around the forward direction (shaded region). Here, $\hat{k}=(\cos\theta,\sin\theta \cos\phi, \sin\theta \sin \phi)$, is the direction of observation, with $\theta$ the angle from the $\hat{x}$-axis (forward direction) and $\phi$ the azimuthal angle in the $y-z$ plane from $\hat{y}$. 
For the circularly-polarized transition, $\hat{e}_{q}=\hat{e}_{+} = - (\hat{x}+ i\hat{y})/\sqrt{2}$, and we get $I_0(\vec{k})\propto(1+\sin^2\theta \sin^2\phi)/2$ (see Appendix~\ref{APP_sec:non_int_model_solid_angle}). As some of the light in the forward direction is filtered out in the experiment to remove the laser light, we average the intensity over an annular region \cite{ZhuPRA2016,Darrick2024}, such that the averaged intensity is expressed as $\int_{\theta_{\rm in}}^{\theta_{\rm out}} d \theta \int_0^{2\pi} d \phi \,\,\sin\theta\,\, I_0(\vec{k}) \sum_{ij} \langle \hat{\sigma}_+ \hat{\sigma}_- \rangle e^{i\vec{k}\cdot\vec{r}_{ij}}/A$, where $\theta_{\rm out} = \pi/7 \pm \pi/50$, $\theta_{\rm in} = \pi/22$, and $A=\int_{\theta_{\rm in}}^{\theta_{\rm out}} d \theta \int_0^{2\pi} d \phi \,\,\sin\theta$ is the normalization factor. The values of $\theta_{\rm out}$ and $\theta_{\rm in}$ are extremely difficult to determine experimentally and hence, we consider them as fitting parameters, while making sure their values remain  within a reasonable range.
\begin{figure}[ht!]
\includegraphics[width=\linewidth]{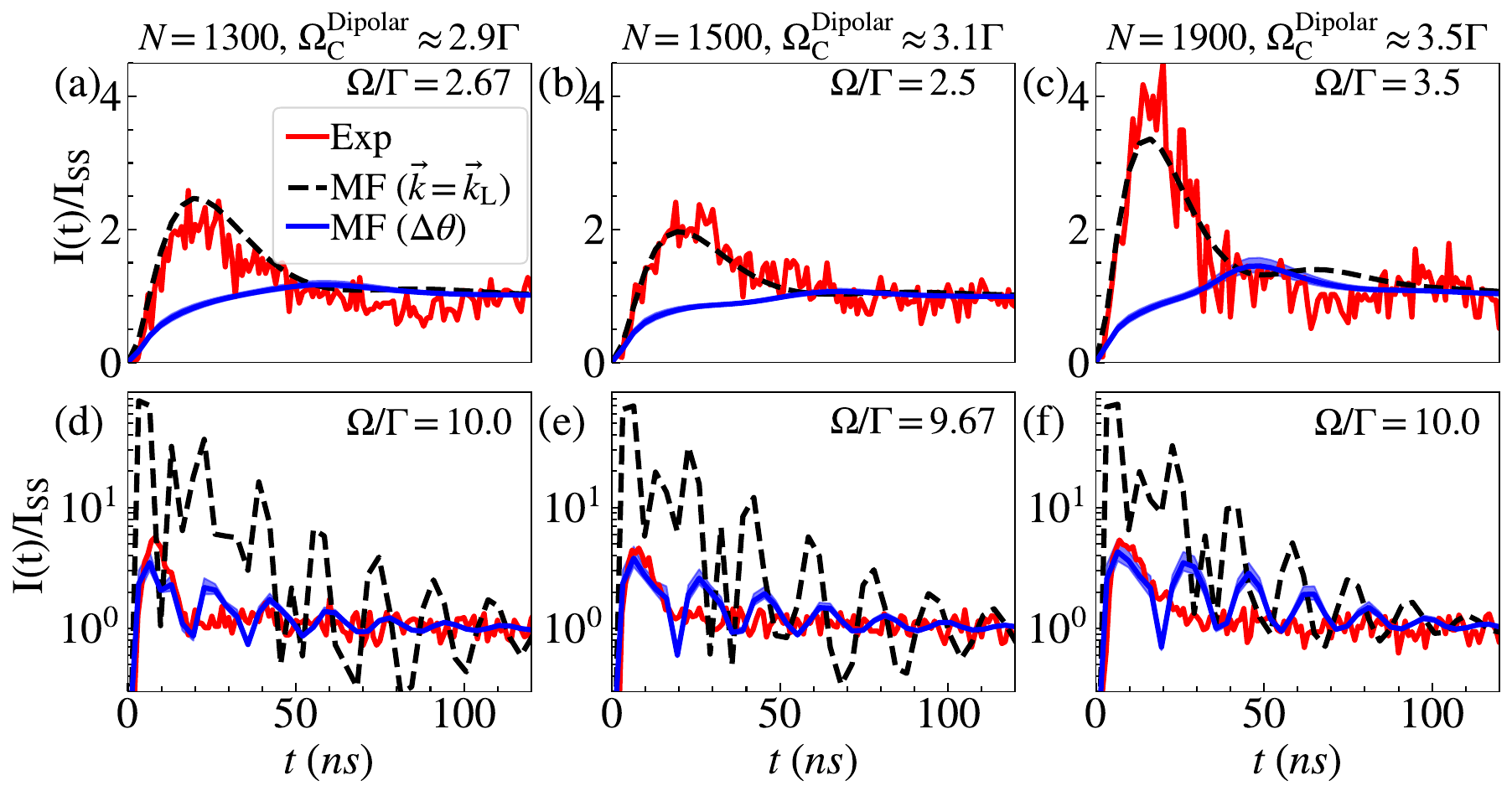}
\caption{Dynamics of intensity for the MF dipolar model in the forward direction (black dashed), averaged over a solid angle ($\Delta\theta=\theta_{\rm out}-\theta_{\rm in}$) about the forward direction with $\theta_{\rm in}=\pi/22$ and $\theta_{\rm out}=\pi/7 \pm \pi/50$ (blue solid, blue shaded region), experimental data (red traces) \cite{Antoine2024}. The intensity $\rm I(t)$ has been scaled by its steady-state value, $\rm  I_{SS}$, for each dataset. (a)-(c) weak and intermediate driving regimes ($\Omega \leq \Omega_{\rm C}^{\rm Dipolar}$), (d)-(f) strong-driving regime ($\Omega > \Omega_{\rm C}^{\rm Dipolar}$).}
\label{fig:Exp_Intensity_vs_t_Om}
\end{figure}

\begin{figure}[ht!]
\includegraphics[width=0.7\linewidth]{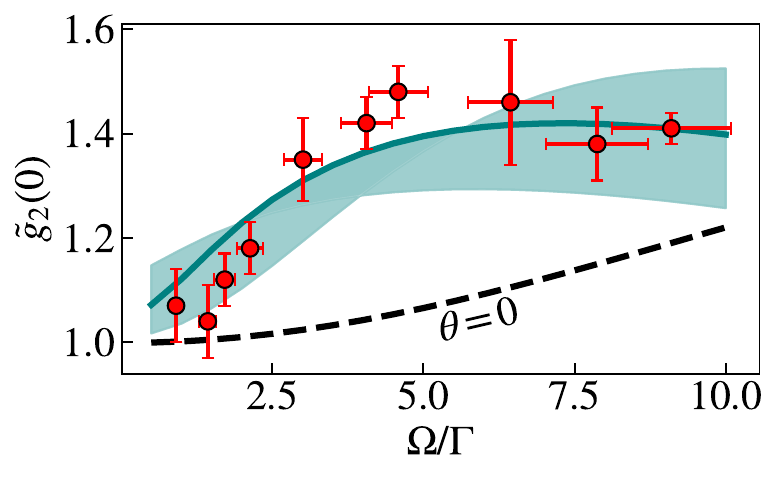}
\caption{$\tilde{g}_2(0)$ in the steady-state for $N=1400$: in the forward direction ($\theta=0$, black dashed), averaged over a solid angle ($\Delta\theta=\theta_{\rm out}-\theta_{\rm in}$) about the forward direction with $\theta_{\rm in}=\pi/22$ and $\theta_{\rm out}=\pi/16 \pm \pi/100$ (green shaded region), experimental data (red dots) from Ref.~\cite{ferioli_non-equilibrium_2023}.}
\label{fig:Exp_g2}
\end{figure}

Furthermore, we have included  direct contributions from the probe light to the intensity, considering that the filter may not be perfect (see Appendix~\ref{APP_sec:incl_laser_light} for details). We use  a $\sim 4\%$ leakage of probe light intensity, which is also a fitting parameter and its value is consistent across Figs.~\ref{fig:Exp_Intensity_Om}, \ref{fig:Exp_Intensity_vs_t_Om}, and \ref{fig:Exp_g2}.
The experimental data, which is the photon rate in arbitrary units, is scaled with a fitting factor, which was used in Ref.~\cite{ferioli_non-equilibrium_2023} to compare their data with the CRF model. To compare our results with the experimental data, we scale the intensity with a fitting factor $I_N$, which is obtained by matching the peak of the experimental curve with the peak of the numerics for each $N$.

In Fig.~\ref{fig:Exp_Intensity_vs_t_Om}, we compare the dynamics of the forward intensity, $\rm I(t)$, measured in the experiment (red traces) \cite{Antoine2024} with the intensity from the MF numerics in the strictly forward direction (black dashed) and the MF numerics data averaged over a solid angle about the forward direction (blue solid) with $\theta_{\rm in}=\pi/22$ and a range of $\theta_{\rm out} = \pi/7 \pm \pi/50$ (blue shaded region). This is the same range of solid angle values discussed earlier for the steady-state intensity. To properly compare the bare photon rate from the experiment with the numerical values of the intensity, we scale the dynamical intensity $\rm I(t)$ by its steady-state value $\rm I_{SS}$, thus making the scaled intensity, $\rm I(t)/I_{SS}$, independent of the distance between the atomic cloud and the detector, which is difficult to determine exactly for the experimental setup. In the weak ($\rm\Omega<\Omega_C^{Dipolar}$) and intermediate ($\rm\Omega\sim\Omega_C^{Dipolar}$) driving regimes, the experiment agrees well with the purely forward intensity of the dipolar model (black dashed) for different $N$, as shown in Figs.~\ref{fig:Exp_Intensity_vs_t_Om}(a)-(c). 
In Figs.~\ref{fig:Exp_Intensity_vs_t_Om}(d)-(f), we see that in the strong-driving regime ($\rm\Omega>\Omega_C^{Dipolar}$), the experiment agrees well with the dipolar model intensity when averaged over the solid angle (blue shaded region) for different $N$. Thus, the solid angle ($\theta_{\rm in},\theta_{\rm out}$), for which the numerics and the experiment agree, depends on the drive strength, $\Omega/\Gamma$. One possible explanation  is the fact that the radiative force of the laser  pushes the atomic cloud closer to the filter and the detector, changing  the solid angle of the detected light.
Of course, the radiative force may also lead to other motional effects such as dephasing, which we have not considered here.

In Fig.~\ref{fig:Exp_g2}, we look at the steady-state equal-time two-point photon correlation function $\tilde{g}_2(0)$, as defined in Eq.~(\ref{eq:g2_k_kp_def}), in the strictly forward direction ($\theta=0$, black dashed) and averaged over an annular region around the forward direction, same as we did for the intensity earlier, with the same $\theta_{\rm in} = \pi/22$ but with a smaller $\theta_{\rm out} = \pi/16\pm \pi/100$ (green shaded region). Again, we find fair agreement between the experimental (red dots) \cite{ferioli_non-equilibrium_2023} and numerical (green shaded region) values, when averaged over the solid angle. We see some discrepancies at intermediate $\Omega/\Gamma$ values, which could arise from the dephasing of coherences due to atomic motion in the experiment. The best-fit $\theta_{\rm out}$ values are not identical to the ones used for the intensity plots but  such changes could occur between different experimental measurements. Here, we plot the numerical values for the non-interacting model instead of the dipolar model, as we find it is quicker to integrate over the solid angle analytically over the non-interacting model's steady-state, than to do a discrete sum in this case. We have already shown that the $\tilde{g}_2(0)$ values for the non-interacting model coincide with those of the MF dipolar model in Fig.~\ref{fig:SS_Dip_4plots}d, so we expect these values to be valid for our system. For further details about the analytical averaging of $\tilde{g}_2(0)$ for the non-interacting model, see Appendix~\ref{APP_sec:non_int_model}.   
\section{\label{sec:conclusions}Conclusions}

We have studied the age-old, albeit complicated, problem of characterizing the emission properties of a spatially-extended ensemble of driven interacting atoms in free-space. What makes this problem complicated is the lack of symmetries such as translational invariance in periodic arrays and permutational invariance in optical cavities. In the absence of symmetries, an exact microscopic treatment of the system is impossible due to its exponentially large Hilbert space ($\sim 2^N$). Nevertheless, we are able to greatly reduce the complexity of the problem by identifying two distinct driving regimes (weak and strong), where a mean-field (MF) treatment is almost exact and the problem becomes at least numerically tractable with just $3N$ degrees of freedom. Furthermore, we find that our MF numerics are able to qualitatively capture the physics across all driving regimes and beyond-MF methods (MACE-MF and cumulant) lead to negligible corrections.

We find a crossover between non-equilibrium phases in the steady-state of our system as a function of the driving strength $\Omega/\Gamma$, akin to the superradiant phase transition in the CRF model. We find that the inhomogeneity of free-space dipolar interactions plays a key role in making our system strikingly different from the CRF model. This is most prominent in the $N$-scaling of the critical order parameter, which is $\propto \sqrt{N}$ in our system and differs from that in the permutationally invariant CRF model ($\propto N$). Moreover, the strongly-driven phase in our model is completely mixed and single-particle-like, whereas the ``normal'' phase of the CRF has collective quantum correlations.  

At the cloud densities considered, our system is governed by the far-field dipolar interaction ($\sim 1/r$), which is long-range in quasi-1D. The collective effects of the long-range interactions are tempered by the dephasing arising from inhomogeneous frequency shifts and free-space emission. Using these physical insights, we propose a heuristic  modified CRF model capable of  describing our system, which includes single-particle dephasing, shearing, and spontaneous emission. This modified CRF is able to qualitatively reproduce the emergent features of our complicated microscopic dipolar model and is analytically solvable at the mean-field level. 

More importantly, our results are able to reproduce most of the experimental findings from a recent work \cite{ferioli_non-equilibrium_2023}. Our work bridges the theoretical gap between the widely studied CRF model and the spatially-extended inhomogeneous atomic ensembles accessible in current experimental setups. While previous work has shed some light on the properties of dilute atomic ensembles, our work extends this knowledge to moderately dense ensembles where collective effects are relevant beyond just frequency shifts and linewidth broadening.

For future work, it would be useful to measure (using Ramsey spectroscopy) the collective frequency shift arising from dipolar interactions, which leads to a non-zero $\langle \hat{\tilde{S}}_x \rangle$ in the dipolar model, as opposed to the CRF and non-interacting models, where $\langle \hat{\tilde{S}}_x \rangle=0$. This is a smoking gun signature of many-body effects in this system. The next step would be studying highly dense atomic ensembles, where the near-field elastic dipolar interactions ($\sim 1/r^3$) dominate the physics and the MF treatment breaks down. The effects of atomic motion would become pertinent in the presence of strong light-matter interactions in dense ensembles. Another avenue worth exploring is the case of multilevel atoms, which includes the internal level structure of alkaline-earth(-like) atoms due to hyperfine splitting. Even in the weak-driving regime, the multilevel system can have quantum correlations in the ground-state manifold \cite{AsenjoPNAS2019} and is distinct from the semi-classical two-level picture. While there has been some recent work on the emission properties of multilevel arrays in free-space, it is largely confined to the early-time dynamics \cite{BilitewskiPRL2022} and inverted arrays \cite{masson2023dicke}. Characterizing the steady-state properties of multilevel ensembles is crucial for current experiments and remains an unsolved problem.


During the completion of our work, we became aware of a related recent work that reaches similar conclusions as us using a complementary theoretical treatment \cite{Darrick2024}.

\begin{acknowledgments}
We thank Jarrod T. Reilly,  Eric Yilun Song, and James K. Thompson for insightful discussions and feedback on our manuscript. We wish to acknowledge support by the VBFF,  AFOSR grant FA9550-18-1-0319,  by the NSF JILA-PFC PHY-2317149, QLCI-OMA-2016244, by the U.S. Department of Energy, Office of Science, National Quantum Information Science Research Centers Quantum Systems Accelerator, and by NIST.

\end{acknowledgments}

\appendix

\section{\label{APP_sec:non_int_model}Non-interacting model}

Here we consider the non-interacting system and its steady-state solution. The master equation describing this system is $\dot{\hat{\rho}} = -i[\hat{H}_0, \hat{\rho}]+\mathcal{L}_0(\hat{\rho})$, where $\hat{H}_{0} = - ({\Omega}/{2}) \sum_k (e^{i\Vec{k}_{\rm L}\cdot \Vec{r}_k} \sigma^+_k + h.c.)$ is the Hamiltonian and $\mathcal{L}_0(\hat{\rho})= (\Gamma/2)\sum_{j} \left( 2 \sigma^-_j  \hat{\rho} \sigma^+_j - \{\sigma^z_j, \hat{\rho}\}/2 - \hat{\rho} \right)$ is the Lindbladian. The atomic equations of motion reduce to the standard optical Bloch equations:
\begin{align}
    \langle{\dot{\sigma}^z_k}\rangle &= - \Gamma \left(\langle{{\sigma}^z_k}\rangle + 1\right) + i\Omega\left(  e^{i\vec{k}_{\rm L}\cdot\vec{r}_k} \langle{{\sigma}^+_k}\rangle - e^{-i\vec{k}_{\rm L}\cdot\vec{r}_k} \langle{{\sigma}^-_k}\rangle \right),  \\
    \langle{\dot{\sigma}^+_k}\rangle &=- \frac{\Gamma}{2}\langle{{\sigma}^+_k}\rangle + i \frac{\Omega}{2}  e^{-i\vec{k}_{\rm L}\cdot\vec{r}_k}   \langle{{\sigma}^z_k}\rangle .
\end{align} and the steady-state can be obtained by setting $\langle{\dot{\sigma}^z_k}\rangle=0$ and $\langle{\dot{\sigma}^+_k}\rangle=0$ for all $k$, as
\begin{align}
    \langle{{\sigma}^z_k}\rangle &= -\frac{\Gamma^2}{\Gamma^2 + 2\Omega^2} \equiv -R \,\,\, \\
    \langle{{{\sigma}}^+_k}\rangle &= -i\frac{\Omega}{\Gamma} e^{-i \vec{k}_{\rm L}\cdot \vec{r}_k}R.
\end{align}
The steady-state intensity in an observed direction $\vec{k}$ is
\begin{align}\label{eq:int_non_int_k_dir_SS}
    \frac{I(\vec{k})}{I_0(\vec{k})} &= \sum_j \left( \frac{\langle{{\sigma}^z_j}\rangle + 1}{2} \right) + \sum_{i\neq j} \langle{{\sigma}^+_i}\rangle\langle{{\sigma}^-_j}\rangle e^{i\vec{k}\cdot(\vec{r}_i - \vec{r}_j)} \nn \\
    &= \frac{(1-R)}{2} N + \frac{\Omega^2}{{\Gamma^2}} R^2 \sum_{i\neq j} e^{i(\vec{k} - \vec{k}_{\rm L})\cdot(\vec{r}_i - \vec{r}_j)}.
\end{align}
The two photon correlation function in an observed direction $\vec{k}$ is
\begin{align}
    \tilde{g}_2(0)(\vec{k}) = \frac{\langle \tilde{S}^+(\vec{k}) \tilde{S}^+ (\vec{k})\tilde{S}^-(\vec{k}) \tilde{S}^-(\vec{k}) \rangle}{\langle \tilde{S}^+(\vec{k}) \tilde{S}^-(\vec{k}) \rangle^2},
\end{align}
where $\tilde{S}_{\pm}(\vec{k}) = \sum_i\sigma_i^\pm e^{\pm i \vec{k}\cdot \vec{r}_i}$. For non-interacting particles, the correlations factor  for different atoms as $\langle \sigma_\alpha^i \sigma_\beta^j\rangle = \langle \sigma_\alpha^i \rangle \langle \sigma_\beta^j\rangle$. For the same atom, we use the commutation relations of the Pauli operators as $\langle \sigma_\alpha^i \sigma_\beta^i\rangle = \delta_{\alpha,\beta} \mathbbm{1} + 2i \varepsilon^{\alpha \beta \gamma} \langle \sigma_\gamma^i \rangle $, where $\alpha,\beta,\gamma \in \{x,y,z\}$. Using this scheme, the two photon correlation function along the measurement direction can be expressed as
\begin{align}\label{eq:g2_def_MF} 
    &\tilde{g}_2(0)(\vec{k}_{\rm L}) = \bigg(\frac{1}{2}\sum_{\langle i j \rangle}\left(\langle\sigma^{z}_i\rangle + 1\right) \left(\langle\sigma^{z}_j\rangle+ 1\right) \nn \\
    &+ 2\sum_{\langle i j k \rangle}\left(\langle\sigma^{z}_i\rangle + 1\right)  \langle\tilde{\sigma}^{+}_j\rangle \langle\tilde{\sigma}^{-}_k\rangle + \sum_{\langle i j k l \rangle} \langle\tilde{\sigma}^{+}_i\rangle \langle \tilde{\sigma}^{+}_j\rangle \langle\tilde{\sigma}^{-}_k\rangle \langle \tilde{\sigma}^{-}_l\rangle\bigg) \nn \\
    &\times  \bigg(\frac{1}{2} \sum_{i}\left(\langle\sigma^{z}_i\rangle + 1\right) + \sum_{\langle i j \rangle} \langle\tilde{\sigma}^{+}_i\rangle \langle\tilde{\sigma}^{-}_j\rangle \bigg)^{-2},
\end{align}
where we have suppressed the notation to denote $\tilde{\sigma}^{+}_j\equiv {\sigma}^{+}_j e^{i\vec{k}_{\rm L}\cdot \vec{r}_j}$ and $\langle i j \dots \rangle$ denotes sum over unlike indices, i.e., $i\neq j\neq \dots$.

\subsection{Weak-driving regime}

In the weak-driving regime ($\Omega \ll \Gamma$), the steady-state solution can be expressed as 
\begin{align}
    \langle\sigma^z_k\rangle &= -\left(1 - \frac{2\Omega^2}{\Gamma^2}\right) \\
    \langle\sigma^+_k\rangle &= i\frac{\Omega}{\Gamma} e^{-i \vec{k}_{\rm L}\cdot \vec{r}_k},
\end{align}
and the intensity along the laser wavevector is 
\begin{align}
    \frac{I(\vec{k}_{\rm L})}{I_0(\vec{k}_{\rm L})} =   \frac{\Omega^2}{\Gamma^2}N\left(N - \frac{1}{2} \right)  \approx   \frac{\Omega^2}{\Gamma^2}N^2  ,
\end{align}
which scales as $N^2$ in the large-$N$ limit due to the constructive interference of coherences along the measurement direction. In this regime, the two-photon correlation function can be expressed as
\begin{align}\label{eq:g2_MF_lowOm}
    \tilde{g}_2(0)(\vec{k}_{\rm L}) = 1 - \frac{2}{N} + \frac{1}{N^2} \Rightarrow  \lim_{N\to \infty}  \tilde{g}_2(0)(\vec{k}_{\rm L}) = 1,
\end{align}
which is consistent with the value for a coherent state. In the extreme weak-driving limit, the atoms are very weakly-excited such that $\langle \sigma_j^z \rangle \to -1$, which along with Eq.~(\ref{eq:g2_def_MF}) gives $\tilde{g}_2(0)(\vec{k}_{\rm L}) \to 1$.

\subsection{Strong-driving regime}

In the strong-driving regime ($\Omega \gg \Gamma$), the steady-state solution can be expressed as 
\begin{align}
    \langle{{\sigma}^z_k}\rangle &= -\frac{\Gamma^2}{2\Omega^2} \\
    \langle{{{\sigma}}^+_k}\rangle &= i\frac{\Gamma}{2\Omega} e^{-i \vec{k}_{\rm L}\cdot \vec{r}_k},
\end{align}
and the intensity along the laser wavevector given by 
\begin{align}
    \frac{I(\vec{k}_{\rm L})}{I_0(\vec{k}_{\rm L})} = \frac{N}{2} + \frac{{\Gamma^2}}{4\Omega^2}  N(N-2).
\end{align}
When the drive is in the regime $\frac{{\Gamma^2}}{4\Omega^2} < 1/N \Rightarrow \Omega > \Gamma \sqrt{N}/2$ for all $N$, the intensity scales as $N$. As the drive gets extremely large, $\Omega \gg N\Gamma/2 $, the state gets fully mixed and $I(\vec{k}_{\rm L})\to\frac{N}{2} $. In this limit, we get
\begin{align}\label{eq:g2_MF_highOm}
    \tilde{g}_2(0) = 2 - \frac{2}{N} \Rightarrow  \lim_{N\to \infty}  \tilde{g}_2(0) = 2,
\end{align} consistent with the fact that  the system is described by a thermal state. In the extreme strong-driving limit, the system is maximally mixed such that $\langle \sigma_j^z \rangle \to 0$ and $\langle \sigma_j^{+,-} \rangle \to 0$. Plugging this into Eq.~(\ref{eq:g2_def_MF}) also gives $\tilde{g}_2(0)(\vec{k}_{\rm L}) \to 2$.

\subsection{\label{APP_sec:non_int_model_solid_angle} Averaging over a solid angle}

For our pencil-shaped cloud (Fig.~\ref{fig:schematic}), the atomic positions are distributed in a Gaussian distribution,
\begin{align}\label{eq:rho_xyz_Gaussian}
    \rho(x,y,z) = \frac{N\exp\left(-\frac{x^2}{2\sigma_{\rm ax}^2}-\frac{y^2+z^2}{2\sigma_{\rm rad}^2}\right)}{({2\pi})^{3/2} \sigma_{\rm ax} \sigma_{\rm rad}^2},
\end{align}
where $\sigma_{\rm ax} = 20 \lambda$ and $\sigma_{\rm rad} = \lambda/2$ are the axial and radial standard deviations of the cloud, respectively. Then, in the large-$N$ limit, we can replace the sum over atoms $i$ and $j$ in the intensity in Eq.~(\ref{eq:int_non_int_k_dir_SS}) by an integral over the positions, weighted by their distribution $\rho(x,y,z)$, as
\begin{align}
    \frac{I(\vec{k})}{I_0(\vec{k})} &\approx \frac{(1-R)}{2} N - \frac{\Omega^2}{{\Gamma^2}} R^2 N \nn \\
    &+ \frac{\Omega^2}{{\Gamma^2}} R^2 \int d^3r \int d^3r' \rho(\vec{r})  \rho(\vec{r}')e^{i(\vec{k} - \vec{k}_{\rm L})\cdot(\vec{r}- \vec{r}')} \nn 
\end{align}
where in the first line we have subtracted the contribution from the $i=j$ term in the integral. We define $\vec{q}= \vec{k}-\vec{k}_{\rm L} = (q_x,q_y,q_z)$. Then, it is easy to do obtain the integral in Cartesian coordinates as
\begin{align}
    A &=  \left | \int_{-\infty}^{\infty} dx \int_{-\infty}^{\infty} dy \int_{-\infty}^{\infty} dz \, \rho(x,y,z)  e^{i\left( q_xx+q_yy+q_zz\right)} \right |^2 \nn \\
    &= N^2 \exp(-(\sigma_{\rm ax}^2 q_x^2 + \sigma_{\rm rad}^2 q_y^2 + \sigma_{\rm rad}^2 q_z^2)).
\end{align}
Now, we substitute $\vec{k}_{\rm L} = (2\pi/\lambda) \hat{x} $ and $\vec{k}=(2\pi/\lambda) $ $(\cos\theta,\sin\theta \cos\phi, \sin\theta \sin \phi)$, with $\theta$ the angle from the $\hat{x}$-axis (forward direction) and $\phi$ the azimuthal angle in the $y-z$ plane from $\hat{y}$. Then, we obtain the intensity as
\begin{align}\label{eq:int_non_int_k_dir_SS_tp}
    \frac{I(\theta,\phi)}{I_0(\vec{k})}  &= \frac{2(\Omega/\Gamma)^4}{(1+2(\Omega/\Gamma)^2)^2}N\bigg[ 1  \nn \\
    &+ \frac{N \exp(-(2\pi)^2(\sigma_{\rm ax}^2 (1-\cos\theta)^2 + \sigma_{\rm rad}^2 \sin^2\theta ))}{2(\Omega/\Gamma)^2} \bigg].
\end{align}
Close to the forward direction $\theta \equiv \delta\theta \approx 0$, the intensity can be expressed as 
\begin{align}\label{eq:int_non_int_k_dir_SS_close_to_forw}
    \frac{I(\delta \theta,\phi)}{I_0(\vec{k})}  &\approx \frac{2(\Omega/\Gamma)^4}{(1+2(\Omega/\Gamma)^2)^2}N\bigg[ 1 + \frac{N e^{-(2\pi)^2(\sigma_{\rm rad}^2 \delta \theta^2 + \sigma_{\rm ax}^2 \delta \theta^4/4) }}{2(\Omega/\Gamma)^2} \bigg],
\end{align}
such that the $N^2$-enhancement falls off exponentially fast as the observation direction deviates from the forward direction.

We need to multiply the intensity above (Eq.~(\ref{eq:int_non_int_k_dir_SS_tp})) by the geometric factor $I_0(\hat{k})$. Due to the large distance of the detector from the cloud, $I_0(\hat{k})$ is the factor associated with the far-field part of the Electromagnetic Green's tensor and appears in the expression of the dipolar intensity pattern as $I_0(\hat{k}) = \hat{e}_q \cdot (\mathbbm{1}-\hat{k}\otimes \hat{k})\cdot (\mathbbm{1}-\hat{k}\otimes \hat{k})\cdot \hat{e}_q^*$, where $\hat{e}_q$ is the polarization of light associated with the atomic transition corresponding to fluorescence. For circularly polarized light, we get $I_0(\hat{k})=(1+\sin^2\theta \sin^2\phi)/2$. 

Then, the intensity can be averaged over an annular region, which has $\rm \theta_{in} = \pi/22$ and $\rm \theta_{out}=\pi/7$ as the inner and outer boundaries, to obtain $\langle I \rangle \equiv \int_0^{2\pi} d\phi \int_{\rm \theta_{in}}^{\rm\theta_{out}} d\theta I (\theta,\phi) \sin\theta (1+\sin^2\theta \sin^2\phi)/(2 A)\approx\frac{2(\Omega/\Gamma)^4}{(1+2(\Omega/\Gamma)^2)^2}N\big[ c_A^{\rm I}   + \frac{c_A^{\rm II} N}{2(\Omega/\Gamma)^2} \big]$, where $A = \int_0^{2\pi} d\phi \int_{\rm \theta_{in}}^{\rm\theta_{out}} d\theta  \sin\theta $. It can be seen that the extra factor of $c_A^{\rm II}/c_A^{\rm I} \approx 0.022$, due to the averaging  over a finite solid angle around the forward direction, reduces the relative strength of the coherent emission and shifts the peak of the intensity to lower $\Omega/\Gamma$. We have substituted $\rm \sigma_{ax}=\lambda/2$ and $\rm \sigma_{rad}=20\lambda$ above, corresponding to our system (Fig.~\ref{fig:schematic}), to get a numerical value.

Similarly, we can also integrate over the atomic positions, weighted by the Gaussian distribution, to obtain the analytical expression for the steady-state two-photon correlation function, $g_2 (\Vec{k},\Vec{k}')$, defined in Eq.~(\ref{eq:g2_k_kp_def}), in terms of $\hat{k}\equiv (\theta,\phi)$ and $\hat{k}'\equiv (\theta',\phi')$.

\section{\label{APP_sec:MF_dipolar_eqns}Mean-field Dipolar Model}

The mean-field equations of motion for the dipolar model can be obtained from the master equation by factoring the  multi-atom correlations  as $\langle {\hat{A}}_k {\hat{B}}_j \rangle = \langle {\hat{A}}_k \rangle \langle {\hat{B}}_j \rangle$, where $j\neq k$. For an atom $k$, this treatment leads to:
\begin{align}
    \langle{\dot{\sigma}^z_k}\rangle &= - \Gamma \left(\langle{{\sigma}^z_k}\rangle + 1\right) + 2 i\bigg(\frac{\Omega}{2}  e^{i\vec{k}_L\cdot\vec{r}_k} + \sum_{\substack{j=1\\j\neq k}}^N {\mathcal{G}_{kj}} \langle{{\sigma}^-_j}\rangle \bigg) \langle{{\sigma}^+_k}\rangle \nn \\
    & - 2 i\bigg(\frac{\Omega}{2}  e^{-i\vec{k}_L\cdot\vec{r}_k} + \sum_{\substack{j=1\\j\neq k}}^N {\mathcal{G}_{kj}^*} \langle{{\sigma}^+_j}\rangle \bigg) \langle{{\sigma}^-_k}\rangle, \\ 
    \langle{\dot{\sigma}^+_k}\rangle &= - \frac{\Gamma}{2} \langle{{\sigma}^+_k}\rangle + i\bigg( \frac{\Omega}{2}  e^{-i\vec{k}_L\cdot\vec{r}_k} + \sum_{\substack{j=1\\j\neq k}}^N {\mathcal{G}_{kj}^*} \langle{{\sigma}^+_j}\rangle   \bigg) \langle{{\sigma}^z_k}\rangle,
\end{align}
where ${\mathcal{G}}_{kj} = {\mathcal{R}}_{kj} + i {\mathcal{I}}_{kj} = {\mathcal{G}}_{jk}$ is the dipolar interaction coefficient and is symmetric with respect to the indices. We can see from the equations above that the dipolar interaction term acts like an effective time-dependent complex drive with a strength that depends on the coherence of other atoms in the array. The elastic and inelastic dipolar coefficients are the real and imaginary parts, respectively, of the free-space electromagnetic Green's function,
\begin{align}
    \mathcal{R}_{kj} &= ({3\Gamma}/{4})\bigg[(1-\cos^2\theta)\frac{\cos(k_0 r)}{k_0 r} \nonumber\\
    &+ (1-3\cos^2\theta)\left(-\frac{ \sin(k_0 r)}{(k_0 r)^2} - \frac{\cos(k_0 r)}{(k_0 r)^3}\right) \bigg] \nonumber\\
    \mathcal{I}_{kj} &= ({3\Gamma}/{4})\bigg[(1-\cos^2\theta)\frac{\sin(k_0 r)}{k_0 r} \nonumber\\
    &+ (1-3\cos^2\theta)\left(\frac{\cos(k_0 r)}{(k_0 r)^2} - \frac{\sin(k_0 r)}{(k_0 r)^3}\right) \bigg] \nonumber
\end{align}
where $r=|\vec{r}_{kj}|$,  $\cos\theta = \hat{r}_{kj}\cdot \hat{e}_q$, $q=0,\pm 1$, $\theta$ is the angle between the polarization of the atomic transition ($\hat{e}_q$, orientation of the transition ``dipole'') and the inter-atomic distance. For the MF dipolar model at the densities considered in this paper, $k_0 \bar{r} \geq 1$ and the physics is dominated by the $1/r$ (far-field) term. The far-field interaction coefficients are somewhat simpler -- 
\begin{align}
    \mathcal{R}_{kj} &\approx ({3\Gamma}/{4})(1-\cos^2\theta)\frac{\cos(k_0 r)}{k_0 r} \nonumber\\
    \mathcal{I}_{kj} &\approx ({3\Gamma}/{4})(1-\cos^2\theta)\frac{\sin(k_0 r)}{k_0 r} \nonumber
\end{align}
and we have ${\mathcal{G}}_{kj} = {\mathcal{R}}_{kj} + i {\mathcal{I}}_{kj} \approx  ({3\Gamma}/{4}) (1-\cos^2\theta) e^{ik_0 r}/(k_0 r)$. By gauging away the phase of the laser, we can rewrite the MF equations as --
\begin{align}
    \langle{\dot{\sigma}^z_k}\rangle &= - \Gamma \left(\langle{{\sigma}^z_k}\rangle + 1\right) + 2 i\bigg(\frac{\Omega}{2} + \sum_{\substack{j=1\\j\neq k}}^N {\tilde{\mathcal{G}}_{kj}} \langle{\tilde{\sigma}^-_j}\rangle \bigg) \langle{\tilde{\sigma}^+_k}\rangle \nn \\
    & - 2 i\bigg(\frac{\Omega}{2}  + \sum_{\substack{j=1\\j\neq k}}^N {\tilde{\mathcal{G}}_{kj}^*} \langle{\tilde{\sigma}^+_j}\rangle \bigg) \langle{\tilde{\sigma}^-_k}\rangle, \\ 
    \langle{\dot{\tilde{\sigma}}^+_k}\rangle &= - \frac{\Gamma}{2} \langle{\tilde{\sigma}^+_k}\rangle + i\bigg( \frac{\Omega}{2}  + \sum_{\substack{j=1\\j\neq k}}^N {\tilde{\mathcal{G}}_{kj}^*} \langle{\tilde{\sigma}^+_j}\rangle   \bigg) \langle{{\sigma}^z_k}\rangle,
\end{align}
where we have included the laser phase in the interaction coefficient as $\tilde{\mathcal{G}}_{kj} = \mathcal{G}_{kj} e^{-i\vec{k}_{\rm L}\cdot\vec{r}_{kj}} \approx  ({3\Gamma}/{4}) (1-\cos^2\theta) e^{i(k_0 r-\vec{k}_{\rm L}\cdot\vec{r}_{kj})}/(k_0 r)$, in the far-field limit. This additional phase leads to constructive interference of interaction terms in the direction of the laser wavevector, i.e., $k_0 r-\vec{k}_{\rm L}\cdot\vec{r}_{kj}=0$. For our setup, $\vec{k}_{\rm L} = 2\pi/\lambda \hat{x} \Rightarrow \vec{k}_{\rm L}\cdot\vec{r}_{kj} = k_0 (x_k-x_j)$ and in the quasi-1D gas, we have $|\vec{r}_{kj}| \approx |x_k-x_j|$ and $\cos^2\theta \approx |\hat{x}\cdot \hat{e}_q|^2$ ($q=0,\pm 1$) for the majority of atomic pairs. For a linearly polarized atomic transition ($\hat{e}_0 = \hat{z}$), $|\hat{x}\cdot \hat{e}_q|^2=0$ and for a circularly polarized transition ($\hat{e}_+ = -(\hat{x}+i\hat{y})/\sqrt{2}$), $|\hat{x}\cdot \hat{e}_q|^2=1/2$. Then, $\tilde{\mathcal{G}}_{kj,k>j} \approx  ({3\Gamma}/{8})/(k_0 r)$ and $\tilde{\mathcal{G}}_{kj,k<j} \approx  ({3\Gamma}/{8})  e^{2ik_0 r}/(k_0 r)$, where we have assumed that the atomic indices are sorted in increasing order of position along the $x$-axis. Unlike in periodic arrays with special lattice spacing such that $e^{2ik_0 r}=1$, in a disordered array the $e^{2ik_0 r}$ phases in $\tilde{\mathcal{G}}_{kj,k<j}$ terms would get washed out in comparison to the  $\tilde{\mathcal{G}}_{kj,k>j}$ terms. In the MF equations of motion, we have constructive interference of terms $\sim \tilde{\mathcal{G}}_{kj,k>j} \langle{\tilde{\sigma}^-_j}\rangle  \langle{\tilde{\sigma}^+_k}\rangle$. This implies that due to the phase matching of dipolar interactions with the laser drive, effects will constructively add up from an atom $k$ further along in the path of the laser absorbing a photon emitted from an atom $j$ earlier in the path of the laser.

\section{\label{APP_sec:CRF_calc} CRF model analysis}

Consider the CRF master equation given by:
\begin{equation}
\frac{\partial\hat{\rho}}{\partial t}=-i\Omega\left[\hat{S}_{x},\hat{\rho}\right]+\Gamma\left(\hat{S}_{-}\hat{\rho}\hat{S}_{+}-\frac{1}{2}\{\hat{S}_{+}\hat{S}_{-},\hat{\rho}\}\right)
\label{eq:master_eq_orig}
\end{equation}
Eq.~(\ref{eq:master_eq_orig}) can be rewritten in the following form:
\begin{align}
\frac{\partial\hat{\rho}}{\partial t} & =\Gamma\left(\hat{O}\hat{\rho}\hat{O}^{\dagger}-\frac{1}{2}\{\hat{O}^{\dagger}\hat{O},\hat{\rho}\}\right),
\label{eq:Rho_terms_Op}
\end{align}
where
\begin{align}
\hat{O}=\hat{S}_{-}+i\beta\frac{N}{2} & =\hat{S}_{x}-i\hat{S}_{y}+i\beta\frac{N}{2},
\label{eq:O_oper}
\end{align}
where $\beta$ is the order parameter and is given by $\beta=\frac{\Omega}{\Omega_{c}}$
with $\Omega_{c}=\frac{N\Gamma}{2}$. From the factorization of Eq.~(\ref{eq:Rho_terms_Op})
we can infer the properties of the steady-state \cite{barberena_driven-dissipative_2019}.

For $\beta<1$ the system is in the superradiant phase, to analyze
the stable state within the polarized phase, we employ a Holstein-Primakoff
expansion centered around the polarization direction. In mean field
theory, the Bloch vector stabilizes at an angle $\theta$,
where $\sin\theta=\beta$. This leads us to consider
a rotated coordinate system. 
\begin{align}
\hat{S}_{x} & =\hat{S}_{x}^{\prime},\\
\hat{S}_{y} & =\hat{S}'_{y}\cos\theta-\hat{S}'_{z}\sin\theta,\\
\hat{S}_{z} & =\hat{S}'_{z}\cos\theta+\hat{S}'_{y}\sin\theta,
\label{eq:Rot_oper}
\end{align}
such that the Bloch vector is aligned along $\ensuremath{-z'}$. We
then do a Holstein-Primakoff (HP) expansion about this direction,
with the (lowest order) replacements $\ensuremath{\hat{S}_{z}^{\prime}\approx-\frac{N}{2}}$,
$\ensuremath{\hat{S}_{x}^{\prime}\approx\hat{x}\sqrt{\frac{N}{2}}}$,
$\ensuremath{\hat{S}_{y}^{\prime}\approx-\hat{p}\sqrt{\frac{N}{2}}}$
with $\hat{x}=\frac{\hat{a}+\hat{a}^{\dagger}}{\sqrt{2}}$ and $\hat{p}=\frac{\hat{a}-\hat{a}^{\dagger}}{\sqrt{2}i}$
which satisfies $\left[\hat{x},\hat{p}\right]=i$. The operator $\hat{O}$ in Eq.~(\ref{eq:O_oper})
in the HP expansion will have the form:
\begin{align}
\hat{O} & =\hat{S}_{x}-i\hat{S}_{y}+i\beta\frac{N}{2}\approx\hat{x}+i\hat{p}\cos\theta.
\end{align}

For $\beta<1$ the state is in the superradiant phase and will remain
coherent, this implies that the density matrix of the steady-state
satisfies $\hat{\rho}=\left|0_{D}\right\rangle \!\!\left\langle 0_{D}\right|$
and $\hat{O}\left|0_{D}\right\rangle =0$, that is, $\hat{O}$ corresponds
to an annihilation operator with unique dark state $\left|0_{D}\right\rangle $.
To normalize properly $\hat{O}$ we recall that the annihilation operator
$\hat{a}_{D}=k\hat{O}$ satisfies the commutation relation $\left[\hat{a}_{D},\hat{a}_{D}^{\dagger}\right]=1$,
replacing we find that $k^{2}\left[\hat{x}+i\hat{p}\cos\theta,\hat{x}-i\hat{p}\cos\theta\right]=k^{2}\cos\theta\left(-2i^{2}\right)=2k^{2}\cos\theta=1$,
so $k=1/\sqrt{2\cos\theta}$ and the annihilation operator reads as
$\hat{a}_{D}=\left(\hat{x}+i\hat{p}\cos\theta\right)/\sqrt{2\cos\theta}$.
In these terms we can define $\hat{x}_{D}=\frac{\hat{a}_{D}+\hat{a}_{D}^{\dagger}}{\sqrt{2}}$
and $\hat{p}_{D}=\frac{\hat{a}_{D}-\hat{a}_{D}^{\dagger}}{\sqrt{2}i}$,
so we can find that $\hat{x}=\sqrt{\cos\theta}\hat{x}_{D}$ and $\hat{p}=\hat{p}_{D}/\sqrt{\cos\theta}$.
These operators will allow to calculate expected values in the superradiant
regime. 

Our aim is to find expressions that represent the behavior of expected
value of operators of interest at the limit of a large number of particles.
Replacing the operators of Eq.~(\ref{eq:Rot_oper}) expanded in HP we find that:

\begin{align}
\hat{S}_{x} & \approx\hat{x}_{D}\sqrt{\frac{N}{2}\cos\theta},\\
\hat{S}_{y} & \approx\frac{N}{2}\sin\theta-\hat{p}_{D}\sqrt{\frac{N}{2}\cos\theta},\\
\hat{S}_{z} & \approx-\frac{N}{2}\cos\theta-\hat{p}_{D}\sqrt{\frac{N}{2\cos\theta}}\sin\theta.
\end{align}

From the mean field analysis we can obtain that $\left\langle \hat{S}_{x}\right\rangle =0$,
$\left\langle \hat{S}_{y}\right\rangle =\frac{N}{2}\sin\theta$ and $\left\langle \hat{S}_{z}\right\rangle =-\frac{N}{2}\cos\theta$ \cite{barberena_driven-dissipative_2019}.
For the expected value $\left\langle \hat{S}_{+}\hat{S}_{-}\right\rangle $
we can calculate using the HP expansion:
\begin{small}
\begin{align}
\hat{S}_{+} & =\hat{S}_{x}+i\hat{S}_{y}\approx\hat{x}_{D}\sqrt{\frac{N}{2}\cos\theta}+i\left(\frac{N}{2}\sin\theta-\hat{p}_{D}\sqrt{\frac{N}{2}\cos\theta}\right)\nonumber \\
 & =i\frac{N}{2}\sin\theta+\sqrt{\frac{N}{2}\cos\theta}\left(\hat{x}_{D}-i\hat{p}_{D}\right)\\
 & =i\frac{N}{2}\sin\theta+\sqrt{N\cos\theta}\hat{a}_{D}^{\dagger}.\nonumber 
\end{align}
\end{small}

So, $\hat{S}_{+}\hat{S}_{-}$ expanded as a function of the HP operators
and its expected value is:
\begin{footnotesize}
\begin{align}
\hat{S}_{+}\hat{S}_{-} & \approx\left(i\frac{N}{2}\sin\theta\!+\!\sqrt{N\cos\theta}\hat{a}_{D}^{\dagger}\right)\!\left(-i\frac{N}{2}\sin\theta\!+\!\sqrt{N\cos\theta}\hat{a}_{D}\right),\nonumber \\
\left\langle \hat{S}_{+}\hat{S}_{-}\right\rangle  & =\left\langle 0_{D}\left|\hat{S}_{+}\hat{S}_{-}\right|0_{D}\right\rangle \approx i\frac{N}{2}\sin\theta\left(-i\frac{N}{2}\sin\theta\right)\nonumber \\
 & =\left(\frac{N}{2}\sin\theta\right)^{2}.
\end{align}
\end{footnotesize}

We used the fact that $\left\langle \hat{a}_{D}^{\dagger}\right\rangle = \left\langle \hat{a}_{D}\right\rangle = \left\langle \hat{a}_{D}^{\dagger}\hat{a}_{D}\right\rangle = 0$. This formula is valid for $\theta<\pi/2$ where the HP approximation
holds.

Now we estimate the expected value $\left\langle \hat{S}_{+}\hat{S}_{+}\hat{S}_{-}\hat{S}_{-}\right\rangle $
as:

\begin{small}
\begin{align}
\left\langle \hat{S}_{+}\hat{S}_{+}\hat{S}_{-}\hat{S}_{-}\right\rangle  & =\left\langle 0_{D}\left|\hat{S}_{+}\hat{S}_{+}\hat{S}_{-}\hat{S}_{-}\right|0_{D}\right\rangle , \nonumber \\
\hat{S}_{-}\left|0_{D}\right\rangle  & \approx-i\frac{N}{2}\sin\theta\left|0_{D}\right\rangle , \nonumber \\
\left\langle \hat{S}_{+}\hat{S}_{+}\hat{S}_{-}\hat{S}_{-}\right\rangle  & \approx\left(i\frac{N}{2}\sin\theta\right)\left(-i\frac{N}{2}\sin\theta\right)\left\langle 0_{D}\left|\hat{S}_{+}\hat{S}_{-}\right|0_{D}\right\rangle  \nonumber \\
 & =\left(\frac{N}{2}\sin\theta\right)^{4}.
\end{align}
\end{small}

In this case we find that $g_{2}\left(0\right)=\frac{\left\langle \hat{S}_{+}\hat{S}_{+}\hat{S}_{-}\hat{S}_{-}\right\rangle }{\left\langle \hat{S}_{+}\hat{S}_{-}\right\rangle ^{2}}$
at the HP limit is $g_{2}\left(0\right)\approx\frac{\left(\frac{N}{2}\sin\theta\right)^{4}}{\left(\left(\frac{N}{2}\sin\theta\right)^{2}\right)^{2}}=1$ for the superradiant phase. Regarding the HP
approximation, this tends to improve at the limit of large number
of particles such that it converges to the exact result \cite{barberena_driven-dissipative_2019}, so
we can expect that for $\beta<1$ the previous result is valid at
$N\rightarrow\infty$ and in these conditions $g_{2}\left(0\right)=1$.

For $\beta>1$ we are in the normal phase, in that case the steady-state
is highly mixed. In general, the steady-state can be written formally
as:
\begin{equation}
\hat{\rho}_{\text{ss}}=\mathcal{N}\frac{1}{\frac{\hat{S}_{-}}{N/2}+i\beta}\cdot\frac{1}{\frac{\hat{S}_{+}}{N/2}-i\beta}
\end{equation}
where $N$ is a normalization constant and $\beta=2\Omega/\left(\Gamma N\right)$. In the normal phase ($\ensuremath{\beta>1}$), the properties of the steady-state are derived through semiclassical analysis. We parameterize the phase space using angles $\ensuremath{\theta}$ and $\ensuremath{\phi}$ as follows $\left(\hat{S}_{x},\hat{S}_{y},\hat{S}_{z}\right)\rightarrow\frac{N}{2}\left(\sin\theta\cos\phi,\sin\theta\sin\phi,\cos\theta\right)$ so $\hat{S}_{+}\rightarrow\frac{N}{2}\sin\theta\text{e}^{i\phi}$, and operator traces are replaced by integrals over the sphere, with the measure $\ensuremath{\frac{N}{4\pi}\sin\theta\,d\theta\,d\phi}$. The normalization constant $\ensuremath{N}$ is determined through this process, and a justification for the semiclassical approximation is provided in \cite{Barberena_averages}. To obtain $\mathcal{N}$ we recall the normalization condition $\text{Tr}\left(\hat{\rho}_{\text{ss}}\right)=1$, so we require to satisfy:
\begin{small}
\begin{align}
1\! & =\!\mathcal{N}\!\int_{0}^{\pi}\!\int_{0}^{2\pi}\!\frac{1}{\left(\sin\theta e^{-i\phi}\!+\!i\beta\right)\!\left(\sin\theta e^{i\phi}\!-\!i\beta\right)}\frac{N}{4\pi}\sin\theta d\phi d\theta.
\label{eq:int_norm}
\end{align}
\end{small}
This integral can be done using calculus of residues around the unit circle $z=e^{i\phi}$ in the counterclockwise direction and we must be aware that $\sin\theta/\beta<1$ in the normal phase, which will be relevant for determining the suitable residues inside the integration curve. In particular, using this substitution in (\ref{eq:int_norm}) leads to:
\begin{small}
\begin{align}
1 &= -\mathcal{N}\frac{N}{4\pi\beta}\int_{0}^{\pi}\left(\oint_{\left|z\right|=1}\frac{dz}{\left(z-i\frac{\sin\theta}{\beta}\right)\left(z-i\frac{\beta}{\sin\theta}\right)}\right)d\theta.
\label{eq:int_norm_complex}
\end{align}
\end{small}
In (\ref{eq:int_norm_complex}) we can identify that the residue $z_{\text{res,1}}=i\frac{\sin\theta}{\beta}$ belongs to the interior of the integration curve whereas $z_{\text{res,2}}=i\frac{\beta}{\sin\theta}$ does not belong because for all $\theta$ we have $\sin\theta<\beta$ in the normal phase. By the Cauchy's integral formula we deduce that:
\begin{align}
1\!&=\!\mathcal{N}\frac{N}{2}\int_{0}^{\pi}\!\frac{\sin\theta}{\beta^{2}-\sin^{2}\theta}d\theta.
\label{eq:int_norm_real}
\end{align}

From here we find that $\mathcal{N}=\frac{\sqrt{\beta^{2}-1}}{N\arctan\left(\left(\sqrt{\beta^{2}-1}\right)^{-1}\right)}$.

The estimation of the expected values can be performed using a similar integration process
such that $\hat{O}=\hat{O}\left(\theta,\phi\right)$ and $\eta=\sqrt{\beta^{2}-1}$
as:

\begin{small}
\begin{align}
\left\langle \hat{O}\right\rangle  & \approx\frac{\eta}{4\pi\arctan\left(1/\eta\right)}\!\int_{0}^{\pi}\!\int_{0}^{2\pi}\!\frac{\hat{O}\left(\theta,\phi\right)\sin\theta d\phi d\theta}{\left(\sin\theta e^{-i\phi}\!+\!i\beta\right)\!\left(\sin\theta e^{i\phi}\!-\!i\beta\right)}.
\end{align}
\label{eq:Expected_O}
\end{small}

For example in the case of $\left\langle \hat{S}_{x}^{2}\right\rangle $
we can observe that $\hat{S}_{x}^{2}\rightarrow\left(\frac{N}{2}\right)^{2}\sin^{2}\theta\cos^{2}\phi$
so in the normal phase we have that:

\begin{small}
\begin{align}
\left\langle \hat{S}_{x}^{2}\right\rangle  & \approx\frac{\eta}{4\pi\arctan\left(1/\eta\right)}\!\int_{0}^{\pi}\!\int_{0}^{2\pi}\!\!\frac{\left(\frac{N}{2}\right)^{2}\sin^{3}\theta\cos^{2}\phi d\phi d\theta}{\left(\sin\theta e^{-i\phi}\!+\!i\beta\right)\!\left(\sin\theta e^{i\phi}\!-\!i\beta\right)}\nonumber \\
 & =\left(\frac{N}{2}\right)^{2}\frac{\eta}{3\beta^{2}\arctan\left(1/\eta\right)}.
\end{align} 
\end{small}

In particular for calculating $g_{2}\left(0\right)$ we require to estimate $\left\langle \hat{S}_{+}\hat{S}_{+}\hat{S}_{-}\hat{S}_{-}\right\rangle$. Given that we are estimating this quantity at the limit of large $N$ then we expand the operator in the semiclassical approximation at the leading order of $N$. This imply that $\hat{S}_{+}\hat{S}_{+}\hat{S}_{-}\hat{S}_{-}=\hat{S}_{+}^{2}\hat{S}_{-}^{2}\rightarrow\left(\frac{N}{2}\sin\theta\text{e}^{i\phi}\right)^{2}\left(\frac{N}{2}\sin\theta\text{e}^{-i\phi}\right)^{2}+O\left(N^{3}\right)\approx\left(\frac{N}{2}\right)^{4}\sin^{4}\theta$ for large $N$. In this case the mathematical expression for finding $g_{2}\left(0\right)$ at the leading order is given by:

\begin{small}
\begin{align}
m_{n} & =\frac{\eta}{4\pi\arctan\left(1/\eta\right)}\!\int_{0}^{\pi}\!\int_{0}^{2\pi}\!\!\frac{\left(\frac{N}{2}\right)^{n}\sin^{n+1}\theta d\phi d\theta}{\left(\sin\theta e^{-i\phi}\!+\!i\beta\right)\left(\sin\theta e^{i\phi}\!-\!i\beta\right)},\\
g_{2}\left(0\right) & \approx\frac{m_{4}}{m_{2}^{2}}\nonumber \\
 & \approx\left(\frac{\beta^{4}}{\eta}\arctan\left(\frac{1}{\eta}\right)-\beta^{2}-\frac{2}{3}\right)\left(\beta^{2}\arctan\left(\frac{1}{\eta}\right)-\eta\right)^{2}\nonumber \\
 & \times\eta\arctan\left(\frac{1}{\eta}\right).
\end{align}
\end{small}

The remanent integrals used to estimate the quantities of interest in Eqs.~\ref{eq:Sx_Dicke}-\ref{eq:g2_Dicke} are the following:

\begin{small}
\begin{align}
\left\langle \hat{S}_{y}\right\rangle  & \approx\frac{\eta}{4\pi\arctan\left(\frac{1}{\eta}\right)}\!\int_{0}^{\pi}\!\!\int_{0}^{2\pi}\!\!\!\!\frac{\frac{N}{2}\sin^{2}\theta\sin\phi d\phi d\theta}{\left(\sin\theta e^{-i\phi}\!+\!i\beta\right)\!\left(\sin\theta e^{i\phi}\!-\!i\beta\right)}\nonumber \\
 & =\frac{N}{2}\left(\beta-\frac{\eta}{\beta\arctan\left(1/\eta\right)}\right),\\
\left\langle \hat{S}_{+}\hat{S}_{-}\right\rangle  & \approx\left\langle \hat{S}_{x}^{2}+\hat{S}_{y}^{2}\right\rangle\nonumber  \\
 & =\frac{\eta}{4\pi\arctan\left(\frac{1}{\eta}\right)}\!\int_{0}^{\pi}\!\!\int_{0}^{2\pi}\!\!\!\!\frac{\left(\frac{N}{2}\right)^{2}\sin^{3}\theta d\phi d\theta}{\left(\sin\theta e^{-i\phi}\!+\!i\beta\right)\!\left(\sin\theta e^{i\phi}\!-\!i\beta\right)}\nonumber \\
 & =\frac{N^{2}}{4}\left(\beta^{2}-\frac{\eta}{\arctan\left(\frac{1}{\eta}\right)}\right).
\end{align}
\end{small}

\section{\label{APP_sec:analytical_model}Modified CRF model}

The main steady-state features of the dipolar model can be recovered from a much simpler model, which can be solved analytically, the Cooperative Resonance Fluorescence model with one-axis twisting, dephasing, and spontaneous emission. Even though  this model has a bistable solution in the superradiant phase. For our chosen initial condition $|g\rangle^{\otimes N}$, we find that the system reaches a steady state very similar to the dipolar model. To see this, first we obtain the mean field steady state, which can be derived from the equations of motion of this model, given by:
\begin{align}
    \langle{\dot{S}_z}\rangle &= -\Omega \langle{{S}_y}\rangle  - \gamma_s \left(\frac{N}{2}+\langle S_z\rangle \right) - \Gamma_{\rm D} (\langle{{S}_x}\rangle^2 +\langle{{S}_y}\rangle^2) , \nonumber \\
    \langle{\dot{S}_y}\rangle &=  {\Omega} \langle{{S}_z}\rangle + {\Gamma_{\rm D}} \langle{{S}_z}\rangle \langle{S_y}\rangle - \frac{(\gamma_s+4\gamma_d)}{2} \langle S_y\rangle \nonumber \\
    &+ 2\chi \langle S_x\rangle \langle S_z\rangle, \nonumber\\
    \langle{\dot{S}_x}\rangle &=  {\Gamma_{\rm D}} \langle{{S}_z}\rangle \langle{S_x}\rangle - \frac{(\gamma_s+4\gamma_d)}{2} \langle S_x\rangle - 2\chi \langle S_y\rangle \langle S_z\rangle. \label{eq:DDM_SpEm_Deph_MF_eqn_S_xyz}
\end{align}
where, as usual, we have used $\langle {\sigma}_j^\alpha {\sigma}_k^\beta \rangle = \langle {\sigma}_j^\alpha \rangle \langle{\sigma}_k^\beta \rangle $ for $j\neq k$ and $\langle {\sigma}_k^\alpha {\sigma}_k^\beta \rangle = 2i\varepsilon_{\alpha \beta \gamma }\langle {\sigma}_k^\gamma \rangle $. For the cooperative emission and one-axis twisting (OAT) terms, we have $\langle {\sigma}_j^\alpha {\sigma}_k^\beta \rangle = \langle {\sigma}_j^\alpha \rangle \langle{\sigma}_k^\beta \rangle $ for all $j,k$ \cite{carmichael_analytical_1980}. 
We can see from the equations above that for $\chi=0$, $\langle{\dot{S}_x}\rangle =  {\Gamma_{\rm D}} \langle{{S}_z}\rangle \langle{S_x}\rangle - \frac{(\gamma_s+4\gamma_d)}{2} \langle S_x\rangle \Rightarrow \langle{\dot{S}_x (t)}\rangle = \langle{S_x}(0)\rangle \int dt \exp\left[ {\Gamma_{\rm D}} \langle{{S}_z}(t)\rangle  - \frac{(\gamma_s+4\gamma_d)}{2} \right] $. Hence, for an initial state with $\langle \hat{S}_x (0) \rangle = 0$ and $\chi=0$, the value of $\langle \hat{S}_x (t) \rangle$ remains zero for the full dynamics of the system. This is because the laser drive is along $\hat{S}_x$, so it commutes with and does not alter $\hat{S}_x$. The other terms are dissipative and they destroy the coherence, $\hat{S}_x$. However, the OAT ($\chi\neq 0$) term causes shearing of the collective Bloch vector about the $z$-axis, which leads to $\hat{S}_x \neq 0$. 

For simplicity, we will here on use the notation $z\equiv\langle{{S}_z}\rangle/(N/2)$ and we define new variables $(r,\phi)$ to express the collective coherence as $\langle S_+ \rangle/N = r e^{i\phi} $, such that $r = \sqrt{\langle{{S}_x}\rangle^2 + \langle{{S}_y}\rangle^2}/N$ is the contrast and $\phi$ is the phase in the $XY$-plane of the collective Bloch sphere \cite{tucker_single-particle_2020}. Then, the MF equations can be expressed in terms of $(z,r,\phi)$ as
\begin{align}
  \dot{z}= & -2\Omega r \sin\phi - \gamma_s (1+z) - 2\Gamma_{\rm D} N r^2,  \label{eq:zdot} \\
  \dot{r}= & \frac{z}{2}(\Omega \sin\phi +\Gamma_{\rm D} N r) - \frac{(\gamma_s+4\gamma_d)}{2}r, \label{eq:rdot}\\
  \dot{\phi}= & \frac{z}{2r}\Omega\cos\phi + \frac{\chi N z}{2} . \label{eq:phidot}
\end{align} 
The steady state can be obtained by setting $\dot{z}=\dot{r}=\dot{\phi}=0$. Then, we multiply $4r/z$ to Eq.~(\ref{eq:rdot}) to get
\begin{align}
     -2\Omega r \sin\phi - 2\Gamma_{\rm D} N r^2 = -2(\gamma_s + 4 \gamma_d) \frac{r^2}{z} \label{eq:rdot2}
\end{align}
We plug Eq.~(\ref{eq:rdot2}) into Eq.~(\ref{eq:zdot}) to get the steady state as
\begin{align}
      &-2(\gamma_s + 4 \gamma_d) \frac{r^2}{z} - \gamma_s (1+z) = 0 \nn \\
      \Rightarrow z &= -\frac{1}{2} \pm \frac{1}{2}\sqrt{1 - 8 \left(1 + \frac{4\gamma_d}{\gamma_s} \right) r^2}.
\end{align}
There is a phase-transition when the term inside the square-root reaches zero and the critical value $r_c$ can be obtained as
\begin{align}\label{eq:rc_zc}
     r_c = \frac{1}{2\sqrt{2(1+4\gamma_d/\gamma_s)}}, \,\, z_c = -1/2.
\end{align}
Now, we want to obtain the critical driving strength. We can square and rewrite Eq.~(\ref{eq:zdot}) and Eq.~(\ref{eq:phidot}) as
\begin{align}
    4\Omega^2 r^2\sin^2\phi  &= \left[ 2\Gamma_{\rm D} N r^2 + \gamma_s (1+z)^2 \right]^2, \label{eq:sqr1} \\
    4\Omega^2 r^2\cos^2\phi  &= 4 \chi^2 N^2 r^4 \label{eq:sqr2}.
\end{align}
We add Eq.~(\ref{eq:sqr1}) and Eq.~(\ref{eq:sqr2}), and keep upto leading order terms in $N$ to obtain
\begin{align}
 \Omega = \pm\sqrt{(\chi^2+\Gamma_{\rm D}^2) N^2r^2 + \Gamma_{\rm D} \gamma_s N(1+z) } \label{eq:omega_rz}.
\end{align}
The sign of $\Omega$ determines the direction of the drive along $\pm \hat{S}_x$. For $\chi=0$, we get back the same MF equations when we flip the signs of $\Omega$ and $\phi$ simultaneously. Thus, the two drive directions $\pm \hat{S}_x$ are physically equivalent. However, a non-zero $\chi$ breaks this symmetry. Nevertheless, we get qualitatively similar steady states for the negative and positive values of $\Omega$, so we will only consider the positive value for illustration here. Now, we plug $z_c$ and $r_c$ from Eq.~(\ref{eq:rc_zc}) into Eq.~(\ref{eq:omega_rz}) to get the critical drive strength as
\begin{align}\label{eq:mod_CRF_crit_Om}
    {\rm \Omega_C^{Mod-CRF}} = \frac{\Gamma_{\rm D}}{2\sqrt{2}} \sqrt{\frac{1 + (\chi/\gamma_d)^2}{1+(4\gamma_d/\gamma_s)} N^2 + \frac{4\gamma_s}{\Gamma_{\rm D}} N}.
\end{align}

If we consider the case without dephasing and OAT, i.e., $\gamma_d = 0$ and $\chi=0$, we recover $\Omega_{\rm C} \approx  {\Gamma_{\rm D} N}/({2 \sqrt{2}})$ \cite{Walls1978,carmichael_analytical_1980,roberts2023exact}. Thus, the inclusion of spontaneous emission in the CRF model only modifies the critical driving strength by a constant factor without altering its $N$-scaling. As discussed in the main text, accounting for the frequency shifts from dipolar interactions is important. For that we set the dephasing rate to be  $4\gamma_d/\gamma_s = c_d N$ ($c_d$=constant). Then, we take  the large-$N$ limit, to obtain the critical point as
\begin{align}\label{eq:OmC_sqrtN_DDM_SpEm_Deph}
     \Omega_{\rm C} \approx  \frac{\Gamma_{\rm D}}{2 \sqrt{2}} \sqrt{N} \sqrt{\frac{1 + (\chi/\Gamma_{\rm D})^2}{c_d} + \frac{4\gamma_s}{\Gamma_{\rm D}}} .
\end{align}

\begin{figure}[ht!]
\includegraphics[width=\linewidth]{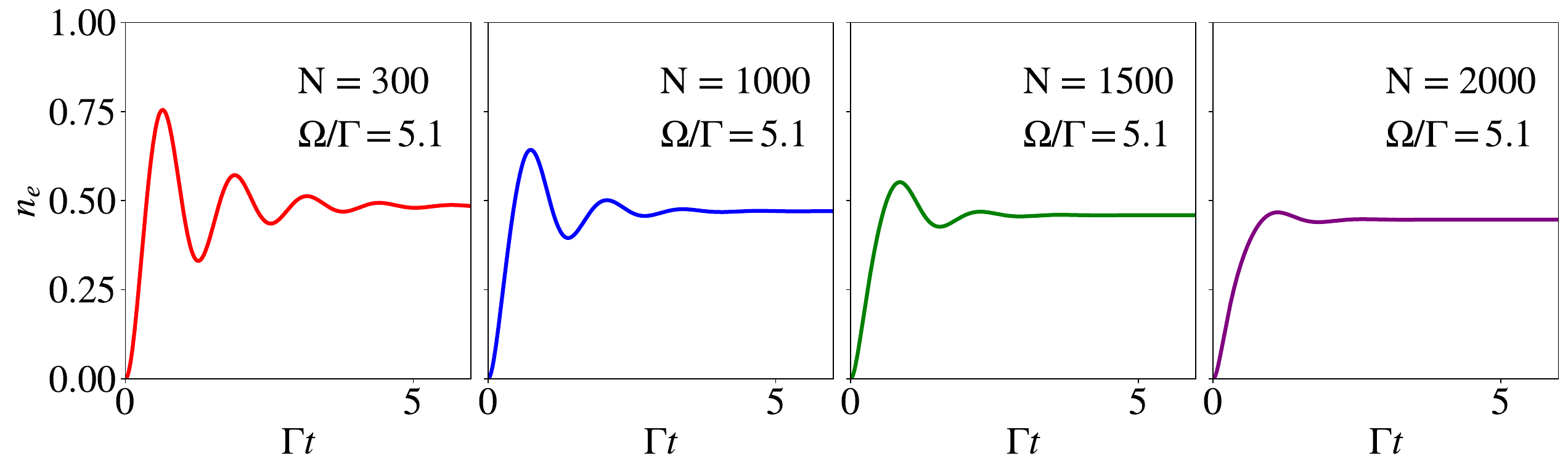}
\caption{(a) Dynamics of the excitation fraction $n_e$ for fixed $\Omega/\Gamma\equiv \Omega/\gamma_s=5.1$ for the CRF model with dephasing ($c_d=0.002$), OAT, and spontaneous emission.}
\label{fig:DDM_SpEm_Deph_ne_vs_t}
\end{figure}

In Fig.~\ref{fig:DDM_SpEm_Deph_ne_vs_t}, we show the dynamics of the excitation fraction $n_e (t)$, using the mean-field equations for the modified CRF model (Eq.~(\ref{eq:DDM_SpEm_Deph_MF_eqn_S_xyz})), with  $c_d=0.002$. This specific value of $c_d$ is not special and we find qualitatively similar results for other values as long as $c_d\ll 1$. We have taken the continuum limit to obtain $\chi = \frac{1}{N}\int d^3r {\rm Re}\tilde{\mathcal{G}} (\vec{r}) \rho(\vec{r}) \approx 0.003$ and $\Gamma_{\rm D} = \frac{1}{N}\int d^3r {\rm Im}\tilde{\mathcal{G}} (\vec{r}) \rho(\vec{r}) \approx 0.002$, where $\rho(\vec{r})$ (Eq.~(\ref{eq:rho_xyz_Gaussian})) is the pencil-shape distribution of the cloud.  Similar to the MF dipolar model, we find that  for fixed $\Omega/\gamma_s$, increasing  $N$, leads to larger dephasing and to a suppression of $n_e$.

\section{\label{APP_sec:incl_laser_light} Contribution of laser light to the intensity}

We consider a coordinate system where the centre of the atomic cloud is the reference point for the origin, $C=(0,0,0)$. All the other vectors/positions are defined with respect to this point. Then, the position of an atom $j$ (with respect to $C$) is given as $\vec{r}_j$. The position of the laser source is $\vec{d}_{\rm L}=(-d,0,0)$. The position of a point on the detector is $\vec{R}=R(\cos\theta,\sin\theta\cos\phi,\sin\theta\sin\phi)$, where $\theta$ is the angle from the $x$-axis and $\phi$ is the azimuthal angle in the $y-z$ plane.

The total electric field and the intensity at a point $\vec{R}$ on the detector are given as
\begin{align}
    \Vec{E}^+_{\rm total}(\vec{R}) = \Vec{E}^+_{\rm drive}(\Vec{R}) + \langle \Vec{E}^+_{\rm atoms}(\Vec{R})\rangle,\\
    {I}_{\rm total}(\vec{R}) = \Vec{E}^-_{\rm total}(\vec{R})\cdot \Vec{E}^+_{\rm total}(\vec{R}),
\end{align}
where the electric field of the driving laser is
\begin{align}
    \Vec{E}^+_{\rm drive}(\Vec{R}) = E(\Vec{R}) \hat{e}_{\rm L}=(\hbar \Omega/|\vec{d}|) \hat{e}_{\rm L} e^{i\vec{k}_{\rm L}\cdot \vec{R}'}
\end{align}
where $\vec{R}'=\vec{R}-\vec{d}_{\rm L} = (R\cos\theta+d,R\sin\theta\cos\phi,R\sin\theta\sin\phi)$ is the distance vector from the laser source to the detector, the laser polarization is $\hat{e}_{\rm L} = \hat{e}_{y}$, and its wavevector is $\vec{k}_{\rm L}=(2\pi/\lambda)\hat{x}$ for our system. The electric field of the atomic dipoles is
\begin{align}
    \Vec{E}^+_{\rm atoms}(\Vec{R}) = \frac{\mu_0\omega_0^3}{3\pi\Gamma c} \sum_j \mathbf{G}(\Vec{R}-\vec{r}_j)\cdot \hat{e}_+ |\vec{d}| \sigma_j^-
\end{align}
where $\vec{R}-\vec{r}_j$ is the distance vector from atom $j$ to the detector.

Then, considering that the detector is very far from the atomic cloud, it is safe to assume that $R\gg |\vec{r}_j|$ and the electric field of the atoms is dominated by the far-field limit terms ($R \gg \lambda \Rightarrow \mathbf{G}\propto 1/(k_0 R)$). Under these assumptions, the total intensity (laser + atomic emission) is obtained as
\begin{align}
    {I}_{\rm total}(\vec{R}) &= \left(\frac{\hbar\Gamma}{k_0 |\vec{R}||\vec{d}|}\right)^2 \bigg[ \frac{\Omega^2}{\Gamma^2} \nn \\
    &+ \frac{3\Omega}{2\sqrt{2}\Gamma} \sum_j \bigg( {\rm Im}\langle\Tilde{\Tilde{\sigma}}_j^-\rangle(1-\sin^2\theta\cos^2\phi) \nn \\ 
    &+ {\rm Re}\langle\Tilde{\Tilde{\sigma}}_j^-\rangle \frac{\sin(2\theta)}{2}\cos\phi \bigg) \nonumber\\
    &+ \frac{9}{16} \sum_{j,k} \frac{(1+\sin^2\theta\cos^2\phi)}{2} \langle\sigma_j^+ \sigma_k^-\rangle e^{i\vec{k}\cdot\vec{r}_{jk}} \bigg],
\end{align}
where $\vec{k}=(2\pi/\lambda) \hat{R}$, and we have defined 
\begin{align}\label{eq:double_tilde_s-}
    \Tilde{\Tilde{\sigma}}_j^- &= \sigma_j^- \exp(i\vec{k}\cdot(\vec{R}-\vec{r}_j)-i\vec{k}_{\rm L}\cdot \vec{R}')\nonumber \\
    &=\sigma_j^- \exp(i\vec{k}\cdot(\vec{R}-\vec{r}_j)-i\vec{k}_{\rm L}\cdot (\vec{R}-\vec{d}_{\rm L}))\nonumber \\ 
    &=\sigma_j^- \exp(-i(\vec{k}\cdot\vec{r}_j-\vec{k}_{\rm L}\cdot \vec{d}_{\rm L})-i(\vec{k}_{\rm L}-\vec{k})\cdot \vec{R})
\end{align}
The spiral basis was defined to absorb the laser-induced phase in the Hamiltonian. Keeping the same definition, in the current coordinate system, the spiral basis is defined as $\tilde{\sigma}_j^- = i\sigma_j^- e^{-i\vec{k}_{\rm L}\cdot(\vec{r}_j-\vec{d}_{\rm L})}$, where the extra phase factor of $i$ comes from the overlap of the circular polarization of the transition with the laser polarization. So we can rewrite Eq.~(\ref{eq:double_tilde_s-}) in terms of the spiral basis operator as
\begin{align}
    \Tilde{\Tilde{\sigma}}_j^- &= -i \tilde{\sigma}_j^- \exp(-i(\vec{k}-\vec{k}_{\rm L})\cdot(\vec{r}_j-\vec{R}))
\end{align}
In the forward direction $\vec{k}=\vec{k}_{\rm L}$, the phases cancel, and we have $\Tilde{\Tilde{\sigma}}_j^- = -i \tilde{\sigma}_j^- $. For simplicity, we assume $(\vec{k}-\vec{k}_{\rm L})\cdot \vec{R} \approx 2 \pi$ as it is difficult to determine its exact value experimentally. Since we are averaging over $\vec{k}$ to obtain the final numerical values, we expect that changing the value of $|\vec{R}|$ will not change the result significantly. 

For simplicity, when including the contribution of the probe light in estimating $\tilde{g}_2(0)$, we have neglected the term accounting for the interference between the probe and the atomic electric fields.

\section{\label{APP_sec:numerical_methods}Numerical Approximations}

\subsection{\label{sec:MF_Cumulant}Mean-field methods}

\begin{figure}[ht!]
\includegraphics[scale=0.13]{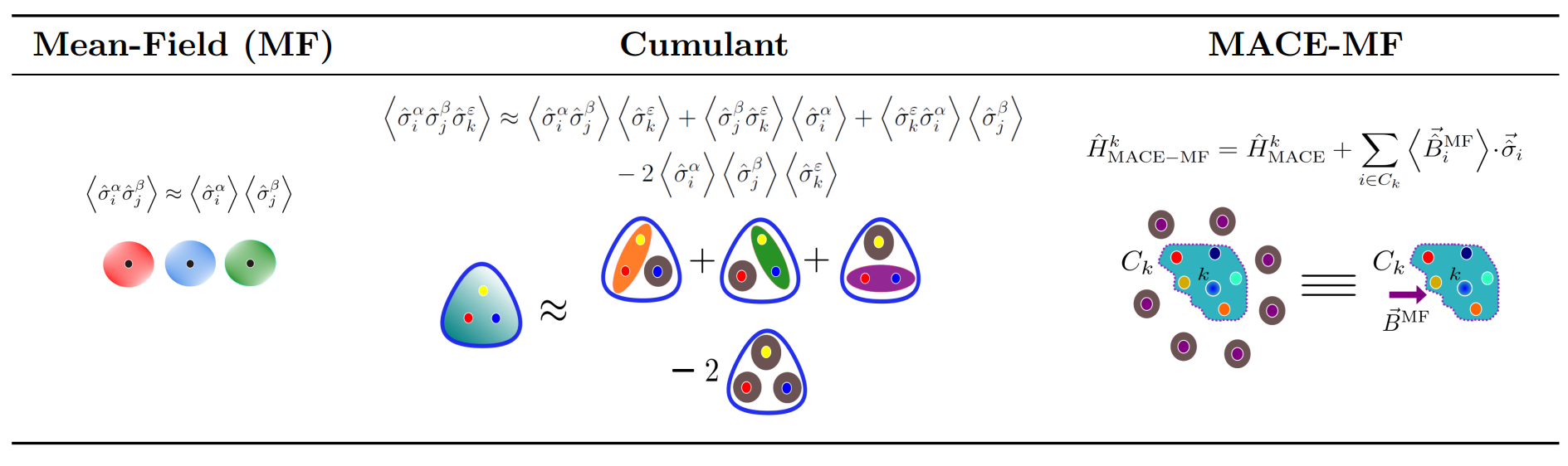}
\caption{Numerical approximations. Mean-Field (Left): Uses non-overlapping clusters, factoring many-body correlations into single-particle values.
Cumulant Approximation (Center): At second order, factors 3-body correlations into two-body and single-body correlations.
MACE-MF (Right): Treats a cluster and its surrounding particles as a single cluster influenced by an external time-dependent magnetic field.}.
\label{fig:num_methods}

\end{figure}

We use mean-field theory to obtain the dynamics and steady-state of the dipolar model. Under mean-field theory, the connected part of the atom-atom correlations is assumed to be negligible. In this case all  correlations can be expressed as products of classical expectation values, as shown in Fig.~\ref{fig:num_methods}. To benchmark the results obtained from mean-field theory, we use the cumulant method, which goes beyond the mean-field approximation by including two-point connected correlations (see Fig.~\ref{fig:num_methods} for an intuitive picture). The MF equations of motion for the dipolar model are given in  Appendix \ref{APP_sec:MF_dipolar_eqns}.

\subsection{\label{APP_sec:MACE-MF}MACE-MF Method}

We introduce an improved  cluster technique which we call  MACE-MF (Moving Average Cluster Expansion + Mean-Field). The method combines dynamics considering both local clusters and mean-field interactions with external particles. This method is based on a cluster approximation, called MACE \cite{Hazzard},  developed to deal with dipolar interactions in dilute arrays in the past. It  exactly solves for the dynamics inside clusters which are not rigid but instead adjusted depending of the observable in consideration. For single point observables $\hat{\sigma}_k$, it chooses a cluster that contains the strongest coupled particles to the $k$-particle in consideration. It disregards the influence of  particles outside  the cluster. In fact, at a cluster size of 1, MACE essentially reduces to single-particle dynamics.

To account for  interaction effects with  the  particles outside a  given cluster, we include them  via a mean-field approach. In this way  the dipole-dipole interaction with the outside particles is mapped  as an effective time-dependent magnetic field acting upon the atoms in the cluster  as shown in Fig.~\ref{fig:num_methods}. So by combining the  exact dynamics within the cluster  with the mean-field contributions we aim to improve the description of the non-equilibrium dynamics of the  system. 

The clusters are chosen and adapted to optimize the single or two particle  observables in consideration by including in the cluster the particles that feature the strongest coupling constants to  the particle or particles the observable is acting on.  

We  generalize it to also include the effect of the external particles to each cluster via the mean-field approximation. 
This means that the interaction of a given cluster with the surrounding particles is reduced to an effective  magnetic field acting on the particles in the cluster.  We denote this  approach  as MACE-Mean-Field (\text{MACEMF)} . Over all, the method aims to include the  best of both  the  MF and the MACE solutions and improve upon them.   In the limit of one particle per cluster the MACEMF  reduces to the MF approximation, and if we  neglect the MF couplings between  clusters then the  MACEMF  reduces to the MACE approximation.

For the $j$ particle  in the array,  the effective Hamiltonian
and Lindblandian associated to the cluster of the particle $j$ read as:

\begin{subequations}
\small
\begin{gather}
H_{{\rm CMF}}^{j} =-\Omega\sum_{i\in C_{j}}\left({\rm e}^{{\rm i}\overrightarrow{k}_{{\rm L}}\cdot\overrightarrow{r}_{i}}\left(e_{{\rm L}}\cdot e_{0}^{*}\right)\sigma_{i}^{+}+{\text{h.c.}}\right)\\
\begin{aligned}
&-\sum_{i,k\in C_{j}}\mathcal{R}_{ik}\sigma_{i}^{+}\sigma_{k}^{-}-\sum_{\substack{i\in C_{j},\\
k\notin C_{j}
}
}\mathcal{R}_{ik}\left(\sigma_{i}^{+}\left\langle \sigma_{k}^{-}\right\rangle +{\text{h.c.}}\right),
\end{aligned}\nonumber{}\\
\label{eq:LinblandianFull}
L\left[\rho\right] =\sum_{i,k}\mathcal{I}_{ik}\left(2\sigma_{k}^{-}\rho\sigma_{i}^{+}-\left\{ \sigma_{i}^{+}\sigma_{k}^{-},\rho\right\} \right)\\
+\sum_{\substack{i\in C_{j},\\
k\notin C_{j}
}
}\mathcal{I}_{ik}\left(\left\langle \sigma_{k}^{-}\right\rangle \left[\rho_{j},\sigma_{i}^{+}\right]-\left\langle \sigma_{k}^{+}\right\rangle \left[\rho_{j},\sigma_{i}^{-}\right]\right)\nonumber{}.
\end{gather}
\label{eq:Dipolar_eq}
\end{subequations} where $C_{j}$ denotes the set of particles that are contained in the $j$-particle cluster  and $\rho_{j}$ is the density matrix that describes this cluster. 

The cluster $C_j$ is selected by choosing the particles that have the strongest interactions with  the $j$-particle ( largest  $|\mathcal{G}_{jk}|$). 

Identifying the second part of the lindblandian \ref{eq:LinblandianFull} as an effective element of the hamiltonian we can rewrite:

\begin{subequations}
\small
\begin{gather}
H_{{\rm CMF}}^{j} =\!-\!\!\sum_{i\in C_{j}}\!\left(\!\left(\!\Omega{\rm e}^{{\rm i}\overrightarrow{k}_{{\rm L}}\cdot\overrightarrow{r}_{k}}\left(e_{{\rm L}}\!\cdot\! e_{0}^{*}\right)+\!\!\sum_{\substack{k\notin C_{j}}
}\!\mathcal{R}_{ik}\left\langle \sigma_{k}^{-}\!\!\right\rangle \right)\!\sigma_{i}^{+}\!+\!\text{h.c.}\!\!\right)\label{eq:S3afi}\\
\begin{aligned}
&-\sum_{i,k\in C_{j}}\mathcal{R}_{ik}\sigma_{i}^{+}\sigma_{k}^{-}-{\rm i}\sum_{\substack{i\in C_{j},\\
k\notin C_{j}
}
}\mathcal{I}_{ik}\left(\left\langle \sigma_{k}^{-}\right\rangle \sigma_{i}^{+}-\left\langle \sigma_{k}^{+}\right\rangle \sigma_{i}^{-}\right),
\end{aligned}\nonumber{}\\
L\left[\rho\right] =\sum_{i,k\in C_{j}}\mathcal{I}_{ik}\left(2\sigma_{k}^{-}\rho_{j}\sigma_{i}^{+}-\left\{ \sigma_{i}^{+}\sigma_{k}^{-},\rho_{j}\right\} \right).
\end{gather}
\label{eq:Dipolar_eq2}
\end{subequations}

Rearranging further the hamiltonian \ref{eq:S3afi} we obtain that: 

{\small{}
\begin{align}
H_{{\rm CMF}}^{j} & =\!-\!\!\sum_{i\in C_{j}}\!\left(\!\left(\!\Omega{\rm e}^{{\rm i}\overrightarrow{k}_{{\rm L}}\cdot\overrightarrow{r}_{k}}\left(e_{{\rm L}}\!\cdot\!e_{0}^{*}\right)+\!\!\sum_{\substack{k\notin C_{j}}
}\!\mathcal{G}_{ik}\left\langle \sigma_{k}^{-}\!\right\rangle \!\right)\!\sigma_{i}^{+}\!+\!\text{h.c.}\!\!\right)\\
 & -\sum_{i,k\in C_{j}}\mathcal{R}_{ik}\sigma_{i}^{+}\sigma_{k}^{-}\nonumber 
\end{align}
}{\small\par}

For estimating second-order correlations of the form $\left<\sigma_j^{\alpha}\sigma_k^{\beta} \right>$, where $\alpha,\beta=x,y,z$ and $j\neq k$, we consider the following cases (recall that by default $i\in C_i$ for all $i$):

{\small{}
\begin{align}
\left\langle \sigma_{j}^{\alpha}\sigma_{k}^{\beta}\right\rangle  & \approx\begin{cases}
\frac{\left\langle \sigma_{j}^{\alpha}\sigma_{k}^{\beta}\right\rangle _{C_{j}}+\left\langle \sigma_{j}^{\alpha}\sigma_{k}^{\beta}\right\rangle _{C_{k}}}{2} & ,\text{if \ensuremath{k\in C_{j}} and \ensuremath{j\in C_{k}}},\\
\frac{\left\langle \sigma_{j}^{\alpha}\sigma_{k}^{\beta}\right\rangle _{C_{j}}+\left\langle \sigma_{j}^{\alpha}\vphantom{\sigma_{k}^{\beta}}\right\rangle \left\langle \vphantom{\sigma_{k}^{\beta}}\sigma_{k}^{\beta}\right\rangle }{2} & ,\text{if \ensuremath{k\in C_{j}} and \ensuremath{j\notin C_{k}}},\\
\frac{\left\langle \sigma_{j}^{\alpha}\vphantom{\sigma_{k}^{\beta}}\right\rangle \left\langle \vphantom{\sigma_{k}^{\beta}}\sigma_{k}^{\beta}\right\rangle +\left\langle \sigma_{j}^{\alpha}\sigma_{k}^{\beta}\right\rangle _{C_{k}}}{2} & ,\text{if \ensuremath{k\notin C_{j}} and \ensuremath{j\in C_{k}}},\\
\left\langle \vphantom{\left\langle \sigma_{k}^{\beta}\right\rangle }\sigma_{j}^{\alpha}\right\rangle \left\langle \vphantom{\left\langle \sigma_{k}^{\beta}\right\rangle }\sigma_{k}^{\beta}\right\rangle  & ,\text{if \ensuremath{k\notin C_{j}} and \ensuremath{j\notin C_{k}}}
\end{cases}
\label{eq:clust_2nd}
\end{align}
}{\small\par}
Here $\left< \mathcal{O} \right>_{C_j}=\text{Tr}(\rho_j\mathcal{O})$, which represents the expected value of the operator $\mathcal{O}$ respect the cluster $C_j$.
Similar  decomposition  can be used  to estimate higher-order correlations.

\section{\label{APP_sec:validity_MF}Validity of Mean-field theory}

In this section, we benchmark the steady-state properties of the MF dipolar model by comparing with the results of beyond mean-field methods such as MACE-MF, exact diagonalization (ED), and the cumulant approximation. We average the data over $\sim 10$ realizations for the MF, MACE-MF, and ED results.

\begin{figure}[ht!]
\includegraphics[width=0.45\linewidth]
{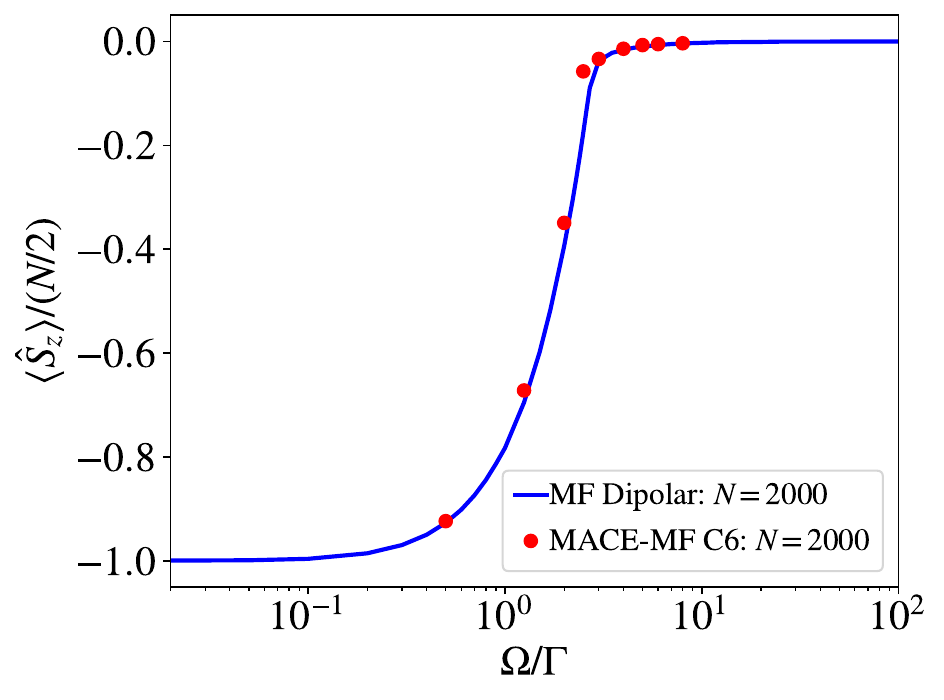}
\includegraphics[width=0.45\linewidth]
{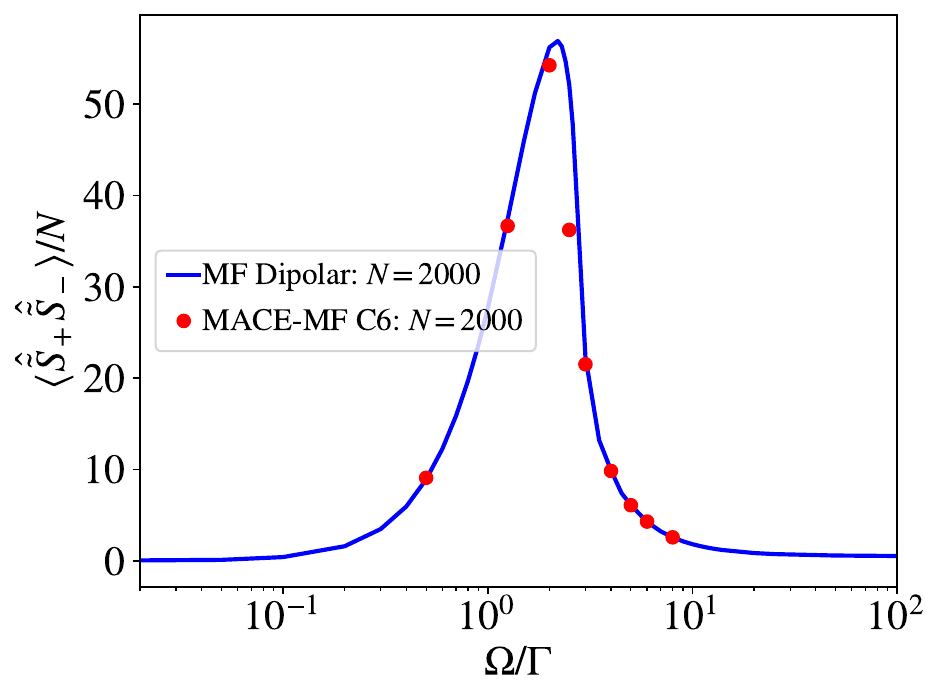}
\caption{Comparing MACE-MF C6, i.e., a cluster-size of 6 atoms, with MF in the steady-state for $N=2000$}
\label{fig:MACE-MF_vs_MF_SS_Sz_Int}
\end{figure}

The MACE-MF includes short-range beyond-MF correlations by treating them exactly (using ED) within a given cluster size. In Fig.~\ref{fig:MACE-MF_vs_MF_SS_Sz_Int}, we show that the MF steady-state atomic inversion and forward intensity for $N=2000$ agree extremely well with those obtained from MACE-MF with cluster size 6, across a range of $\Omega/\Gamma$ spanning the different dynamical phases. We find that the MACE-MF results do not change appreciably as we increase cluster size beyond 6. The agreement between MF and MACE-MF suggests that the effect of short-range interactions is minimal in the steady-state and gets washed out at these densities. At higher densities, the inter-atomic distances would be shorter and $1/r^3$ interactions would dominate, causing MF to break down.

\begin{figure}[ht!]
\includegraphics[width=0.45\linewidth]
{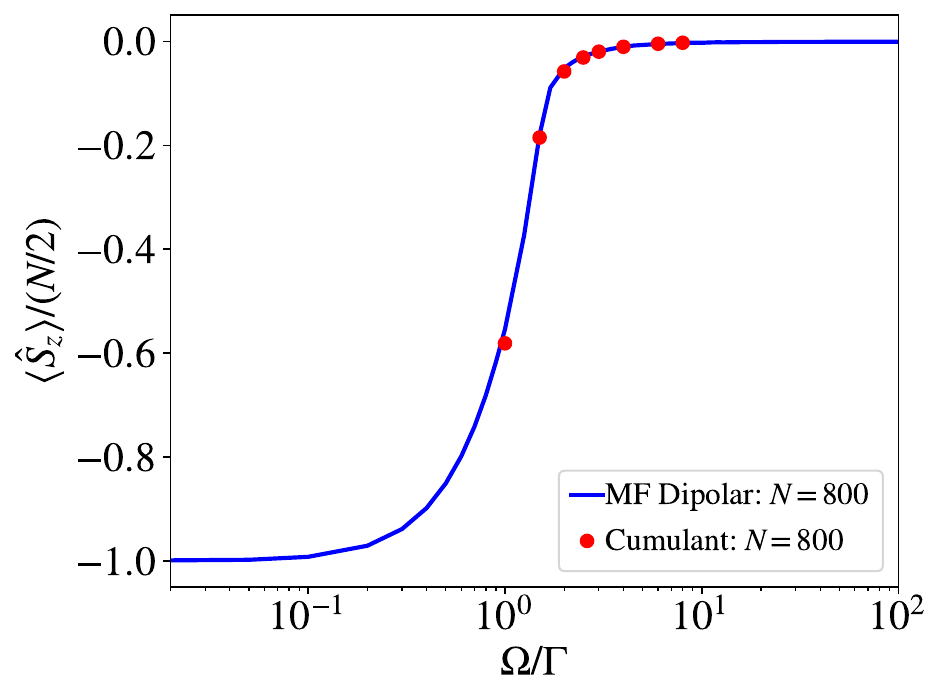}
\includegraphics[width=0.45\linewidth]
{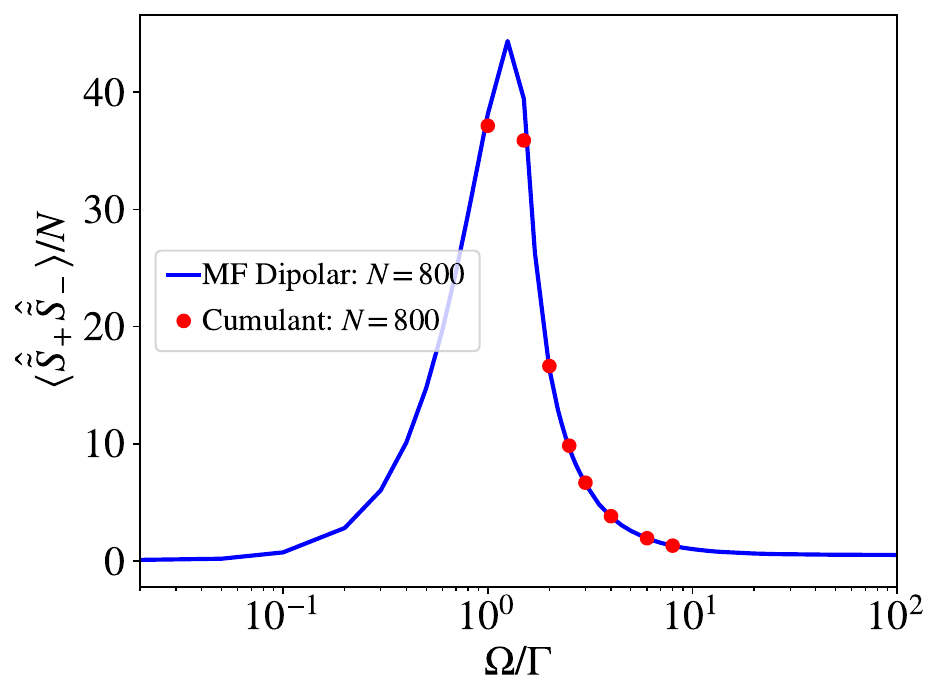}
\caption{Comparing Cumulant (single realization) with MF (averaged over 10 realizations) in the steady-state for $N=800$}
\label{fig:Cum_vs_MF_SS_Sz_Int}
\end{figure}

The cumulant method includes long-range beyond-MF two-point correlations. In Fig.~\ref{fig:Cum_vs_MF_SS_Sz_Int}, we show that the MF steady-state atomic inversion and forward intensity for $N=800$ agree extremely well with those obtained from cumulant, across a range of $\Omega/\Gamma$ in the intermediate driving regime, where we expect the effect of correlations to be most apparent, compared to the weak and strong driving regimes. From the excellent agreement between MF and cumulant, we can see that the effect of long-range correlations is negligible in the steady-state of our system. Cumulant numerics with higher densities and larger-$N$ take extremely long runtimes, restricting us up to $N=800$ here. However, as we already see good agreement with MACE-MF at $N=2000$, we do not expect the cumulant results to change at least qualitatively up to $N=2000$.

\begin{figure}[ht!]
\includegraphics[width=0.45\linewidth]
{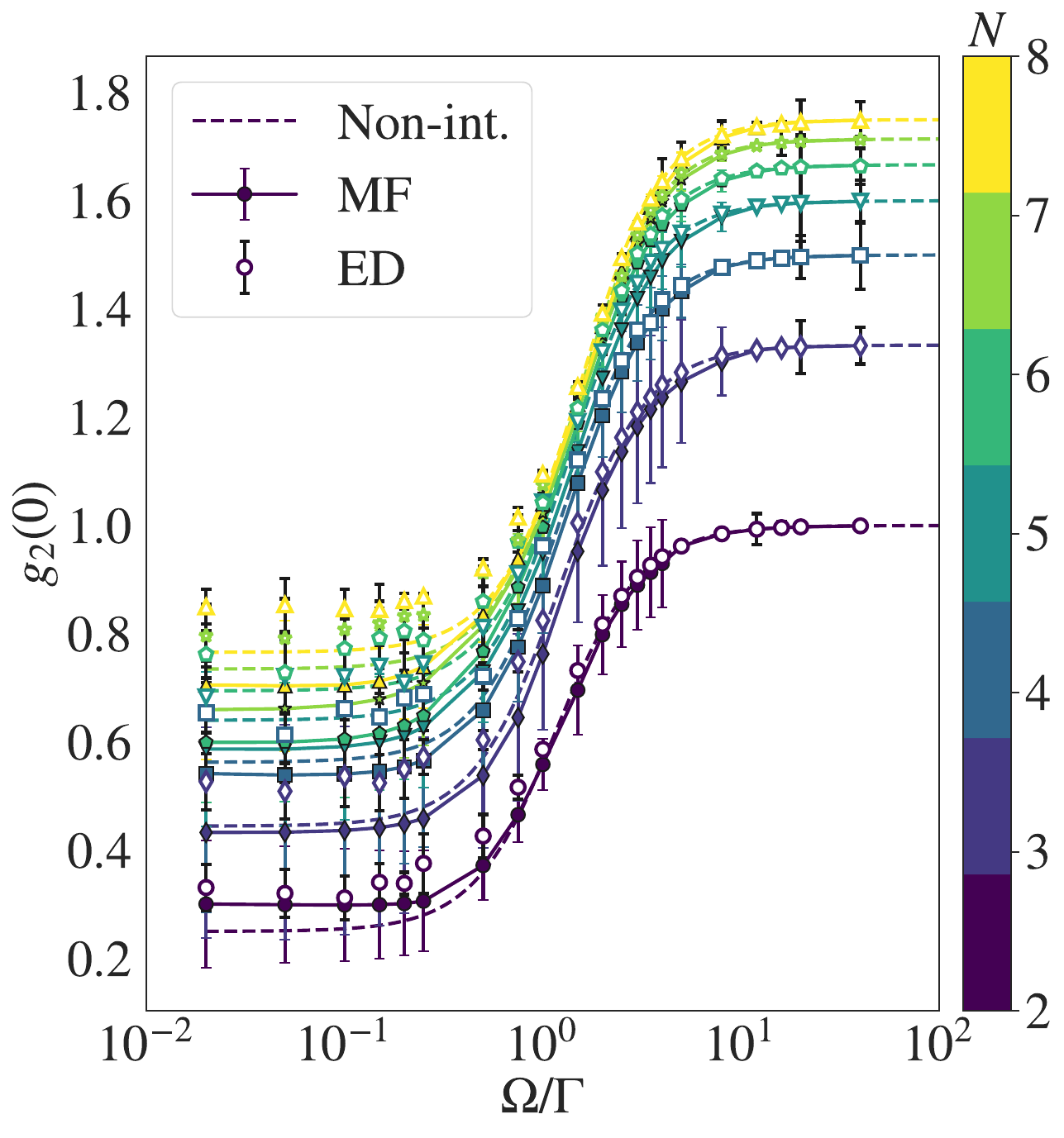}
\caption{Comparing ED with MF for the two-photon correlation function in the steady-state $\tilde{g}_2(0)$}
\label{fig:ED_vs_MF_g2}
\end{figure}

In Fig.~\ref{fig:ED_vs_MF_g2}, we compare the two-photon correlation function $\tilde{g}_2(0)$ in the steady-state of the MF dipolar model with that obtained from ED. We choose small $N$ for this comparison as the Hilbert space grows exponentially with $N$ and ED becomes numerically intractable for larger $N$. We keep the OD fixed while reducing $N$ so that the interaction strength is still similar to the $N=2000$ system. We find that at small $\Omega/\Gamma$, there is a lot of variance in $\tilde{g}_2(0)$ across realizations and within these error bars, there seems to be a fair agreement between ED and MF. At large $\Omega/\Gamma$, the system becomes single-particle-like and error bars get smaller as interactions do not play a role and the change in atomic positions across realizations stops affecting the physics. Hence, there is a much better agreement between ED and MF in this regime. Due to finite $N$ corrections, $\tilde{g}_2(0)$ does not go from exactly 1 to 2, as the drive strength is increased.


\bibliographystyle{apsrev}
\bibliography{main.bbl}

\end{document}